\documentclass[final,3p,times]{elsarticle}

\makeatletter
\def\ps@pprintTitle{%
  \let\@oddhead\@empty
  \let\@evenhead\@empty
  \def\@oddfoot{\hbox to \textwidth{\hfil\@date\hfil}}%
}
\makeatother

\RequirePackage{anyfontsize}

\usepackage{lmodern}
\usepackage{amsmath}

\usepackage{xparse} 

\usepackage{xcolor}
\usepackage{algorithm}
\usepackage{algpseudocode}

\usepackage{booktabs}
\usepackage{graphicx}
\usepackage{subcaption}
\usepackage{cancel}

\usepackage{amssymb}
\usepackage{amsmath}
\usepackage{dashrule}
\usepackage[pdftex, pdfborderstyle={/S/U/W 1}]{hyperref}
\usepackage{stmaryrd}

\usepackage{caption}

\makeatletter
\newcommand{\oset}[2]{%
  {\mathop{#2}\limits^{\vbox to -.5\ex@{\kern-\tw@\ex@
   \hbox{\scriptsize #1}\vss}}}}
\makeatother
\newcommand{\rvec}[1]{\overset{\shortrightarrow}{#1}}
\newcommand{\lvec}[1]{\overset{\shortleftarrow}{#1}}

\newcommand{\bs}[1]{{\boldsymbol{#1}}}
\newcommand{\bft}[1]{{\mathbf{#1}}}
\newcommand{\bb}[1]{{\mathbb{#1}}}

\newcommand{\id}{\mathrm{d}}

\newcommand{\op}[1]{\operatorname{#1}}
\newcommand{\bnabla}{\bs{\nabla}}
\newcommand{\nablav}{\bs{\nabla}}

\newcommand{\parenthesis}[1]{\left({#1}\right)}

\newcommand{\set}[1]{\left\{{#1}\right\}}

\newcommand{\norm}[1]{\left\Vert{#1}\right\Vert}

\newcommand{\abs}[1]{\lvert{#1}\rvert}

\newcommand{\scalar}[2]{\langle{#1},{#2}\rangle}

\usepackage{color}

\definecolor{DarkGreen}{RGB}{0,100,0}

\newcommand{\eref}[1]{Eq.~(\ref{#1})}
\newcommand{\sref}[1]{Sect.~\ref{#1}}
\newcommand{\tref}[1]{Table~\ref{#1}}
\newcommand{\fref}[1]{Figure~\ref{#1}}

\newcommand{\demi}{\tfrac{1}{2}}
\newcommand{\iexp}{\mathrm{e}}
\newcommand{\ci}{\mathrm{i}}
\newcommand{\itr}{\mathsf{T}}
\newcommand{\zerov}{\bft{0}}
\newcommand{\Iv}{\bs{I}}

\newcommand{\idata}{\text{data}}
\newcommand{\imin}{\text{min}}
\newcommand{\imax}{\text{max}}
\newcommand{\idiv}{\text{div}}
\newcommand{\ikf}{\text{f}}
\newcommand{\iin}{\text{in}}

\newcommand{\Nset}{\mathbb N}
\newcommand{\Rset}{\mathbb R}
\newcommand{\Cset}{\mathbb C}
\newcommand{\Zset}{\mathbb Z}
\newcommand{\torus}{\mathbb T}
\newcommand{\dataset}{\mathcal D}
\newcommand{\kset}{\mathcal K}

\newcommand{\esp}{\mathbb E}
\newcommand{\Var}{\bb{V}}

\newcommand{\Normal}{{\mathcal N}}
\newcommand{\CNormal}{\mathcal{CN}}

\newcommand{\Loss}{{\mathcal L}}

\newcommand{\rj}{r}
\newcommand{\uj}{u}
\newcommand{\vj}{v}
\newcommand{\Uj}{U}

\newcommand{\uv}{\bft{\uj}}
\newcommand{\vv}{\bft{\vj}}

\newcommand{\Uv}{\bft{\Uj}}
\newcommand{\rv}{\bs{\rj}}
\newcommand{\xv}{\bft{x}}
\newcommand{\yv}{\bft{y}}
\newcommand{\zv}{\bs{z}}

\newcommand{\pres}{\text{p}}

\newcommand{\density}{\varrho}

\newcommand{\viscosity}{\nu}
\newcommand{\friction}{\alpha}
\newcommand{\TKES}{E}

\newcommand{\pdf}{p}

\newcommand{\PP}{\bb{P}}
\newcommand{\measure}{\mu}

\newcommand{\parav}{\bs{\theta}}
\newcommand{\sj}{s}
\newcommand{\score}{\bft{\sj}}

\newcommand{\denoiser}{\bs{D}}

\newcommand{\ddim}{d}
\newcommand{\ndim}{n}
\newcommand{\Ndim}{N}
\newcommand{\nsample}{S}
\newcommand{\nstep}{T}

\newcommand{\T}{\mathcal{T}}
\newcommand{\NIC}{N_\text{IC}}
\newcommand{\MIC}{M_\text{IC}}

\newcommand{\MS}{M_\text{S}}
\newcommand{\Ntotal}{N_\text{T}}
\newcommand{\Nframes}{N_\text{t}}
\newcommand{\Mframes}{M_\text{t}}
\newcommand{\Pearson}{\rho_\text{P}}

\newcommand{\FFT}[1]{\widehat{#1}}
\newcommand{\unit}[1]{\hat{#1}}

\newcommand{\cjg}[1]{\overline{#1}}
\newcommand{\opt}[1]{{#1}^\star}

\newcommand{\kj}{k}
\newcommand{\kv}{\bs{\kj}}
\newcommand{\hkv}{\unit{\kv}}

\newcommand{\fv}{\bft{f}}

\newcommand{\vortj}{\omega}

\newcommand{\curl}{\bft{curl}}

\newcommand{\Laplace}{\Delta}

\newcommand{\Hilbert}{\mathcal{H}}
\newcommand{\Uilbert}{\mathcal{U}}

\newcommand{\Hdiv}{\Hilbert}
\newcommand{\stokesbasis}{\mathcal{E}}

\newcommand{\Leray}{\mathcal{P}}
\newcommand{\leray}{\bft{P}}

\newcommand{\Stokes}{\mathcal{A}}
\newcommand{\stokes}{\bft{A}}

\newcommand{\Convect}{\mathcal{B}}
\newcommand{\convect}{\bft{B}}

\newcommand{\medium}{\mathcal{O}}
\newcommand{\filtration}{\mathcal{F}}
\newcommand{\projector}{\bs{\pi}}

\newcommand{\ev}{\bs{e}}
\newcommand{\pv}{\bs{p}}
\newcommand{\qv}{\bs{q}}
\newcommand{\ef}{f} 

\newcommand{\Ufwd}{\rvec{\Uj}}
\newcommand{\Ubwd}{\lvec{\Uj}}
\newcommand{\Wienerj}{W}

\newcommand{\Wfwd}{\rvec{\Wienerj}}
\newcommand{\Wbwd}{\lvec{\Wienerj}}

\newcommand{\dift}{t}
\newcommand{\cond}{|}
\newcommand{\cv}{\bft{c}}
\newcommand{\shape}{s}
\newcommand{\noise}{\sigma}
\newcommand{\corrector}{\lvec{D}}

\newcommand{\drift}{A}
\newcommand{\diff}{C}
\newcommand{\Drift}{\mathcal{\drift}}
\newcommand{\Diff}{\mathcal{\diff}}

\newcommand{\NSEinputs}{\opt{\bs{\xi}}}

\newcommand{\Mvanilla}{($\mathcal{V}$)}

\newcommand{\Mmanifold}{(\ref{eq:EDMdivfree})}
\newcommand{\Mdenoiser}{(\ref{eq:denoiser-Leray})}
\newcommand{\Mresidual}{(\ref{eq:PINN-loss})}
\newcommand{\Mautoreg}{(\ref{eq:sample-correction})}
\newcommand{\Mcorrector}{(\ref{eq:PC-Leray})}

\newcommand{\blackdash}{\textcolor{black}{\rule[1.5pt]{0.1cm}{1pt}\;\rule[1.5pt]{0.1cm}{1pt}}}
\newcommand{\reddash}{\textcolor{red}{\rule[1.5pt]{0.1cm}{1pt}\;\rule[1.5pt]{0.1cm}{1pt}}}
\newcommand{\orangedash}{\textcolor{orange}{\rule[1.5pt]{0.1cm}{1pt}\;\rule[1.5pt]{0.1cm}{1pt}}}

\begin{document}
\begin{frontmatter}

\title{Divergence-Free Diffusion Models for Incompressible Fluid Flows}

\author[label1,label2]{Wilfried Genuist}
\author[label1]{\'Eric Savin}
\author[label2]{Filippo Gatti}
\author[label2]{Didier Clouteau}

\affiliation[label1]{organization={DTIS, ONERA, Université Paris-Saclay},
            city={Palaiseau},
            postcode={91120}, 
            country={France}}

\affiliation[label2]{organization={LMPS-Laboratoire de Mécanique Paris-Saclay, Université Paris-Saclay, ENS Paris-Saclay, CentraleSupélec, CNRS},
            city={Gif-sur-Yvette},
            postcode={91190}, 
            country={France}}

\begin{abstract}

Generative diffusion models are extensively used in unsupervised and self-supervised machine learning with the aim to generate new samples from a probability distribution estimated with a set of known samples. They have demonstrated impressive results in replicating dense, real-world contents such as images, musical pieces, or human languages. This work investigates their application to the numerical simulation of incompressible fluid flows, with a view toward incorporating physical constraints such as incompressibility in the probabilistic forecasting framework enabled by generative networks. For that purpose, we explore different conditional, score-based diffusion models where the divergence-free constraint is imposed by the Leray spectral projector, and autoregressive conditioning is aimed at stabilizing the forecasted flow snapshots at distant time horizons. The proposed models are run on a benchmark turbulence problem, namely a Kolmogorov flow, which allows for a fairly detailed analytical and numerical treatment and thus simplifies the evaluation of the numerical methods used to simulate it. Numerical experiments of increasing complexity are performed in order to compare the advantages and limitations of the diffusion models we have implemented and appraise their performances, including: (i) in-distribution assessment over the same time horizons and for similar physical conditions as the ones seen during training; (ii) rollout predictions over time horizons unseen during training; and (iii) out-of-distribution tests for forecasting flows markedly different from those seen during training. In particular, these results illustrate the ability of diffusion models to reproduce the main statistical characteristics of Kolmogorov turbulence in scenarios departing from the ones they were trained on.

\end{abstract}



\begin{keyword}
Computational fluid dynamics \sep Kolmogorov flow \sep Scientific machine learning \sep Diffusion models \sep  Score-based generative models \sep Stochastic differential equations.

\end{keyword}

\end{frontmatter}




\section{Introduction} \label{introduction}
\subsection{Context and aim of the paper}

This research is concerned with the construction of numerical solutions $\smash{\opt{\uv}}$ of the incompressible Navier-Stokes Equations ($\text{NSE}$) given some input parameters $\smash{\opt{\bs{\xi}}}$, namely find $\smash{\opt{\uv}}$ such that:
\begin{equation}\label{eq:NSE0}
\opt{\uv}=\text{NSE}(\NSEinputs)\,.
\end{equation}
The parameters $\smash{\opt{\bs{\xi}}}$ are the initial conditions, the boundary conditions, the viscosity, the external forcing, the friction coefficient if any, \emph{etc}... We must stress at the outset that no rigorous mathematical definition exists so far of what ``solutions'' of Navier-Stokes equations could be, since the problem of ``solving'' the latter remains broadly open \cite{FOI01,TEM01}. The main known results for three-dimensional flows are that weak solutions of finite energy exist for all times but may not be unique. On the other hand, strong solutions of finite enstrophy are unique but exist only for short times; see for example \cite{ROB20} for a recent account on these fundamental topics. Now beyond those theoretical considerations, and for the purpose of obtaining numerical solutions satisfying (or approximating) \eqref{eq:NSE0}, we may have access to different types of data:
\begin{itemize}
\item A set of $\nsample$ input parameters $\{\bs{\xi}_s\}_{s=1}^\nsample$: then \eref{eq:NSE0} may be solved for each $\smash{\bs{\xi}_s}$ by classical numerical methods in computational fluid dynamics (CFD) \cite{CAN88,HIR07,POP00} and reduced basis methods \cite{QUA16}, for example, may be used to obtain $\smash{\opt{\uv}}$ for new input parameters $\smash{\opt{\bs{\xi}}}$.
\item Input parameters are random with a known probability measure $\smash{\measure_\iin}$ on a suitable probability space: then one may construct the pushforward measure of the solutions of \eqref{eq:NSE0} and sample from it, given the input parameters $\smash{\opt{\bs{\xi}}\sim\measure_\iin}$, or resort to collocation or spectral methods such as the polynomial chaos expansion \cite{OLM10}, for example. The formulation of Navier-Stokes equations as stochastic differential equations with random driving forces has also relevance in various engineering and geophysical applications, among others. From a mathematical point of view, this formulation allows to prove properties such as the continuous dependence on initial data or the convergence to equilibrium; see \emph{e.g.} \cite{FLA08} and references therein.
\item A set of $\nsample$ input-output observations $\{\bs{\xi}_s,\uv_s\}_{s=1}^\nsample$: this is a supervised learning framework, as widely adopted in scientific machine learning \cite{BAC24}. It has been addressed by numerous authors using \emph{e.g.}, Gaussian process regression \cite {RAS05}, physics-informed neural networks \cite{LAG98,raissi2019physics}, or operator networks \cite{GNN09,LI21FNO,CHE95,LU21}, to name only a few of the recent advances in this direction. In this regard, CFD has benefited from the important developments in high-performance computing and experimental measurement capabilities achieved in the last decades, resulting in ever increasing volumes of data for data-driven problem solving approaches \cite{DUR19,brunton2020machine}. 
\item A set of $\nsample$ output observations $\{\uv_s\}_{s=1}^\nsample$: this is an unsupervised learning framework, where the objective is to infer $\opt{\uv}$ following the very probability distribution of which these observations are assumed to be known samples. Generative adversarial networks (GANs) \cite{GAN2014}, variational auto-encoders (VAEs) \cite{Kingma2013}, or denoising diffusion probabilistic models (DDPMs) \cite{SOH15,DDPM}, among others, aim at computing and sampling complex, multi-modal and possibly conditional distributions on (very) high-dimensional sets. Unsupervised problems actually needs to be somehow conditioned to be solved, so that they become supervised. The aforementioned generative models for example train themselves after initialization with labeled training data, and are thus coined self-supervised. Extracting lower-dimensional representations from experimental data and large-scale simulations is another way of learning flow features that may be mapped back to the higher-dimensional set of actual flows in order to infer new realizations thereof. Proper orthogonal decomposition, or principal component analysis, was for example introduced in the context of turbulence in \cite{LUM67}.
\end{itemize}
The aim of the paper is to address \eqref{eq:NSE0} from the perspective of self-supervised learning using autoregressive diffusion models. In the continuation of earlier work in \cite{GEN25}, past predictions from generative diffusion networks are used to condition the inference of solutions at future time steps--hence the terminology autoregressive from \emph{e.g.} \cite{benchmarking}. The issue of how to account for physical constraints in diffusion models is further addressed. Incompressible flows with divergence-free velocity fields are more particularly considered.

\subsection{Related works}

The rapid growth in computational power and advances in theoretical and numerical analysis have made accurate fluid mechanics simulation a central challenge in modern science and engineering. For a long time, this challenge has been addressed with CFD, which relies on the numerical solution of the governing Partial Differential Equations (PDE) using finite difference, finite volume, or spectral discretizations, enabling high-fidelity approaches such as Direct Numerical Simulation (DNS) and reduced-cost strategies as Large Eddy Simulation (LES) \cite{CAN88,HIR07,POP00,DNS_base,SCH22}. While these methods provide strong theoretical guarantees and physical interpretability, their high computational cost and limited scalability have driven the rise of physics-based machine learning, or scientific machine learning, as a complementary paradigm \cite{DUR19,brunton2020machine,BUZ23,LIN23,VIN22}. By embedding physical principles, conservation laws, and PDE structure directly into the learning or generative pipeline, scientific machine learning aims to preserve the underlying physics and interpretability while lowering the computational cost of classical solvers through statical patterns, offering new pathways for efficient, data-informed fluid flow prediction and analysis. 

While several existing machine learning approaches aim for a unified modeling frameworks that jointly encode physical constraints and inference, such formulations can become restrictive or difficult to scale for high-dimensional data \cite{raissi2019physics,karniadakis2021physics}. In contrast, diffusion and score-based generative models \cite{SOH15,DDPM,SMSDE} offer a flexible probabilistic alternative, particularly well suited to learning high-dimensional distributions and enabling conditional generation under uncertainty. Their foundations trace back to the interpretation of generative modeling as a thermodynamic process, where data are progressively corrupted by noise and subsequently reconstructed by reversing the diffusion dynamics \cite{SOH15}. This perspective was made practical by denoising diffusion probabilistic models \cite{DDPM}, which introduced a simple discrete-time objective that scales to high-dimensional data while maintaining excellent sample quality \cite{attention,song2020denoising}. In parallel, score-based modeling reframed generation as learning the gradient of the log-density (\emph{i.e.} the score-function) via denoising score-matching \cite{Song2019}, later unified with diffusion models through continuous-time Stochastic Differential Equations (SDE) \cite{SMSDE}. This unification clarified that diffusion models admit both stochastic reverse-time dynamics, grounded in classical results on reverse diffusion \cite{Anderson,haussmann1986time}, and deterministic probability flow Ordinary Differential Equations (ODE), enabling accurate log-likelihood computation and flexible sampling trade-offs \cite{song2020denoising,chen2023probability,karras2022elucidating}. Novel work has substantially improved the efficiency, stability, and controllability of diffusion models. Variational formulations recast diffusion as latent-variable models with learnable noise schedules \cite{kingma2021variational}, while improved parameterizations and noise discretizations have led to better likelihoods and higher-fidelity samples \cite{nichol2021improved}. Conditioning strategies such as classifier-free guidance \cite{ho2022classifierfree} enable conditional generation using a single model, eliminating the need for separately trained classifiers. A broad range of alternative conditional generation techniques has since been explored (see \emph{e.g.} \cite{benchmarking,shysheya2024conditional}). Other works have pursued key design choices, including samplers and noise schedules, and have shown that many empirical improvements arise from theoretical considerations rather than from architectural novelty alone \cite{karras2022elucidating}. Finally, acceleration methods such as implicit sampling \cite{song2020denoising}, progressive distillation \cite{salimans2022progressive}, and consistency-based training \cite{song2023consistency} further reduce inference cost, often at the expense of theoretical guarantees.

Beyond mere sample generation, diffusion models have been extended to sequential and temporal data, where output consistency is crucial. Early adaptations introduced conditional diffusion for time-series forecasting, explicitly modeling uncertainty over missing values \cite{tashiro2021csdi,alcaraz2022diffusion}. Autoregressive diffusion models demonstrated that diffusion can be integrated into sequential forecasting pipelines, capturing complex temporal dependencies while retaining probabilistic outputs \cite{GEN25,benchmarking,shysheya2024conditional,rasul2021autoregressive}. 
In parallel, diffusion models have been increasingly interpreted through the lens of stochastic control and optimal transport. Schr\"odinger bridge formulations connect diffusion training to entropy-regularized transport between data and noise distributions, providing a formal framework for conditional generation, interpolation, and control \cite{de2021diffusion}. These ideas underpin modern approaches to inverse problems and extend beyond temporal modeling. Furthermore, score-based posterior sampling enables efficient reconstruction without retraining, extending diffusion models to ill-posed inverse problems with strong empirical performance \cite{song2021solving,chung2022diffusion}. While recent work has begun to establish robustness guarantees \cite{xu2024provably}, conditional time-series diffusion models often lack explicit enforcement of governing physical laws and flexible architecture. 

The relevance of diffusion models to physics-based systems has therefore attracted growing attention, particularly for complex dynamical systems governed by PDEs. In fluid mechanics and geophysical forecasting, operator learning methods such as Fourier Neural Operators and large-scale graph-based models have demonstrated impressive predictive skills at unprecedented scales \cite{wang2024prediction,pathak2022fourcastnet,keisler2022forecasting,lam2023learning}. However, these approaches are fundamentally deterministic and may struggle to represent uncertainty, multi-modality, and rare events contrary to diffusion models. Recent work has shown that diffusion can generate high-fidelity turbulent flows and perform super-resolution by leveraging spectral decompositions and multi-scale priors \cite{sardar2024spectrally,lienen2023zero}. Autoregressive conditional diffusion models have further enabled long-horizon flow simulation, but stability, error accumulation, and physical consistency remain open challenges \cite{benchmarking,shysheya2024conditional}. To address these issues, several physics-informed diffusion models incorporate PDE constraints either during training or sampling. Approaches such as Pi-Fusion and residual diffusion enforce physics through soft penalties or corrective steps \cite{qiu2024pi,shu2023physics}. While these methods improve physical plausibility, they typically rely on approximate enforcement and do not guarantee exact conservation laws. This limitation echoes broader challenges in physics-informed machine learning, where soft constraints can lead to optimization stiffness and poor generalization \cite{karniadakis2021physics,cuomo2022scientific}. 

A broad and complementary line of research seeks to improve learning efficiency and stability by embedding geometric, structural, and physical priors directly into model architectures. Approaches such as equivariant networks with respect to symmetry groups \cite{cohen2016group}, geometric deep learning frameworks \cite{bronstein2021geometric}, and energy-preserving neural dynamics \cite{greydanus2019hamiltonian} demonstrate that explicitly respecting underlying structures can significantly enhance data efficiency and robustness of the model. Within generative modeling, these ideas have motivated extensions of diffusion processes to manifolds and function spaces, enabling sampling over constrained geometries and infinite-dimensional objects such as PDE solution spaces \cite{de2022riemannian,huang2022riemannian,lim2025score,wang2025fundiff}. In parallel, structure-preserving constraints, most notably divergence-free conditions for incompressible flows, have been enforced through architectural mechanisms such as divergence-free neural fields and PDE-constrained convolutional operators \cite{richter2022neural,jiang2020enforcing}. Despite the diversity of these techniques, existing methods remain fragmented: manifold-based diffusion typically relies on soft or approximate constraints and involves substantial computational overhead due to discretization \cite{de2022riemannian,elhag2023manifold}, while architectural approaches enforce physical structure but are seldom integrated into current diffusion frameworks. To our knowledge, no diffusion-based model unifies autoregressive temporal modeling, principled diffusion design (\emph{e.g.} EDM-style noise and sampling schedules \cite{karras2022elucidating}), and hard architectural enforcement of divergence-free constraints as contemplated in this work. Autoregressive diffusion models capture temporal dependencies without strict physical guarantees \cite{rasul2021autoregressive}, whereas physics-informed diffusion approaches incorporate constraints only approximately through penalty terms \cite{qiu2024pi,shu2023physics}, highlighting a clear gap between geometric rigor, physical fidelity, and scalable generative modeling.

\subsection{Objective and main contributions}

Bridging this gap, the present work develops a theoretically grounded, physics-aware generative framework for incompressible fluid dynamics, focusing on the integration of divergence-free constraints within a diffusion model in order to address \eqref{eq:NSE0}. Rather than pursuing an exhaustive empirical comparison across machine learning network architectures, we build upon a strong diffusion modeling baseline \cite{karras2022elucidating} and compare several design choices for incompressible-flow enforcement. The evaluation prioritizes the statistical accuracy of reconstructed flows under decaying turbulence, as well as the coherence of flow structures. While many generative models and operator learning approaches achieve strong empirical performance in data driven settings, their application to physics constrained problems typically requires problem specific design choices and fine tuning. To avoid confounding such effects, we adopt the canonical two-dimensional Kolmogorov flow, widely used in the machine learning literature (see \emph{e.g.} \cite{shu2023physics,wang2025fundiff,shan2024pird,oommen2024integrating,whittaker2024turbulence,amoros2025guiding} and more) as a controlled and flexible benchmark that is expected to generalize to a broad class of turbulence problem while providing a simple framework, enabling to explore new modeling ideas and further development. 

The main contribution of this work is the construction of a diffusion-based generative framework for fluid dynamics time series that unifies modern diffusion design with strict divergence-free structure. More specifically, we have (i) leveraged state-of-the-art diffusion principles for stability and sampling efficiency, following the EDM framework \cite{karras2022elucidating}; (ii) incorporated autoregressive temporal modeling to enable time-coherent forecasting; and (iii) enforced divergence-free constraints through spectral projection, leveraging the Leray decomposition as a finite-dimensional prior in a relevant functional setting. In particular, the proposed framework is designed for diffusion-based modeling of fluid systems, emphasizing a fast and modular structure that can be readily adapted to a wide range of other problems. We present numerical experiments designed to progressively challenge our approach in terms of accuracy, consistency, and statistical faithfulness. 

\subsection{Outline of the paper}

The paper is structured as follow. In \sref{sec:2}, we provide the building blocks of the model \eqref{eq:NSE0} with applications to two-dimensional Kolmogorov flows. The Leray projection is introduced as well as the Fourier-Galerkin approximation of solutions. The construction of the dataset used to train the various diffusion models developed in this work is also depicted. The EDM generative framework is recalled in \sref{sec:Diffusion}. Unconditional diffusion models are first outlined. Auto-regressive diffusion models are then introduced as conditional versions of the former. Based on this formulation, several divergence-free approaches are proposed in \sref{sec:EDMincompressible}, encompassing both hard and soft constrained methods. They draw inspiration from a range of established methods in physically constrained and physics-informed machine learning \cite{raissi2019physics,karras2022elucidating,de2022riemannian,jiang2020enforcing,rochman2025enforcing,jayaramlinearly}. We compute and analyze in \sref{SEC:Results} the performance of the hard and soft constrained methods with various numerical experiments. We discuss the advantages and limitations of their implementation in generative modeling, and further emphasize the value of physically informed generative approaches for accurate statistical turbulence reconstruction. Some conclusions and perspectives are finally drawn in \sref{SEC:conclusion}.

Besides, several intermediate developments are diverted to appendices. Details of the numerical implementation of the Fourier-Galerkin method are gathered in \ref{APP:SNSE_digest}. Details on the implementation of the EDM framework and its training objective are gathered in \ref{APP:EDM_loss}. The correctors of diffusion models implemented to mitigate stiff artifacts in long-horizon rollouts and regulate the energy level in auto-regressive predictions are outlined in \ref{APP:correctors}. The machine learning configuration and corresponding training and sampling hyperparameters are summarized in \ref{APP:ML_params}, while the model design and architecture are described in \ref{APP:model_design}. The construction of divergence-free diffusion processes is elaborated in \ref{APP:div free manifold}. Plots of additional numerical experiments are gathered in \ref{APP:IN_results}, \ref{APP:ROLL_results}, and \ref{APP:OOD}. 


\section{Incompressible Navier-Stokes equations and dataset generation} \label{sec:2}

We first introduce in \sref{sec:incompressible} the incompressible Navier-Stokes equations written as non-linear ordinary differential equations in an \emph{adhoc} functional setting. This is classically done in terms of the Leray projector on the set of divergence-free, finite energy functions, the linear Stokes operator, and the bilinear convection operator \cite{FOI01,TEM01}. Then in \sref{sec:Fourier_incompressible} we consider periodic boundary conditions on the torus and a Fourier-Galerkin spectral method \cite{CAN88} to compute solutions $\smash{\opt{\uv}}$ of \eref{eq:NSE0} for an arbitrary set of parameters $\smash{\opt{\bs{\xi}}}$ in this setting. This allows us to construct in \sref{sec:database_gen} the dataset used to train and test the various generative diffusion models developed in this work.

\subsection{Incompressible Navier-Stokes equations}\label{sec:incompressible}
In the deterministic setting, the incompressible Navier–Stokes equations serve as the fundamental model for viscous fluid flows. 
They are well-posed in two dimensions under standard regularity conditions for which global existence and uniqueness of weak solutions are established. However in three dimensions only global existence of the latter has been proved so far; see \cite{ROB20} and references therein. Let $\ddim$ be the spatial dimension, we study an incompressible flow in a $\ddim$-dimensional bounded medium $\medium\subset\Rset^\ddim$:
\begin{equation} \label{NSE}
    \begin{cases}
    \displaystyle\partial_t \uv + (\uv\cdot\nablav)\uv = -\frac{1}{\density}\nablav\pres+\viscosity\Laplace\uv + \fv \,, \\
    \nablav\cdot\uv = 0 \,, 
    \end{cases}
\end{equation}
where $\uv(\rv,t)$ stands for the velocity field, $\pres(\rv,t)$ stands for the pressure field, $\rv,t\in\medium\times[0,\T]$ for some final time $\T>0$, $\density$ is the fluid density, and $\viscosity$ is its kinematic viscosity. Also $\nablav$ is the gradient vector, $\Laplace=\nablav\cdot\nablav$ is the Laplacian operator, and $\bs{a}\cdot\bs{b}$ is the usual dot product of two vectors $\bs{a}$ and $\bs{b}$. The motion of the fluid is triggered by the initial condition $\uv(\rv,0)=\uv_0(\rv)$ and the density $\fv(\rv,t)$ of the external and/or friction forces acting on $\medium\times[0,\T]$. In our case, friction forces of the form $\fv=-\friction\uv$ can be seen as a crude representation of three-dimensional effects on two-dimensional flows, for example air friction on a thin layer of soap, or layer-layer friction in stratified fluids; see \emph{e.g.} \cite{boffetta2010evidence,BOF12}. In geophysical fluid dynamics, they are referred to as Ekman damping, or Ekman drag \cite{HOP87}, which will be the model considered here. Next we introduce the Leray projector $\Leray$ formally defined by \cite{FOI01,TEM01}:
\begin{equation}\label{eq:Leray}
\Leray\vv= \vv-\nablav\Laplace^{-1}(\nablav\cdot\vv)\,,
\end{equation}
where $\vv\in L^2(\medium,\Rset^\ddim)$, say, the set of $\Rset^\ddim$-valued square-integrable functions on $\medium$. It has the properties $\Leray\circ\Leray=\Leray^2=\Leray$ (hence it is indeed a projector); $\nablav\cdot\Leray\vv=0$; $\Leray\vv=\vv$ whenever $\nablav\cdot\vv=0$; and $\Leray\nablav\phi=0$ for any scalar potential $\phi$, in particular $\Leray\nablav\pres=0$. Thus, it is the orthogonal projection of functions in $L^2(\medium,\Rset^\ddim)$ onto the subset $\Hdiv$ of such functions which are in addition divergence-free (or solenoidal). The functional set $\Hdiv$ also embeds boundary conditions on $\partial\medium$, such as a vanishing velocity on an aerodynamic profile in external flows, or walls in internal flows. Applying $\Leray$ to the Navier-Stokes equations \eqref{NSE} yields \cite{FOI01}:
\begin{equation}\label{abstractNSE}
\begin{cases}
    \displaystyle\frac{\id\uv}{\id t} + \nu \Stokes\uv + \Convect(\uv) = \Leray\fv\,, \\
    \uv(0) = \uv_0\in \Hdiv\,,
    \end{cases}
\end{equation}
where we have introduced the linear Stokes operator $\Stokes$ defined by:
\begin{equation*}
\Stokes\uv = \Leray(-\Laplace\uv)\,,
\end{equation*}
and the bilinear convective operator $\Convect$ defined by:
\begin{equation*}
\Convect(\uv,\vv)=\Leray((\uv\cdot\nablav)\vv)\,,\quad\Convect(\uv)=\Convect(\uv,\uv)\,.
\end{equation*}
As a result of the Leray projector and its properties, the pressure gradient is no longer needed in \eqref{abstractNSE}, and we have $\Leray\uv=\uv$ since $\uv$ is divergence-free. We also assume that $\smash{\uv_0}$ is, by default, divergence-free. Once the velocity field $\uv$ is known from \eref{abstractNSE}, one may recover the pressure $\pres$ from a Poisson equation but solving it is out of the range of this paper. With reference to \eref{eq:NSE0}, the parameters $\NSEinputs$ in \eref{abstractNSE} are the initial condition $\uv_0$, the driving load $\fv$, and the kinematic viscosity $\viscosity$, while the boundary conditions are enforced by the choice of $\Hdiv$. Since our objectives are mainly to focus on the enforcement of the divergence-free physical constraint in generative diffusion models, we choose to solve \eqref{abstractNSE} for periodic boundary conditions. This setting allows us to use a Fourier-Galerkin spectral method, of which numerical analysis is well established \cite{CAN88}. It is also advantageous to study the symmetries and conservation properties of the solutions of \eref{abstractNSE} \cite{FRI95}. The Fourier-Galerkin approach is outlined in the next section.

\subsection{Fourier-Galerkin approximation}\label{sec:Fourier_incompressible}
One thus considers periodic boundary conditions in the space variable $\rv\in\medium=\smash{[0,L]^\ddim}$, for some spatial period $L>0$, such that:
\begin{equation*}
\uv(\rv+\bs{n} L,t)=\uv(\rv,t)\,,
\end{equation*}
for all times $t\geq 0$, all $\rv\in\medium$, and all signed integers $\bs{n}=(n_1,\dots n_\ddim)\in\Zset^\ddim$. Then \eref{abstractNSE} is reduced to a finite-dimensional system through a Galerkin projection most conveniently carried out in the Fourier basis \cite{CAN88}, which leverages the periodicity and spectral structure of \eqref{abstractNSE}. A suitable orthonormal basis of $L^2(\medium,\Rset^d)$ is given by the Fourier exponentials $\stokesbasis_\ddim := \{\ef_\kv(\rv)\}_{\kv \in \Zset^\ddim}$ (paired with Fourier coefficients):
\begin{equation}\label{eq:FourierModes}
    \ef_\kv(\rv)=\iexp^{\ci\frac{2\pi}{L}\kv\cdot\rv}\,.
\end{equation}
Then the real-valued, zero-mean velocity field $\uv$ admits the Fourier expansion:
\begin{equation*}
\uv(\rv,t) =\sum_{\kv \in \Zset^\ddim_*}\uv_\kv(t)\ef_\kv(\rv)\,,
\end{equation*}
where its Fourier coefficients $\uv_\kv\in\Cset^d$ shall satisfy $\kv\cdot\uv_\kv=0$ (divergence-free condition) and $\cjg{\uv_\kv}=\uv_{-\kv}$ (real-valuedness of $\uv$, $\cjg{\uv_\kv}$ standing for the complex conjugate of $\uv_\kv$); also $\smash{\Zset^\ddim_*}=\Zset^d\setminus\{\zerov\}$ (no mean flow). The finite dimensional approximation limits the number of Fourier modes $\kv$ to the following subset:
\begin{equation*}
\kset_\Ndim = \left\{\kv \in \Zset^\ddim_*;\, \abs{\kv}_\infty=\max_{1\leq i\leq\ddim}\abs{\kj_i}\leq\Ndim\right\}\,,
\end{equation*}
where $\Ndim$ stands for the cut-off wavenumber fixed below the Nyquist limit to prevent aliasing. Given some projector $\smash{\projector_\Ndim}=\smash{\sum_{\kv\in\kset_\Ndim}\ef_\kv\otimes\ef_\kv}$, the corresponding finite-dimensional Fourier approximation of \eref{abstractNSE} onto the set $\smash{\Hdiv_\Ndim} = \smash{\operatorname{span}\{\ef_\kv;\,\kv\in\kset_\Ndim\}}\subset\Hdiv$ reads:
\begin{equation}\label{eq:FourierNSE}
\begin{cases}
    \displaystyle\frac{\id\uv_\Ndim}{\id t} + \nu \stokes_\Ndim\uv_\Ndim  + \convect_\Ndim(\uv_\Ndim) =\fv_\Ndim \,, \\
    \uv_\Ndim(0) = \bs{\xi}_{\Ndim} \in \Hdiv_\Ndim \,,
\end{cases}
\end{equation}
for the finite-dimensional Fourier expansion:
\begin{equation*}
\uv_\Ndim(t) = \projector_\Ndim\uv(t)=\sum_{\kv\in\kset_\Ndim}\uv_\kv(t)\ef_\kv\,.
\end{equation*}
The projected Stokes operator in \eref{eq:FourierNSE} is $\stokes_\Ndim  = \projector_\Ndim\circ\Stokes\circ\projector_\Ndim$, the projected convection operator is $\convect_\Ndim(\cdot) =\projector_\Ndim \circ \Convect(\projector_\Ndim\cdot,\projector_\Ndim\cdot)$, the projected initial condition is $\uv_{N}(0)=\projector_\Ndim\uv_0$, and the projected forcing is $\fv_\Ndim = \projector_\Ndim\fv$. One may notice that in the Fourier setting, the Leray projection of some field $\vv$ reads:
\begin{equation}\label{eq:Leray-Fourier}
\Leray\vv = \sum_{\kv \in \Zset^d_*} \leray_\kv \vv_\kv \ef_\kv\,, \quad \leray_\kv =\Iv-\hkv\otimes\hkv\,,
\end{equation}
where $\Iv$ is the identity, $\hkv = \smash{\frac{\kv}{\norm{\kv}}}$ whenever $\kv \neq \zerov$, and $\bs{a}\otimes\bs{b}$ is the usual tensor product of two vectors $\bs{a}$ and $\bs{b}$ such that $(\bs{a}\otimes\bs{b})\bs{c}=(\bs{b}\cdot\bs{c})\bs{a}$ for any vector $\bs{c}$. Applying $\leray_\kv$ to $\smash{\vv_\kv}$ amounts to remove the component of $\smash{\vv_\kv}$ parallel to $\kv$ so that $\kv\cdot(\leray_\kv\vv_\kv)=0$. This formulation enables fast, parallelized tensor products (Leray components $\leray_\kv$ being precomputed) and evades the need to solve a Poisson equation to enforce incompressibility. As a side remark, only half of the spectral modes $\kset_\Ndim$ are explicitly evolved (according to the lexicographic ordering, \emph{i.e.} $\smash{\#\kset_\Ndim} = \smash{\demi((2\Ndim+1)^\ddim-1)}$), while the remaining half are recovered from the Hermitian symmetry $\cjg{\uv_\kv} = \uv_{-\kv}$, a property that holds both physically and numerically in machine-learning implementations. The method and the overall simulation procedure are described in \ref{APP:SNSE_digest}, along with the choices of external parameters. In particular, an explicit Euler scheme is considered to discretize \eref{eq:FourierNSE} in time; see \eref{eq:explicit_Euler}.

\subsection{Random dataset generation with decaying turbulence} \label{sec:database_gen}
  




\begin{figure}
    \centering\includegraphics[width=1.\linewidth]{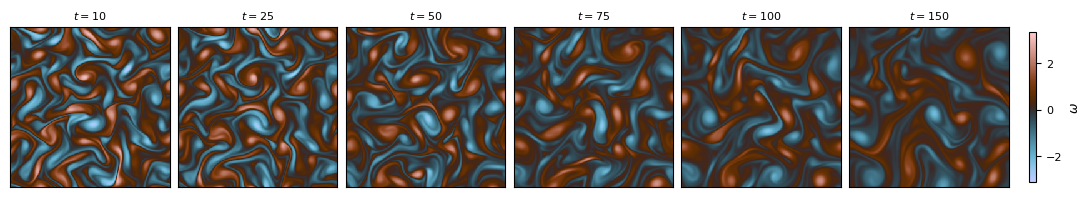}
    \caption{Snapshots of the vorticity field (rollout up to $\smash{j_\imax}=150$).}\label{fig:gen_example_vorticity}
\end{figure}

\begin{figure}
    \centering\includegraphics[width=1.\linewidth]{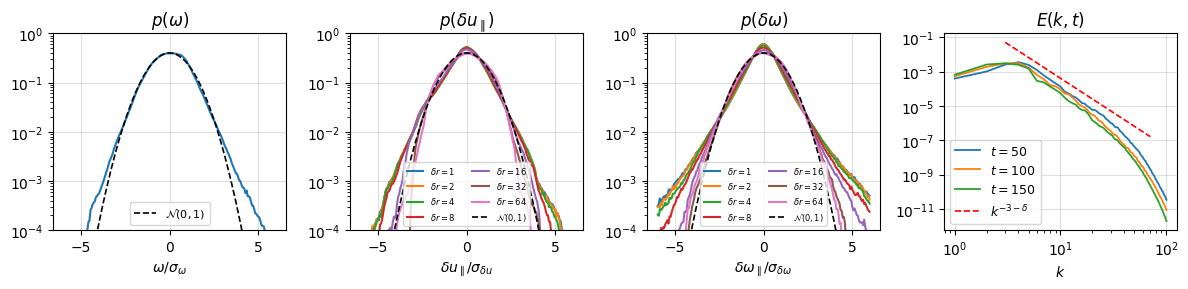}
    \caption{First panel from left to right: empirical density $\pdf_\vortj(z;\delta\rj)$. Second panel: empirical density $\pdf_{\delta\uj_\parallel}(z;\delta\rj)$. Third panel: empirical density $\pdf_{\delta\vortj_\parallel}(z;\delta\rj)$. Fourth panel: TKE spectra $\kj\mapsto\TKES(\kj)$ from a sampled numerical simulation. The plotted metrics are described in \sref{sec:rollout_results}.}\label{fig:gen_example_pdf}
\end{figure}

A key issue in this study is to investigate how different strategies for enforcing incompressibility influence the performance of generative diffusion models. To this end, we construct a randomized training set designed to generate two-dimensional ($\ddim=2$) velocity fields with constant energy across samples. With proper generation, this ensures that the training data are both physically consistent and statistically diverse. The rationale behind this approach is to evaluate the robustness of diffusion models considered in different conditions. In particular, starting from a non-specialized, randomized, and physically viable training set, we assess generalization at three levels:
\begin{itemize}
    \item In-distribution generalization: predictions on new randomized samples drawn from the same distribution as the training set;
    \item Temporal generalization (rollouts): predictions over longer time horizons, under relaxed evaluation criteria;
    \item Out-of-distribution (OOD) generalization: predictions on samples deliberately constructed to deviate from the training set.
\end{itemize}
As a side note, OOD samples are obtained by introducing symmetries and structural constraints that are absent from the randomized training samples, as later elaborated in \sref{sec:OOD_res}. The construction of the training set begins with the sampling of a random vorticity $\vortj$ (a scalar function in two dimensions) according to:
\begin{equation}\label{eq:IC}
    \begin{cases}
    \vortj(\rv) \sim \Normal(0,1)\,, & \rv = (x,y)\,, \\
    \FFT{\vortj}(\kv) \sim \CNormal(0,1)\,, & \kv = (k_x,k_y)\,,
    \end{cases}
\end{equation}
where $\vortj$ is a Gaussian white noise in physical space and $\FFT{\vortj}$ is its Fourier transform, corresponding to a complex Gaussian white noise in Fourier space. Throughout the paper $\Normal(0,\sigma^2)$ stands for the Gaussian distribution on $\Rset$ with zero mean and standard deviation $\sigma$, and $\CNormal(0,\sigma^2)$ is its complex counterpart. To enforce physical consistency with two-dimensional turbulence, we refer to the numerical evidences of \cite{boffetta2010evidence}, where the turbulent kinetic energy (TKE) spectrum $\TKES(\norm{\kv})$ within the inertial range $\smash{\kj_\ikf < \norm{\kv} < \kj_\viscosity}$ of direct enstrophy cascade follows:
$$
\TKES(\norm{\kv}) \propto \norm{\kv}^{-(3+\delta)} \,, 
$$
with a finite correction $\delta \approx 0.35$ observed even in the simulations with the highest resolution. Here $\smash{\ell_\ikf=\frac{\pi}{\kj_\ikf}}$ is the characteristic forcing scale and $\smash{\ell_\viscosity=\frac{\pi}{\kj_\viscosity}}$ is the characteristic dissipation scale. To achieve this, we propose applying a smooth radial filter $\FFT{\vortj}_F(\kv)= F(\kv)\FFT{\vortj}(\kv)$ with:
\begin{equation*}
    F(\kv) =
    \left(\frac{\norm{\kv}}{\kj_\ikf}\right)^{-\frac{\beta}{2}}
    \left(1-\iexp^{-\left(\frac{\norm{\kv}}{\kj_\ikf}\right)^4}\right)
    \iexp^{-\left(\frac{\norm{\kv}}{\kj_\nu}\right)^8} \,,
\end{equation*}
where the power law $\smash{\norm{\kv}^{-\frac{\beta}{2}}}$ enforces the spectral scaling $\abs{F(\kv)}^2 \sim \norm{\kv}^{-\beta}$, while the other terms ensure a clean cutoff outside the range $\smash{[\kj_\ikf,\kj_\nu]}$. Since the vorticity spectrum is related to the velocity spectrum by the Biot-Savart law:
\begin{equation}\label{eq:Biot-Savart}
\FFT{\uv}(\kv) = \frac{\ci\kv^\perp}{\norm{\kv}^2} \FFT{\vortj}(\kv) \,, \quad \kv^\perp=(-\kj_y,\kj_x) \,,
\end{equation}
we have that $ \abs{\FFT{\vortj}(\kv)}^2 \sim \norm{\kv}^2\norm{\smash{\FFT{\uv}(\kv)}}^2$ and a choice of $\beta=1.35$ yields a TKE spectrum $\TKES(\norm{\kv}) \sim\smash{\norm{\kv}^{-\beta-2}} = \norm{\kv}^{-3.35}$, consistent with the reported slope in two-dimensional turbulence \cite{boffetta2010evidence}. While the modes outside the range $\smash{[\kj_\ikf,\kj_\nu]}$ are dissipated, the mean vorticity is removed by enforcing $\FFT{\vortj}_F(\zerov)=0$. With the Biot-Savart law, the real-space velocity field $\uv$ is retrieved and then rescaled to match a desired amplitude $A$:
\begin{equation*}
\uv_{\text{rescaled}} = \frac{A}{\uj_\text{E}} \times \uv \,,
\end{equation*}
where $\smash{\uj_\text{E}^2}=\smash{\langle \norm{\uv}^2 \rangle}=2\TKES$ is the squared averaged velocity field for $\langle \cdot \rangle$ being the spatial average, and $\TKES=\smash{\int_0^{+\infty}\TKES(\kj)\id\kj}$. Besides, we limit ourselves to the externally unforced case considering only a linear friction term $\fv=-\friction\uv$ with friction coefficient $\friction$. This friction force contributes to damp energy at large scales and makes the inverse energy cascade statistically stationary at late times. In Fourier space, its expansion coefficients are $\fv_\kv(t)=-\friction_\kv\uv_\kv(t)$ where:
\begin{equation}\label{eq:friction_k}
\friction_\kv=
\begin{cases}
\friction\quad\text{if}\,\norm{\kv}\leq\kj_\text{d}\,, \\
0\quad\text{otherwise}\,,
\end{cases}
\end{equation}
with $\smash{\kj_\text{d}}=2$ 
targeting large-scale vortices. This damping prevents fast low-frequency energy condensation, below $\smash{k_\text{d}}$ (see \ref{APP:SNSE_digest_PARAMS}). After a short transient, the field relaxes towards an isotropic regime characterized by smooth yet dynamically active vorticity fields. A toy example of this deterministic, decaying turbulence is given in \fref{fig:gen_example_vorticity} together with the structure function distributions in \fref{fig:gen_example_pdf}.

The dataset $\dataset$ is then constituted as follows. We consider $\smash{\NIC}$ samples of the initial condition \eqref{eq:IC} and then derive the corresponding initial velocity fields from the Biot-Savart law \eqref{eq:Biot-Savart} with the filtered vorticity $\smash{\FFT{\vortj}_F^i(\kv)}$, $\smash{1\leq i\leq\NIC}$. Then for each initial condition we run the forward Euler scheme of \eref{eq:explicit_Euler} up to $\smash{\Ntotal}$ time steps. Among them only $\smash{\Nframes}$ regularly spaced time instances are kept in $\dataset$, which is therefore constituted by $\smash{\NIC\times\Nframes}$ frames (snapshots) in $\smash{\Rset^\ndim}$ for each component of the velocity field $\uv=(\uj,\vj)$ in two dimensions, where $n=2\times H\times W$ for $H$ and $W$ being the height and width, respectively, of those snapshots. Thus $\dataset=\smash{\{\uv_i^{(j)},1\leq i\leq\NIC,1\leq j\leq\Nframes\}}$ with $\smash{\uv_i^{(j)}}=\smash{\{\uv_i^{(j)}(\rv_p)\}_{1\leq p\leq H\times W}}$ and:
\begin{equation*}
\uv_i^{(j)}(\rv_p)=\sum_{\kv\in\kset_\Ndim}\uv^{(j)}_{i\kv}\ef_\kv(\rv_p)\,,
\end{equation*}
where $\rv_p$ is the location of the $p$-th pixel in $\medium$, and $\smash{\{\uv_{i\kv}^{(j)},1\leq j\leq\Nframes\}}$ is a sub-sampling of the sequence $\{\uv_{i\kv}^{(j')},1\leq j'\leq\Ntotal\}$ of Fourier coefficients in $\Cset^2$ solving \eref{eq:explicit_Euler} with the $i$-th initial condition \eqref{eq:Biot-Savart}:
\begin{equation*}
\uv_{i\kv}^{(0)}=\frac{\ci\kv^\perp}{\norm{\kv}^2}\FFT{\vortj}_F^i(\kv)\,.
\end{equation*}
Actually sub-sampling is set up past the short transient $\smash{j_\imin}>0$ (which afterwards, is shifted to $0$ by numerical convention) after which the two-dimensional vortical structures of the flow start to fully develop towards an isotropic regime, as mentioned above. This delay is considered as the initial time step for both the physical PDE \eqref{eq:FourierNSE} and its generative counterpart outlined in the next section.



\section{Diffusion models}\label{sec:Diffusion}




In this section the vanilla generative diffusion models used to infer solutions of \eref{eq:FourierNSE} from the dataset constructed in \sref{sec:database_gen} are outlined. The objective is to obtain such solutions for initial conditions $\opt{\uv}_0$ (the parameters $\NSEinputs$) that have not been seen in the training phase. We start by considering unconditional diffusion in \sref{sec:EDM} before we turn to conditional diffusion aimed at guiding the generative processes toward physically sound solutions in \sref{sec:ACDM}. We note that since those generative processes are transient, here and in the subsequent sections the time variable $\dift$ now refers to the diffusion schedule. The associated forward and backward diffusion processes are denoted by $\smash{(\Ufwd_\dift,\dift\geq 0)}$ and $\smash{(\Ubwd_\dift,\dift\geq 0)}$, respectively. They shall not be confused with the velocity field snapshots $\smash{\{\opt{(\uv^{(j)})},1\leq j\leq j_\imax\}}$ that we ultimately wish to infer.


\subsection{EDM generative framework}\label{sec:EDM}
Generative models aim to sample $\opt{\uv}\in\Rset^\ndim$ from an unknown data distribution $\smash{\measure_\idata}$ supported on $\smash{\Rset^\ndim}$ given a dataset $\dataset=\smash{\{\uv_s\}_{s=1}^\nsample}$ of $\nsample$ known samples which are assumed to be distributed according to this very distribution, $\smash{\uv_s}\sim\measure_\idata$. Diffusion models \cite{SOH15,DDPM,Song2019,SMSDE} achieve this by corrupting (noising) a clean data sample $\smash{\uv_0\sim\measure_\idata}$ in $\dataset$ through a forward Markov process $\smash{(\Ufwd_\dift,\dift\geq 0)}$ with known transition probabilities $\smash{\pdf_{\dift|0}(\Ufwd_\dift|\Ufwd_0=\uv_0)}$ at the time steps $\dift$, and then evaluating a $\Rset^\ndim$-valued function $\smash{\uv,\dift\mapsto\score(\uv,\dift}):=\smash{\nablav_\uv \log \pdf_\dift(\Ufwd_\dift=\uv)}$, the so-called Stein's score function, to reverse this process backward in time. However the marginal:
\begin{equation*}
\pdf_t(\uv)=\int_{\Rset^\ndim}\pdf_{\dift|0}(\uv|\uv_0)\measure_\idata(\id\uv_0)
\end{equation*}
is actually unknown since $\measure_\idata$ is unknown, and a parametric neural model $\smash{\uv,t\mapsto\score_\parav(\uv,\dift)}$ is trained instead to predict the score function of the corrupted samples $\smash{\Ufwd_\dift}$. Minimizing the resulting denoising score-matching objective yields a reverse-diffusion dynamics capable of generating new samples from $\smash{\measure_\idata}$, as shown in \cite{SOH15,DDPM,SMSDE,Song2019}. The corresponding training objective reads \cite{vincent2011connection}:
\begin{equation*}
    \opt{\parav}\in\arg\min_\parav
\int_{0}^{T}\!\lambda(t)\;
\esp_{\uv_0\sim\measure_\idata} \esp_{\uv\sim\pdf_{t|0}(\uv|\uv_0)}
\norm{
\score_\parav(\uv,t)
-
\nablav_\uv\log\pdf_{t|0}(\uv|\uv_0)}^2_2\,
\id t \,,
\end{equation*}
to find optimal neural parameters $\opt{\parav}$ such that $\smash{\uv,t\mapsto\score_{\opt{\parav}}(\uv,t)}$ matches $\uv,t\mapsto\score(\uv,t)$. Here $t\mapsto\lambda(t)$ is some weighting function, $T$ is the time horizon of the forward noising process, $\smash{\norm{\bs{a}}_2}=\smash{(\sum_{k=1}^\ndim a_k^2)^{1/2}}$ is the $\ell^2$ norm of $\bs{a}\in\smash{\Rset^\ndim}$, and $\esp$ is the ensemble average, or mathematical expectation. Diffusion models with discrete-time Markov chains were first considered in \cite{SOH15,DDPM,Song2019} and extended to continuous-time Markov processes solving stochastic differential equations (SDEs) in \cite{SMSDE}. \citet{karras2022elucidating} have further expanded this SDE generative framework by introducing a general class of Ornstein–Uhlenbeck (OU) processes to improve the vanilla method of \cite{SOH15,DDPM,Song2019}. In their so-called EDM formulation the forward process $\smash{(\Ufwd,t\geq 0)}$ solves:
\begin{equation}\label{eq:EDM-FSDE}
\id \Ufwd_\dift = \displaystyle\frac{\dot{\shape}(\dift)}{\shape(\dift)} \Ufwd_t \id \dift + \shape(\dift) \sqrt{2\noise(\dift)\dot{\noise}(\dift)} \id \Wfwd_\dift\,,\quad\Ufwd_0\sim\measure_\idata\,,
\end{equation}
with time-varying noise and scale functions $\dift\mapsto\noise(\dift)$ and $\dift\mapsto\shape(\dift)$, respectively, such that $\shape(\dift)>0$, with $\shape(0)=1$\ and $\noise(0)=0$. Also $\smash{(\Wfwd_t,t \geq 0)}$ is a $\ndim$-dimensional Wiener process, and $\dot{\shape}=\frac{\id\shape}{\id t}$, $\dot{\noise}=\frac{\id\noise}{\id t}$. The solution of \eref{eq:EDM-FSDE} thus reads:
\begin{equation}\label{eq:EDMforward}
\Ufwd_\dift = \shape(\dift)\left(\Ufwd_0+\int_0^\dift\sqrt{2 \noise(\tau)\dot{\noise}(\tau)}\id\Wfwd_\tau\right)\,,
\end{equation}
such that the conditioned random vector $\Ufwd_\dift\cond\Ufwd_0$ is Gaussian with conditional expectation and variance given by:
\begin{equation*}
\esp[\smash{\Ufwd_\dift\cond\Ufwd_0}]=\shape(\dift)\Ufwd_0\,,\quad\Var[\smash{\Ufwd_\dift\cond\Ufwd_0}]=(\shape(\dift)\noise(\dift))^2\Iv\,,
\end{equation*}
respectively. Starting from $\smash{\Ubwd_0}=\smash{\Ufwd_T}$, the backward process $\smash{\Ubwd_\dift}=\smash{\Ufwd_{T-\dift}}$ is then \cite{Anderson,haussmann1986time}:
\begin{equation}\label{eq:EDM-BSDE}
\id \Ubwd_\dift = \displaystyle\left(-\frac{\dot{\shape}(\dift)}{\shape(\dift)} \Ubwd_\dift + 2\shape(\dift)^2\noise(\dift)\dot{\noise}(\dift)\score(\Ubwd_\dift,T-\dift)\right)\id \dift + \shape(\dift)\sqrt{2 \noise(\dift)\dot{\noise}(\dift)}\id\Wbwd_\dift\,.
\end{equation}
Running \eref{eq:EDM-BSDE} up to $\dift=T$ yields some frame $\smash{\Ubwd_T\sim\measure_\idata}$ to which the desired new sample $\opt{\uv}\in\Rset^\ndim$ is identified. An equivalent deterministic form (in the sense that their solutions have the same marginal distributions at each time step $\dift$) is the probability-flow ordinary differential equation (PF-ODE) introduced in \cite{SMSDE}:
\begin{equation}\label{eq:EDM-PF-ODE}
\frac{\id\Ubwd_\dift}{\id\dift} = \displaystyle -\frac{\dot{\shape}(\dift)}{\shape(\dift)} \Ubwd_\dift + \shape(\dift)^2\noise(\dift)\dot{\noise}(\dift)\score(\Ubwd_\dift,T-\dift)\,.
\end{equation}
The score function $\score(\uv,t)=\smash{\nablav_\uv\log\pdf_t(\uv)}$ in both \eref{eq:EDM-BSDE} and \eref{eq:EDM-PF-ODE} is known to explode when $\dift\to 0$, as observed in practice for image synthesis in \emph{e.g.} \cite{song2020improved}. There it is proposed to stop the backward process at some earlier time $T-\varepsilon$ for a small $\varepsilon>0$ and consider $\smash{\Ubwd_{T-\varepsilon}\sim\measure_\idata}$ instead. A soft probabilistic truncation has been proposed in \cite{KIM22} to replace this hard thresholding. Here the score function is rather parameterized in terms of the denoiser $\smash{\esp[\Ufwd_0\cond\Ufwd_t]}$ for \eref{eq:EDMforward}, which owing to Tweedie's formula yields \cite{EFR11}: 
\begin{equation*}
\score(\uv,t)=-\frac{\uv-\shape(t)\esp[\Ufwd_0\cond\Ufwd_t=\uv]}{(\shape(t)\noise(t))^2}\,,\quad t>0\,.
\end{equation*}
The PF-ODE \eqref{eq:EDM-PF-ODE} then reads:
\begin{equation}\label{eq:EDM-PF-ODE2}
\frac{\id\Ubwd_\dift}{\id\dift} = \displaystyle-\left(\frac{\dot{\shape}(\dift)}{\shape(\dift)}+ \frac{\dot{\noise}(\dift)}{\noise(\dift)}\right)\Ubwd_\dift + \shape(\dift)\frac{\dot{\noise}(\dift)}{\noise(\dift)}\esp[\Ufwd_0\cond\Ubwd_\dift]\,.
\end{equation}
Following \cite{karras2022elucidating}, the foregoing PF-ODE can be further simplified if a constant scale function $\shape(\dift)=1$ is considered. Indeed using the relation $\id \dift = \id \sigma/\dot{\sigma}$ in \eref{eq:EDM-PF-ODE2}, the latter becomes:
\begin{equation}\label{eq:EDM-PF-ODE3}
\frac{\id\Ubwd_\noise}{\id\noise} = \frac{1}{\noise}\left(-\Ubwd_\noise+\esp[\Ufwd_0\cond\Ubwd_\noise]\right)\,,\quad\noise>0\,.
\end{equation}

Now the denoiser function $\uv\mapsto\esp[\Ufwd_0\cond\uv]$ involved in \eref{eq:EDM-PF-ODE2} or \eref{eq:EDM-PF-ODE3} has yet to be constructed for a selected noise schedule $\noise$. It is approximated by a parametric function $\denoiser_\parav(\uv;\noise)$, typically a U-Net \cite{UNet,UNet3D} of which parameters $\parav$ are optimized by considering the following EDM training loss:
\begin{equation} \label{eq:EDM_loss}
\Loss_\text{EDM}(\parav)=
\esp_{\uv_0\sim\measure_\idata}
\esp_{\noise\sim\pdf(\noise)}
\esp_{\zv\sim\Normal(\zerov,\Iv)}
\big[
\lambda(\noise)\norm{\denoiser_\parav(\uv_0+\noise\zv;\noise)-\uv_0}^2_2 \big] \,.
\end{equation}
Here $\pdf(\noise)$ stands for the probability distribution considered for sampling the noise schedule $\noise$, $\noise\mapsto\lambda(\noise)$ is an associated weighting function. Once the parametric denoiser $\smash{\denoiser_\parav}$ has been trained with optimal parameters $\smash{\opt{\parav}}$, new samples from $\smash{\measure_\idata}$ are generated by discretizing the PF-ODE \eqref{eq:EDM-PF-ODE2} with a classical Euler scheme and an additional Heun’s second-order correction \cite{ASC98}. Indeed this scheme was shown to offer a good tradeoff between truncation error and the required number of evaluations of the trained denoiser $\smash{\denoiser_{\opt{\parav}}}$ in \emph{e.g.} \cite{JOL21}. Starting from a latent random variable $\smash{\Ubwd_0}\sim\Normal(\zerov,(\shape(T)\noise(T))^2\Iv)$, each inference step consists in jointly applying the following backward predictor-corrector iterates over a discrete time schedule $\{\dift_i\}_{0 \leq i \leq \nstep}$: 
\begin{equation}\label{eq:pre-cor}
\begin{cases} 
\Ubwd_i \leftarrow \Ubwd_i + (\dift_{i+1} - \dift_i) \corrector_i \,, \\
\Ubwd_i \leftarrow \Ubwd_i + \displaystyle\demi\left(\dift_{i+1} - \dift_{i}\right)\big(\corrector_i+\corrector_{i+1}\big) \,,
\end{cases}
\end{equation}
where the corrector $\smash{\corrector_i}$ is:
\begin{equation*}
\corrector_i = -\left(\frac{\dot{\shape}_i}{\shape_i}+\frac{\dot{\noise}_i}{\noise_i}\right)\Ubwd_i +\shape_i\frac{\dot{\noise}_i}{\noise_i}\denoiser_{\opt{\parav}}(\Ubwd_i;\noise_i)\,,
\end{equation*}
and $\shape_i=\shape(\dift_i)$, $\dot{\shape}_i=\dot{\shape}(\dift_i)$, $\noise_i=\noise(\dift_i)$, and $\dot{\noise}_i=\dot{\noise}(\dift_i)$. The second-order correction step in \eqref{eq:pre-cor} is applied unless $\noise_{i+1}=0$. Here in accordance with \cite{karras2022elucidating} we choose $\shape_i=1$ and $\dot{\shape}_i=0$, corresponding to $\shape(t)=1$ at all times in the continuous setting. Then we get rid of the discrete time schedule, as well as $\dot{\noise}$, and according to \eref{eq:EDM-PF-ODE3} the predictor-corrector iterates \eqref{eq:pre-cor} rewrite:
\begin{equation}\label{eq:prediction}
\begin{cases} 
\Ubwd_i \leftarrow \Ubwd_i+ (\noise_{i+1} - \noise_i)\corrector_i \,, \\
\Ubwd_i \leftarrow \Ubwd_i + \displaystyle\demi (\noise_{i+1} - \noise_i)\big(\corrector_i+\corrector_{i+1}\big) \,,
\end{cases}
\end{equation}
where again the second-order correction step in \eqref{eq:prediction} is applied unless $\noise_{i+1}=0$, and $\smash{\corrector_i}$ simplifies to:
\begin{equation}\label{eq:corrector}
\corrector_i = \frac{1}{\noise_i}\big(-\Ubwd_i + \denoiser_{\opt{\parav}}(\Ubwd_i; \noise_i)\big)\,.
\end{equation}
Further details on the implementation of the EDM pipeline are given in \ref{APP:EDM_loss}. The backward iterates of \eref{eq:prediction} (or alternatively \eref{eq:pre-cor} if the scale function $\dift\mapsto\shape(\dift)$ is non constant) end up in a sample:
\begin{equation}
\opt{\uv}\equiv\Ubwd_T\sim\measure_{\opt{\parav}}\,,
\end{equation}
where $\smash{\measure_\parav}$ is a parametric approximation of $\smash{\measure_\idata}$ with parameters $\parav$ from the EDM pipeline. The extent to which $\smash{\measure_\parav}$ deviates from $\smash{\measure_\idata}$ in an \emph{adhoc} measure distance (as well as their supports possibly) is out of the scope of this paper, however this question has been addressed recently in \emph{e.g.} \cite{STE25} for neural network-based score estimators; see also references therein.


\subsection{Auto-regressive conditional generation} \label{sec:ACDM}

The foregoing EDM framework can be extended to the conditional generation of a new sample $\opt{\uv}\in\Rset^\ndim$ from a conditional distribution $\measure_\idata(\cdot\cond\cv)$, where $\cv$ stands for the condition imposed to the sampling process \emph{e.g.} observations or labels \cite{ho2022classifierfree,condDDPM}. This is achieved by considering the $\Rset^\ndim$-valued conditional score function $\smash{\uv,\cv,\dift\mapsto\score(\uv,\cv,\dift}):=\smash{\nablav_\uv\log\pdf_\dift(\Ufwd_\dift=\uv\cond\cv)}$, with the conditional probability:
\begin{equation*}
\pdf_t(\uv\cond\cv)=\int_{\Rset^\ndim}\pdf_{\dift|0}(\uv|\uv_0)\measure_\idata(\id\uv_0\cond\cv)\,.
\end{equation*}
The backward SDE used to evolve the backward process $\smash{(\Ubwd_\dift,\dift\geq 0)}$ is the counterpart of \eref{eq:EDM-BSDE} with the above conditional score:
\begin{equation}\label{eq:BSDE-cond}
\id \Ubwd_\dift = \displaystyle\left(-\frac{\dot{\shape}(\dift)}{\shape(\dift)} \Ubwd_\dift + 2\shape(\dift)^2\noise(\dift)\dot{\noise}(\dift)\score(\Ubwd_\dift,\cv,T-\dift)\right)\id \dift + \shape(\dift)\sqrt{2 \noise(\dift)\dot{\noise}(\dift)}\id\Wbwd_\dift\,.
\end{equation}
Running \eref{eq:BSDE-cond} up to $\dift=T$ yields some frame $\smash{\Ubwd_T\sim\measure_\idata}(\cdot\cond\cv)$ to which the desired new sample $\opt{\uv}\in\Rset^\ndim$ conditioned on $\cv$ is identified. The corresponding conditional denoiser function is $\smash{\uv,\cv\mapsto\esp[\Ufwd_0\cond\uv,\cv]}$ and the conditional PF-ODE is the counterpart of \eref{eq:EDM-PF-ODE2}:
\begin{equation*}\label{eq:EDM-PF-ODE2-cond}
\frac{\id\Ubwd_\dift}{\id\dift} = \displaystyle-\left(\frac{\dot{\shape}(\dift)}{\shape(\dift)}+ \frac{\dot{\noise}(\dift)}{\noise(\dift)}\right)\Ubwd_\dift + \shape(\dift)\frac{\dot{\noise}(\dift)}{\noise(\dift)}\esp[\Ufwd_0\cond\Ubwd_\dift,\cv]\,,
\end{equation*}
or alternatively the counterpart of \eref{eq:EDM-PF-ODE3} if the scale function $\shape(\dift)$ is constant:
\begin{equation}\label{eq:EDM-PF-ODE3-cond}
\frac{\id\Ubwd_\noise}{\id\noise} = \frac{1}{\noise}\left(-\Ubwd_\noise+\esp[\Ufwd_0\cond\Ubwd_\noise,\cv]\right)\,,\quad\noise>0\,.
\end{equation}
As for the unconditional case of \sref{sec:EDM}, discrete backward iterates of \eref{eq:EDM-PF-ODE3-cond} of the form \eqref{eq:prediction} end up in a sample:
\begin{equation}
\opt{\uv}\equiv\Ubwd_T\sim\measure_{\opt{\parav}}(\cdot\cond\cv)\,,
\end{equation}
where $\smash{\measure_\parav}(\cdot\cond\cv)$ is now a parametric approximation of $\smash{\measure_\idata(\cdot\cond\cv)}$ with parameters $\parav$ from the conditional EDM pipeline.


Now in view of our objective to generate rollouts over time of fluid flow velocity fields, a relevant conditioning consists in choosing past predictions to forecast future states. Unlike several learning-based simulators that condition each prediction on long temporal sequences and/or additional auxiliary parameters, as for example \cite{shysheya2024conditional,benchmarking,GenCast25}, we employ a strictly autoregressive formulation in which the next state is generated solely from the knowledge of the immediately preceding one. This design is fundamentally justified by the Markov property of SDEs in finite dimensional Hilbert space of the form \eqref{eq:FourierNSE} when the driving force is a Brownian motion \cite[Theorem 3.1]{FLA08}. Owing to this property, conditioning on a single frame is sufficient to encode all information required for physically consistent evolution. Beyond its theoretical grounds, this choice enables efficient next-step generation: the model performs only one forward pass per time step, without maintaining or processing long time histories, and also, enables fast correction mechanisms (see \ref{APP:correctors}). As a result, the autoregressive structure has a low computational cost, allowing the surrogate to run faster than traditional solvers and to support rapid generation although certainly not the best of all \cite{ho2022classifierfree,shysheya2024conditional}. In this sense, the proposed formulation is just a simplification, though a principled and computationally advantageous exploitation of some inherent Markov dynamics. In the remainder of the paper, it is thus implicitly assumed that each future state $\uv^{(j+1)}$ of the velocity field at physical time step $j+1$ is generated by conditioning the EDM pipeline with its immediately preceding state $\uv^{(j)}$ at physical time step $j$ according to the autoregressive setting:
\begin{equation}\label{eq:EDM_autoregressive}
    \opt{(\uv^{(j+1)})} \sim \measure_{\opt{\parav}}
    \big(\,\cdot\,\bigr\cond\opt{(\uv^{(j)})}\big) \,.
\end{equation}
Here we explicitly distinguish the physical time for the evolution of the flow field $\uv$ with steps indexed by $j$, from the diffusion time $\dift$ of the generative process by the backward SDE \eqref{eq:BSDE-cond} (or its PF-ODE counterpart \eqref{eq:EDM-PF-ODE3-cond} in terms of the noise schedule $\noise$). It is implicit that, to generate a complete flow, one must start from the state $\opt{(\uv^{(0)})}$, which is actually provided by the dataset $\dataset$ itself and serves as the only ground-truth input for generating the entire future trajectory, with the model using its own past predictions as conditioning thereafter.

This setting, in turn, affects the vanilla EDM loss $\smash{\Loss_\text{EDM}}$ of \eref{eq:EDM_loss}. It is modified to $\smash{\Loss_\text{EDMc}}$ below so as to train a conditional denoiser $\smash{\denoiser_\parav(\uv,\cv;\noise)}$ from paired samples $\smash{\uv^{(j)},\uv^{(j-1)} \sim \measure_\idata}$ instead of the unconditional training samples $\uv^{(j)}\sim\measure_\idata$ ignoring any time sequencing in the dataset: 
\begin{equation} \label{eq:EDMc_loss}
\Loss_\text{EDMc}(\parav)=
\esp_{\uv^{(j)},\uv^{(j-1)}\sim\measure_\idata}
\esp_{\noise\sim\pdf(\noise)}
\esp_{\zv\sim\Normal(\zerov,\Iv)}
\big[
\lambda(\noise)\norm{\denoiser_\parav(\uv^{(j)}+\noise\zv,\uv^{(j-1)};\noise)-\uv^{(j)}}^2_2 \big] \,.
\end{equation}
This leads to several considerations for the design of the denoiser architecture, extending beyond a simple adaptation of the U-Net variants used in baseline diffusion models. Accordingly, our U-Net variant is constructed with the following key principles in mind:
\begin{itemize}
    \item A conditional architecture enabling controlled conditional generation;
    \item Circular depthwise-separable convolutions enforcing periodic boundary behavior on each field;
    \item A low-$\noise$ refinement branch that improves high-resolution fidelity in the denoising limit \cite{karras2022elucidating};
    \item Conditional modulation using FiLM (Feature-wise Linear Modulation \cite{FiLM17}) parameters derived from the input and the noise schedule $\noise$.
\end{itemize}
The conditioning is done by concatenation of the current and previous states \cite{GEN25}. Further specifications are given in \ref{APP:model_design} to highlight the adaptation of the U-Net architecture. These issues are however not the main scope of this paper. We rather focus on how to extend the generative framework to account for physical constraints, such as incompressibility of the forecasted flow. The consideration of such a physics-based conditioning is addressed in the next section.


\section{Divergence-free constrained diffusion models}\label{sec:EDMincompressible}

Enforcing a divergence-free constraint can be achieved in several ways, ranging from explicit projection operators, \emph{e.g.} the Leray projector $\Leray$ expressed in Fourier space by \eref{eq:Leray-Fourier}, to implicit regularization via divergence-penalizing losses, or by parameterizing the solution space directly through a divergence-free subspace or manifold \cite{FOI01,TEM01}. Each approach comes with its own trade-offs in terms of numerical stability, approximation quality, and compatibility with machine learning architectures. In \sref{sec:DivFreeManifold} the divergence-free condition is enforced by projecting the forward-backward PF-ODE system, while in \sref{sec:soft-div-free} various settings leveraging the network architecture, its training, or the inference process are proposed.

\subsection{Divergence-free manifold}\label{sec:DivFreeManifold}


Many machine learning techniques have been proposed to enforce incompressibility, including denoiser-based corrections, adversarial regularization, or divergence-aware architectures (such as convolutional layers adapted to preserve solenoidal structure); see \emph{e.g.} \cite{richter2022neural,jiang2020enforcing,JIN21,BEN22,MUC25,SCH25}. However, a principled strategy is to constrain the dynamics to evolve strictly within the divergence-free manifold, thereby ensuring physical consistency by construction. Based on \ref{APP:div free manifold}, we formalize the notion of a divergence-free manifold \cite{de2022riemannian,VAL25}, which provides a formal functional setting for incompressible velocity fields and diffusion modeling. Within this framework, OU noising processes of the form \eqref{eq:EDM-FSDE} are naturally well-defined. Since the EDM construction is essentially a finite-dimensional OU process equipped with an optimal noise schedule and preconditioning, it can be interpreted as a refined instance of this general setting. 
For the EDM setup \eqref{eq:EDM-FSDE}-\eqref{eq:EDM-PF-ODE} with the dataset $\dataset$ of frames in $\Rset^\ndim$ which are in addition instances of divergence-free fields projected in a finite-dimensional subset of $\Hdiv$, 
the drift term vanishes as long as $\shape(\dift)=1$ is chosen, and the diffusion coefficient is linear. This reduces the forward SDE–backward PF-ODE system to a simplified projected form as follows:
\begin{equation}\label{eq:EDMdivfree}\tag{$\mathcal{M}$}
\begin{cases}
\id \Ufwd_\dift = \sqrt{2 \noise(\dift)\dot{\noise}(\dift)} \Leray\id \Wfwd_\dift \,, \\
\id \Ubwd_\dift = \displaystyle\frac{\dot{\noise}(\dift)}{\noise(\dift)} \Leray \left(-\Ubwd_\dift + \esp[\Ufwd_0\cond\Ubwd_\dift]\right) \id \dift \,,
\end{cases}
\end{equation}
with $\smash{\Ufwd_0\sim\measure_\idata}$ in $\smash{\Rset^\ndim}$. This corresponds to the following forward sampling:
\begin{equation*}
\Ufwd_\noise = \Ufwd_0 + \noise\Leray\rvec{Z}\,,\quad\rvec{Z}\sim\Normal(\zerov,\Iv_\ndim)\,,
\end{equation*}
where $\smash{\Iv_n}$ is the $\ndim\times\ndim$ identity matrix, and a backward inference starting from $\smash{\Ubwd_0}=\Leray\lvec{Z}$ with $\lvec{Z}\sim\smash{\Normal(\zerov,\noise_\imax^2\Iv_\ndim)}$ and using the trained denoiser projected on the divergence-free manifold:
\begin{equation*}
\Leray\left(\Ufwd_0-\esp[\Ufwd_0\cond\Ubwd_\noise=\uv]\right)\simeq\Ufwd_0-\Leray\denoiser_{\opt{\parav}}(\uv;\noise)\,,
\end{equation*}
which translate to that very Leray projection dynamics during the whole inference process. The extension of this approach to the conditional setting of \eref{eq:EDM_autoregressive} using a conditional denoiser $\smash{\denoiser_{\opt{\parav}}(\uv,\cv;\noise)}$ is straightforward.


\subsection{Soft divergence-free enforcement}\label{sec:soft-div-free}
Other divergence-free methods can be considered and thus compared altogether. Classical machine learning methods contain a number of techniques at their core, designed to address a constraint through optimization. The simplest of them is the training of diffusion models on the divergence-free dataset itself, which serve as the baseline of comparisons (vanilla method). Since diffusion models learn complex data structures, we can evaluate the matching of the divergence-free criteria on prediction made by a properly trained model. Another method that comes to mind is a correction at the model layer \cite{jiang2020enforcing}. First, the Leray projection can be applied at the output layer of the denoiser model $\denoiser_\parav$ of \eref{eq:skipped-denoiser}:
\begin{equation}\label{eq:denoiser-Leray}\tag{$\mathcal{D}$}
\denoiser_{\parav, \idiv}(\uv;\noise) = c_\text{skip}(\noise)\Leray\uv + c_\text{out}(\noise)
\Leray\bs{F}_\parav(c_\text{in}(\noise)\uv,c_\text{noise}(\noise))\,,
\end{equation}
such that the divergence-free condition is enforced at the structural design level, and influences the training phase. Again, the extension of this approach (and the proposed subsequent ones) to a conditional denoiser is straightforward.

Second, working with an autoregressive formulation, the Leray projection can be applied at each frame of the autoregressive prediction \emph{i.e.} samples are projected once the inference is made \cite{rochman2025enforcing}, at a physical time step $j$ in the autoregressive pipeline \eqref{eq:EDM_autoregressive}. With optimal parameters $\smash{\opt{\parav}}$ from the minimization of the loss \eqref{eq:EDMc_loss} the inference at physical time step $j+1$ reads:  
\begin{equation}\label{eq:sample-correction}\tag{SC}
\opt{(\uv^{(j+1)})} \sim \measure_{\opt{\parav}}
    \big(\,\cdot\,\bigr\cond\Leray\opt{(\uv^{(j)})}\big) \,.
\end{equation}
This is typically the approach followed in \cite{MUC25} where the generative process is based on stochastic interpolants. It enforces compliance with physical principles by projecting model outputs onto the manifold defined by the underlying laws \cite{VAL25}.

Another alternative is to exploit the diffusion modeling sampling architecture by applying a Leray projection as an additional corrector step in the predictor-corrector iterates \eqref{eq:prediction}. This correction, which can be seen as the adaptation of \cite{jayaramlinearly,MUC25} within the Leray-projection framework, is aimed towards lower noise levels $\noise$ (coarse solver for close-to-clean samples, \emph{e.g.} annealed correction during backward integration):
\begin{equation}\label{eq:PC-Leray}\tag{PC}
\begin{cases}
\Ubwd_i \leftarrow \Ubwd_i+ (\noise_{i+1} - \noise_i)\corrector_i\,, \\
\Ubwd_i \leftarrow \Ubwd_i + \displaystyle\demi(\noise_{i+1} - \noise_i)\big(\corrector_i+\corrector_{i+1}\big)\,, \\
\Ubwd_i \leftarrow (1-\iexp^{-a\noise_i})\Ubwd_i + \iexp^{-a\noise_i}\Leray\Ubwd_i\,,
\end{cases}
\end{equation}
for some parameter $a>0$ (we fix $a=10$, which was found to be optimal empirically) and $\smash{\corrector_i}$ given by \eref{eq:corrector}. Note that the same step of divergence-free correction is applied to the mean-shift corrector iterates \eqref{eq:mean_shift_corrector} for a fair comparison.

Finally, one can adopt a PINN-like approach \cite{raissi2019physics} by imposing the divergence-free condition through a residual computation during training:
\begin{equation}\label{eq:PINN-loss}\tag{$\Loss$}
\Loss(\parav) = \Loss_\text{EDMc}(\parav) + \lambda_\idiv\Loss_\text{div}(\parav)\,,
\end{equation}
where:
\begin{equation*}
\Loss_\idiv(\parav)=
\esp_{\uv_0\sim\measure_\idata}
\esp_{\noise\sim\pdf(\noise)}
\esp_{\zv\sim\Normal(\zerov,\Iv)}
\big[\norm{(\Iv-\Leray)\denoiser_\parav(\uv_0+\noise\zv,\cv;\noise)}^2_2
\big]\,.
\end{equation*}
$\smash{\lambda_\idiv}$ is a scaling coefficient adaptively calibrated such that the gradient magnitudes of the physical regularization loss $\smash{\Loss_\idiv}$ and the baseline conditional EDM loss $\smash{\Loss_\text{EDMc}}$ are initially comparable:
\begin{equation*}
\lambda_\idiv = \frac{\norm{\nablav_\parav \Loss_\text{EDMc}(\parav)}}{\norm{\nablav_\parav\Loss_\text{div}(\parav)} + \varepsilon} \,,
\end{equation*}
for some matching parameter $0<\varepsilon\ll 1$ (typically $\varepsilon=10^{-7}$ to evade numerical instabilities). This gradient-based normalization ensures that the PINN-like regularization loss contributes meaningfully during the early training phase without dominating the optimization dynamics. 

Each of these methods is represented by its own symbol in \tref{tab:methods}: vanilla model \Mvanilla\ when no divergence-free conditioning is enforced at all (the plain EDM framework of \sref{sec:EDM}), divergence-free diffusion manifold \eqref{eq:EDMdivfree}, denoiser architecture correction \eqref{eq:denoiser-Leray}, sample correction \eqref{eq:sample-correction}, predictor-corrector iterates \eqref{eq:PC-Leray}, and residual computation \eqref{eq:PINN-loss}. Among them, four require an independent training procedure, namely \Mvanilla, \eqref{eq:EDMdivfree}, \eqref{eq:denoiser-Leray}, and \eqref{eq:PINN-loss}. The remaining two evaluations by \eqref{eq:sample-correction} and \eqref{eq:PC-Leray} are derived from the vanilla model ($\mathcal{V}$) once it has been trained. The strength of divergence-free enforcement is determined by whether it is strictly imposed (up to numerical precision) or statistically dependent (e.g., during a training or heuristic sampling phase). Each training method is allocated the same computational budget (\emph{i.e.} the same total number of optimization steps or epochs) \cite{li2019budgeted}, ensuring that all methods are compared under equal resource constraints and can, in principle, converge to their best possible performance under that budget. To maintain fairness, we also use an identical learning-rate scheduler and other hyperparameters (\emph{e.g.} batch size, regularization) across all methods. With this setup, no method is advantaged by allowing more iterations or a more favorable schedule; thus the comparison isolates differences in model architecture, objective, or training algorithm, rather than resources disparity. Further detail on both the training and inference parameters are given in \ref{APP:SNSE_digest_PARAMS}.

\begin{table}[t]
\centering
\begin{tabular}{|c|c|c|c|}
\hline
\textbf{Numerical method} & \textbf{Symbol} & \textbf{Description} & \textbf{Strength} \\ \hline
vanilla  & \Mvanilla & Classical diffusion model. & None \\ \hline
div\_free\_manifold & \Mmanifold & Diffusion model on divergence-free manifold. & \emph{strict} \\ \hline
div\_free\_denoiser & \Mdenoiser & Projection at model output level (architectural). & \emph{strict} \\ \hline
div\_free\_residual & \Mresidual & Residual computation of divergence. & \emph{soft} \\ \hline
div\_free\_autoReg  & \Mautoreg  & Projection at each frame. & \emph{soft} \\ \hline
div\_free\_correction & \Mcorrector & Gradual projection correction. & \emph{strict} \\ \hline
\end{tabular}
\caption{Summary of the proposed divergence-free constraint techniques based on diffusion model structural design.}\label{tab:methods}
\end{table}

\section{Results} \label{SEC:Results}

In this section, we demonstrate the effectiveness and robustness of the proposed divergence-free diffusion methods across a hierarchy of test scenarios. These tests are designed to progressively challenge the models along two complementary axes: short-term predictive accuracy and long-term physical consistency. The former assesses the ability of the model to reproduce fine-scale flow structures and instantaneous spatial correlations within a limited temporal horizon, while the latter evaluates whether the generated dynamics remain statistically faithful to the reference turbulent regime as time advances or as the test conditions depart from the training distribution. The training and sampling parameters for these numerical tests are gathered in \ref{APP:ML_params}.

\paragraph{In-distribution assessment}
The first and most straightforward setting is the in-distribution test, where predictions are performed over the same temporal window as the one considered during training, but with unseen initial conditions \eqref{eq:IC}. We typically use three independent realizations of the latter, which we find sufficient for statistical stability. Because this setup closely matches the training regime, we employ stringent pixel-wise metrics such as the relative mean-squared error (MSE), the $\smash{\ell^2}$ norm, and the accuracy of the TKE spectrum $\kj\mapsto\TKES(\kj)$.

\paragraph{Rollout predictions}
In this setting, the model is evaluated over extended temporal horizons, beyond its training range. While we still report quantitative measures such as the relative $\smash{\ell^2}$ error to gauge possible drift, the focus shifts toward statistical fidelity, whether the generated dynamics preserve relevant physical and turbulence statistics as the rollout progresses.

\paragraph{Out-of-distribution tests}
Finally, we consider out-of-distribution (OOD) scenarios, where the model is exposed to initial or physical conditions markedly different from those seen during training. Here, the evaluation is primarily qualitative, supplemented by selected quantitative indicators, to assess the robustness and generalization capacity of the methods.

\subsection{In-distribution assessment} \label{sec:ID_results}

\begin{figure}[t]
    \centering
    \begin{subfigure}{0.74\linewidth}
        \centering
        \includegraphics[width=\linewidth]{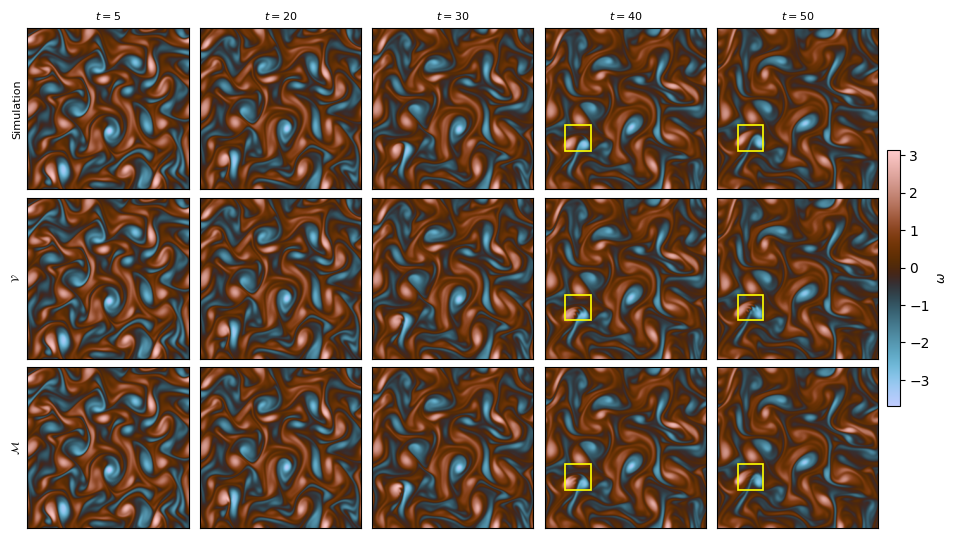}
    \end{subfigure}
    \hfill
    \begin{subfigure}{0.24\linewidth}
        \centering
        \caption*{Zoom-in of the $\color{yellow!70!black}{\square}$ regions}
        \includegraphics[width=\linewidth]{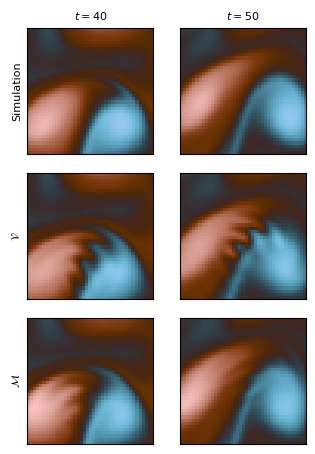}
    \end{subfigure}
    \caption{In-distribution assessment. Left panel: snapshots of the ground-truth solution (vorticity, top row) against snapshots inferred by vanilla EDM diffusion model \Mvanilla\ (middle row) and by diffusion model on divergence-free manifold \Mmanifold\ (bottom row). Right panel: zoomed-in regions of vorticity.}
    \label{fig:curlID}
\end{figure}

\paragraph{Metrics} For the in-distribution assessment, we generate $\smash{\MIC}$ randomly drawn initial conditions \eqref{eq:IC} and forecast the corresponding velocity field snapshots $\smash{\opt{(\uv^{(j)})}\in\Rset^\ndim}$ by the different methods proposed in \tref{tab:methods} up to $\smash{\Mframes}=\smash{\Nframes/2}$. While the training set actually contains $\smash{\Nframes}$ snapshots for each different initial condition, this simple test evaluates the pointwise accuracy in the early phase of the prediction. In addition, for each initial condition the generative process is repeated $\smash{\MS}$ times, such that $\smash{\MIC\times\MS\times\Mframes}$ snapshots in $\smash{\Rset^\ndim}$ are indeed generated for each method in \tref{tab:methods}. As a side note, visual (qualitative) assessments will rely on vorticity plot, as it combines the $2$-dimensional velocity field in one scalar representation, which simplify the plotting aspect. For $\uv(\rv,t) = (\uj(\rv,t), \vj(\rv,t))$ being the flow velocity field with $\rv=(x,y) \in \medium\subset\smash{\Rset^2}$, the scalar vorticity field is $\vortj(\rv,t)=\curl\cdot\uv(\rv,t)=-\partial_y\uj(\rv,t)+\partial_x\vj(\rv,t)$, where $\curl:=\smash{(-\partial_y,\partial_x)^\itr}$. An example is shown on \fref{fig:curlID}, where snapshots $\smash{\uv^{(j)}\in\Rset^\ndim}$ of the ground-truth solution are displayed together with snapshots $\smash{\opt{(\uv^{(j)})}\in\Rset^\ndim}$ predicted by the vanilla EDM diffusion model \Mvanilla\ and the proposed diffusion model on divergence-free manifold \Mmanifold.

\begin{figure}[t]
    \centering
    \begin{subfigure}{0.24\linewidth}
        \centering
        \includegraphics[width=\linewidth]{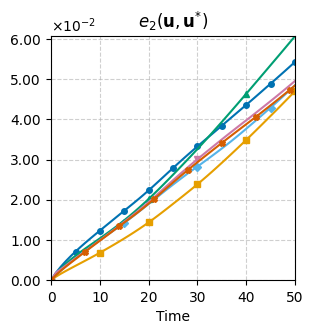}
    \end{subfigure}
    \hfill
    \begin{subfigure}{0.24\linewidth}
        \centering
        \includegraphics[width=\linewidth]{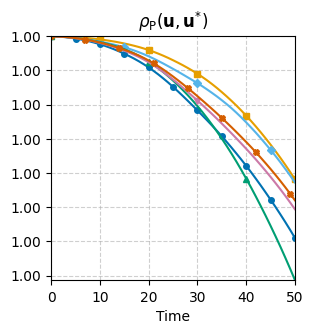}
    \end{subfigure}
    \hfill
    \begin{subfigure}{0.24\linewidth}
        \centering
        \includegraphics[width=\linewidth]{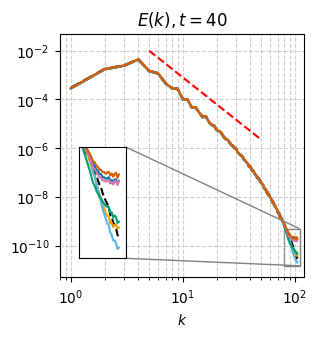}
    \end{subfigure}
    \hfill
    \begin{subfigure}{0.24\linewidth}
        \centering
        \includegraphics[width=\linewidth]{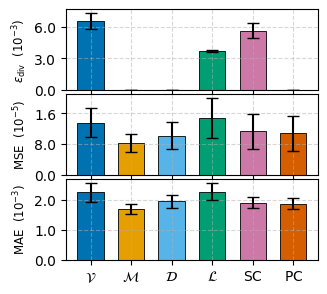}
        \vspace{.05em}
    \end{subfigure}

    \caption{In-distribution assessment. First panel from left to right: relative $\ell^2$ error $\smash{e_2}$ (lower is better). 
    Second panel: Pearson correlation coefficient $\Pearson$ (higher is better).
    Third panel: TKE spectra $\kj\mapsto\TKES(\kj)$.
    Fourth panel: $\varepsilon_\text{div}$, $\op{MSE}$ and $\op{MAE}$ (lower is better). Legends:
    \Mvanilla~\textcolor[HTML]{0072B2}{$\bullet$},\, 
    \Mmanifold~\textcolor[HTML]{E69F00}{$\blacksquare$},\,
    \Mdenoiser~\textcolor[HTML]{56B4E9}{$\blacklozenge$},\,
    \Mresidual~\textcolor[HTML]{009E73}{$\blacktriangle$},\,
    \Mautoreg~\textcolor[HTML]{CC79A7}{$\blacktriangledown$},\,
    \Mcorrector~\textcolor[HTML]{F04000}{$\bft{\xv}$}.
    TKE slopes legend: \blackdash~is ground truth spectrum, \reddash~is $\kj^{-3-\delta}$ slope.
    }
    \label{fig:PTwiseID}
\end{figure}

Across all methods, the primary objective remains the preservation of the divergence-free condition, while simultaneously ensuring accurate quantitative predictions of the flow fields. The degree of incompressibility is assessed through the spectral divergence error $\smash{\varepsilon_\text{div}}$, defined as the normalized scalar product of the inferred velocity field $\smash{\opt{\uv}}$ with its wave vectors in Fourier space: 
$$
\varepsilon_\text{div}=
\frac{\displaystyle
    \sum_{\kv\in\kset_\Ndim} \abs{\kv \cdot\opt{(\uv_\kv^{(j)})}}^2
}{\displaystyle
    \sum_{\kv\in\kset_\Ndim} \norm{\kv}^2 \norm{\smash{\opt{(\uv_\kv^{(j)})}}}^2
} \,,\quad 1\leq j\leq\Mframes\,.
$$
A perfectly divergence-free field yields $\varepsilon_{\text{div}} = 0$ (near $10^{-6}$ with float-$32$ bits numerical precision), while larger values indicate increasing deviation from the solenoidal subspace. To quantify the accuracy of predictions over the computational domain $\medium$, some classical pixel-wise metrics are used, which are recalled in \ref{APP:ID_metrics}. They provide a direct and robust assessment of reconstruction fidelity, which is particularly relevant for short-time, in-distribution evaluations. 
In addition, two temporal metrics are reported in \fref{fig:PTwiseID}, namely the relative $\smash{\ell^2}$ errors $\smash{e_2}(\uv^{(j)},\opt{(\uv^{(j)})})$ for different physical time steps $j$, where:
\begin{equation*}
e_2(\uv,\opt{\uv}) = \frac{\norm{\uv - \opt{\uv}}_2}{\norm{\uv}_2} \,,
\end{equation*}
and the Pearson correlation coefficients $\Pearson(\uv^{(j)},\opt{(\uv^{(j)})})$, where: 
\begin{equation*}
\Pearson(\uv,\opt{\uv}) = \frac{\langle\uv - \bar{\uv},\opt{\uv} - \opt{\bar{\uv}}\rangle}{\norm{\uv - \bar{\uv}}_2\norm{\opt{\uv} - \opt{\bar{\uv}}}_2} \,,
\end{equation*}
for $\langle\bs{a},\bs{b}\rangle$ being the scalar product of vectors $\bs{a},\bs{b}\in\smash{\Rset^n}$, and $\bar{\bs{a}}$ being the component-wise average of $\bs{a}\in\smash{\Rset^\ndim}$. The relative $\smash{\ell^2}$ error $\smash{e_2}$ provides a scale-independent measure of reconstruction quality by normalizing the $\smash{\ell^2}$ norm of the residual with respect to the $\smash{\ell^2}$ norm of the ground truth. Therefore it quantifies the overall proportion of signal energy that is not recovered. Complementarily, the Pearson correlation coefficient $\smash{\Pearson}$ assesses the linear association between predicted and true values, reflecting how well the model reproduces the underlying trends regardless of absolute scaling. Together, these metrics quantify both magnitude accuracy and relational consistency.



\paragraph{Results} Overall, the methods achieve highly accurate reconstructions across all evaluated settings. As shown in \fref{fig:curlID}, and further detailed in the comprehensive set of experiments in \ref{APP:ID_results}, the reconstructed fields are almost perfectly aligned with the ground truth and minimal deviations are observed. These additional results corroborate the initial observation that all models recover the coarse-scale structure with minimal error.

However, when inspecting localized regions at higher magnification, some artifacts begin to emerge. In particular, both soft and strict divergence-free constrained approaches of \sref{sec:soft-div-free} consistently display fine-scale wrinkles or wave-like distortions along regions where the flow exhibits sharp curvature changes. These artifacts, though small in amplitude, indicate a limitation in how well the penalty translates in an autoregressive framework. However, one of the proposed method \Mmanifold\ effectively suppresses these distortions over time, producing smoother and more physically consistent reconstructions even in the most challenging regions. As a side note, this defect was the only one seen across the $\smash{\MS}$ runs of the corresponding generative methods, the other ones being not displayed in this study since our goal is to discriminate the methods as sharply as possible.

The pointwise evaluation (\emph{e.g.} $e_2$, $\varepsilon_\text{div}$, $\op{MSE}$ and $\op{MAE}$) presented in \fref{fig:PTwiseID} further reinforces this trend. Across all metrics, \Mmanifold\ yields the lowest reconstruction errors, demonstrating the clear benefit of enforcing the constraint at the manifold level rather than relying on penalty terms. A general pattern also emerges: the stricter the imposed physical constraint, the better the reconstruction quality, particularly for short time horizons. Indeed, both \Mmanifold\ and \Mdenoiser\ show consistent spectra near Nyquist regions (\emph{e.g.} magnified zones in \fref{fig:PTwiseID}), \Mcorrector\ being the exception to the rule. 

Nevertheless, the temporal analysis reveals an important nuance. As the evolution time step $j$ increases, \Mmanifold\ gradually converges toward the behavior of the divergence-free denoiser model \Mdenoiser\ and the soft-constrained methods; see Pearson correlation coefficients $\Pearson$ in \fref{fig:PTwiseID}. In fact, its error growth rate accelerates over longer rollout intervals, suggesting that the advantages observed at early times may diminish as accumulated approximation errors propagate forward. Depending on the task or integration horizon, this effect may ultimately shift the performance ranking. This will be the purpose of the statistical study of rollout evaluation in the next \sref{sec:rollout_results}. Finally, all methods but \Mresidual\ have consistently beaten the vanilla method \Mvanilla, both in terms of accuracy ($e_2$ error) and divergence-free constraint ($\varepsilon_\text{div}$ report). 

\subsection{Rollout predictions} \label{sec:rollout_results}

Now we study the performances of each method of \tref{tab:methods} on a longer time horizon, \emph{i.e.} rollout predictions. Since further prediction time steps (in addition to the autoregressive formulation \eqref{eq:EDM_autoregressive}) induce longer error accumulation, we focus on statistical assessment of the fluid velocity structure. We first introduce the longitudinal velocity increment $\delta\uj_\parallel(\rv,t;\delta\rj)$ with spatial increment $\delta\rj$ along an arbitrary unit vector $\ev$ given by \cite[Chapter 6]{FRI95}:
\begin{equation} \label{eq:velocity_increment}
\delta\uj_\parallel(\rv,t;\delta\rj) = \displaystyle\left(\uv(\rv+\delta\rj\ev,t) - \uv(\rv,t)\right)\cdot\ev\,.
\end{equation}
Likewise, we define the vorticity increment $\delta\vortj_\parallel(\rv,t;\delta\rj)$ by:
\begin{equation} \label{eq:vorticity_increment}
\delta\vortj_\parallel(\rv,t;\delta\rj) = \vortj(\rv+\delta\rj\ev,t) - \vortj(\rv,t) \,.
\end{equation} 

Besides, we refer to \cite[Chapter 7]{FRI95} to compute the turnover time $\tau_\ell$, or circulation time associated with the scale $\ell$, describing the characteristic distortion time: the period over which a structure of size $\ell$ is substantially deformed by relative motion, \emph{i.e.} its rollout timescale. Taking $\ell \sim L/6$ as an upper-bound estimate, the corresponding large-scale eddy turnover time formally reads (for $\ell \leq L/2 \sim \ell_0$ the integral scale):
\begin{equation*}
    \tau_\ell = \frac{\ell}{\uj_\ell} \,, \quad \uj_\ell = 
    \sqrt{\esp\{(\delta \uj_\parallel(\ell))^2\}} \,,
\end{equation*}
where $\delta \uj_\parallel(\delta\rj)$ is the longitudinal velocity increment defined above in \eref{eq:velocity_increment}, where we drop the dependence on $t$ and $\rv$ invoking the usual assumptions of statistical time stationarity and spatial homogeneity of a turbulent velocity field $\uv$ \cite{FLA08}. Under the loose assumption that:
\begin{equation*}
\esp\{(\delta \uj_\parallel(\ell))^2\} \approx \esp\{\norm{\uv_\Ndim(0)}^2\} \,,
\end{equation*}
where $\uv_\Ndim(0)$ is as in \eref{eq:FourierNSE}, we may treat the velocity fluctuations at scale $\ell$ as being of the same order as the average velocity, thereby allowing larger spatial scales to be effectively taken into account over time. The associated eddy turnover time, which we adopt as the rollout objective, is about $\smash{\tau_\ell}\simeq 7.5$ s. 
This value is deliberately is rounded up (\emph{i.e.}, ceiled) to match $j_\imax=150$ (exactly $3/2$ of the training dataset final time $\smash{\Nframes}=100$). Choosing a larger time horizon corresponds to a more challenging prediction task and avoids any bias.

\subsubsection{Short-horizon rollout accuracy} \label{sec:rollout_results_short}

\begin{figure}[htbp]
    \centering
    \includegraphics[width=\linewidth]{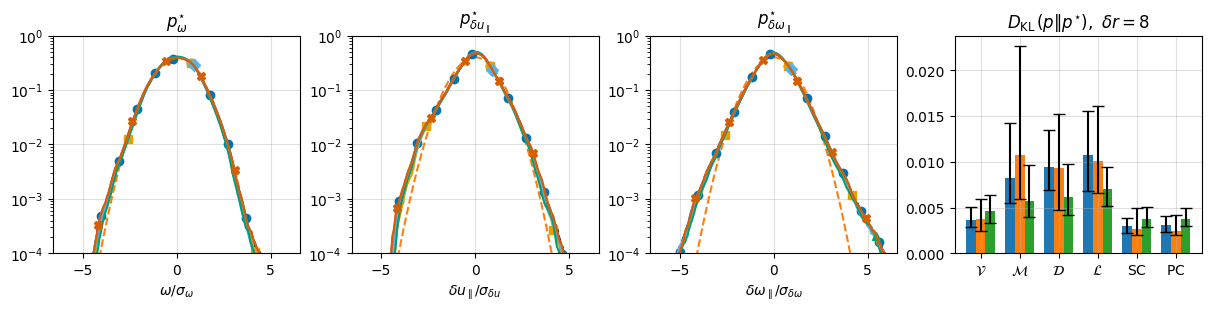}
    \vspace{-2em}
    \caption{Rollout predictions. First panel from left to right: empirical density ${\opt{\pdf}_\vortj(z;\delta\rj)}$. Second panel: empirical density ${\opt{\pdf}_{\delta\uj_\parallel}(z;\delta\rj)}$. Third panel: empirical density ${\opt{\pdf}_{\delta\vortj_\parallel}(z;\delta\rj)}$. Fourth panel: Kullback-Leibler divergence for $\delta\rj=8$.
    Density plots legends:
    \blackdash~ground truth, \orangedash~$\Normal(0,1)$,\,
    \Mvanilla~\textcolor[HTML]{0072B2}{$\bullet$},\, 
    \Mmanifold~\textcolor[HTML]{E69F00}{$\blacksquare$},\,
    \Mdenoiser~\textcolor[HTML]{56B4E9}{$\blacklozenge$},\,
    \Mresidual~\textcolor[HTML]{009E73}{$\blacktriangle$},\,
    \Mautoreg~\textcolor[HTML]{CC79A7}{$\blacktriangledown$},\,
    \Mcorrector~\textcolor[HTML]{F04000}{$\bft{\xv}$}.
    Box plot legends: 
    $D_{\op{KL}}(\pdf_\vortj\lVert \opt{\pdf}_\vortj;\delta\rj)$~\textcolor[HTML]{0072B2}{$\blacksquare$},\,
    $D_{\op{KL}}(\pdf_{\delta\uj_\parallel} \lVert \opt{\pdf}_{\delta\uj_\parallel};\delta\rj)$~\textcolor[HTML]{E69F00}{$\blacksquare$},\,
    $D_{\op{KL}}(\pdf_{\delta\vortj_\parallel} \lVert \opt{\pdf}_{\delta\vortj_\parallel};\delta\rj)$~\textcolor[HTML]{009E73}{$\blacksquare$}. Whiskers indicate the interquartile range.
    }
    \label{fig:pdfs}
\end{figure}

\paragraph{Metrics} For a given generative method in \tref{tab:methods} used to infer the velocity field snapshots $\smash{\opt{(\uv^{(j)})}}$, $\smash{j_\imin}\leq j\leq \smash{j_\imax}$, the probability density functions (PDF) of the spatial increments \eqref{eq:velocity_increment} and \eqref{eq:vorticity_increment} are first computed over all spatial locations $\smash{\rv_p}$, $1\leq p\leq H\times W$, of the computational domain $\medium$ for each time step $j$, yielding per-snapshot empirical distributions. These PDFs are then aggregated over the $\smash{\MIC}$ initial conditions, $\smash{\MS}$ runs of the generative method (with different initial noise seeds), and time steps $\smash{j_\imin\leq j\leq j_\imax}$ to form the overall empirical distribution. Aggregation is performed using an empirical mean to obtain a robust estimate that is less sensitive to intermittencies or outlier realizations, preserving the overall trends observed on the plots. For $I=\smash{(i-1)\times\MS+s}$ with $s=\smash{1,\dots\MS}$, $i=\smash{1,\dots\MIC}$, and $\smash{\opt{(\uv_I^{(j)})}}$ being the corresponding inferred velocity snapshots, the velocity increments \eqref{eq:velocity_increment} read:
\begin{equation*}
\opt{(\delta\uj_\parallel^I(\rv,j;\delta\rj))}=\big(\opt{(\uv_I^{(j)}(\rv+\delta\rj\ev))} - \opt{(\uv_I^{(j)}(\rv))}\big)\cdot\ev\,.
\end{equation*}
Their aggregated empirical PDF then reads:
\begin{equation*}
\opt{\pdf}_{\delta\uj_\parallel}(z;\delta\rj)=\frac{1}{Z}\sum_{j=j_\imin}^{j_\imax}\sum_{p=1}^{H\times W}\sum_{s=1}^{\MS}\sum_{i=1}^{\MIC}\delta\big(z-\opt{(\delta\uj_\parallel^I(\rv_p,j;\delta\rj))}\big)\,,
\end{equation*}
where $z\mapsto\delta(z)$ stands for the Dirac delta function, and $Z=\MIC\times\MS\times H\times W\times(j_\imax-j_\imin+1)$ is the normalization constant. 
The empirical PDF $\smash{\opt{\pdf}_{\delta\vortj_\parallel}(z;\delta\rj)}$ of the vorticity increments \eqref{eq:vorticity_increment} is defined analogously, as well as the empirical PDF $\smash{\opt{\pdf}_\vortj(z;\delta\rj)}$ of the inferred vorticity snapshots $\smash{\opt{(\vortj^{(j)})}}$ themselves. They are displayed on \fref{fig:pdfs}, together with
the corresponding empirical PDFs of the ground-truth solutions $\smash{\uv^{(j)}}$ and $\smash{\vortj^{(j)}}$ denoted by $\smash{\pdf_{\delta\uj_\parallel}(z;\delta\rj)}$, $\smash{\pdf_{\delta\vortj_\parallel}(z;\delta\rj)}$, and $\smash{\pdf_\vortj(z;\delta\rj)}$, respectively. While PDFs capture the statistical behavior of the fields and thus reflect rollout consistency, they are not visually discriminative. We therefore also report the Kullback–Leibler divergence for quantitative comparison. For each method in \tref{tab:methods}, the discrepancy between the predicted and ground-truth PDFs is quantified via:
\begin{equation*}
D_{\op{KL}}(\pdf_\varphi\lVert\opt{\pdf}_\varphi;\delta\rj)=\int\pdf_\varphi(z;\delta\rj)\log\frac{\pdf_\varphi(z;\delta\rj)}{\opt{\pdf}_\varphi(z;\delta\rj)}\id z\,,\quad\varphi\in\{\delta\uj_\parallel, \delta\vortj_\parallel,\vortj\}\,.
\end{equation*}
The box plot in \fref{fig:pdfs} then displays the divergence $\smash{D_{\op{KL}}(\pdf_\varphi\lVert\opt{\pdf}_\varphi;\delta\rj)}$ (with median and interquartile range) for each $\varphi\in\{\delta\uj_\parallel, \delta\vortj_\parallel,\vortj\}$ and for all methods in \tref{tab:methods} with $\delta\rj=8$ pixels (with other displacements available in \ref{APP:ROLL_results}). Apart from the density estimation, we also report a range of metrics given in \sref{sec:ID_results} and, as a broader statistical evaluation, the relative amplitude $\text{r}A$, defined in \ref{APP:ID_metrics}.

\begin{figure}[t]
    \centering
    \begin{subfigure}{0.74\linewidth}
        \centering
        \includegraphics[width=\linewidth]{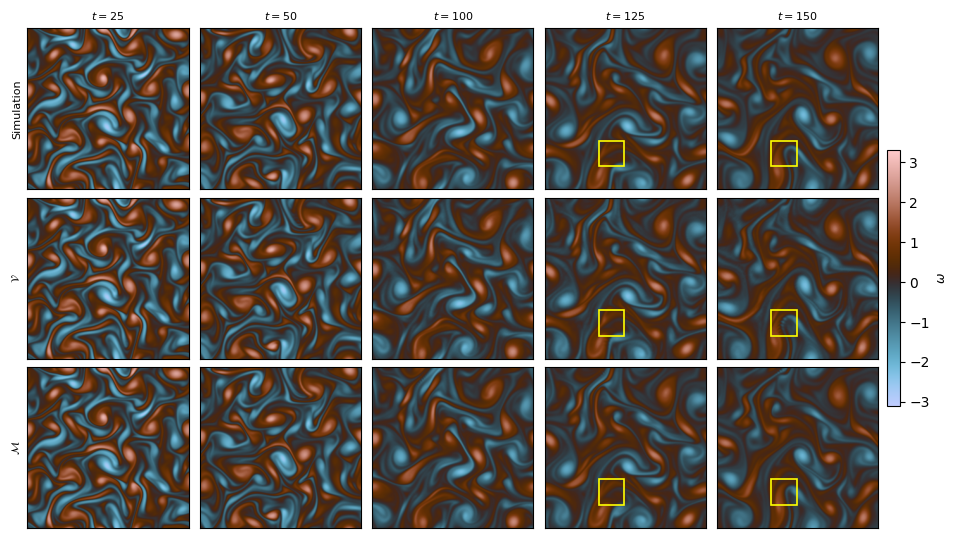}
    \end{subfigure}
    \hfill
    \begin{subfigure}{0.24\linewidth}
        \centering
        \caption*{Zoom-in of the $\color{yellow!70!black}{\square}$ regions}
        \includegraphics[width=\linewidth]{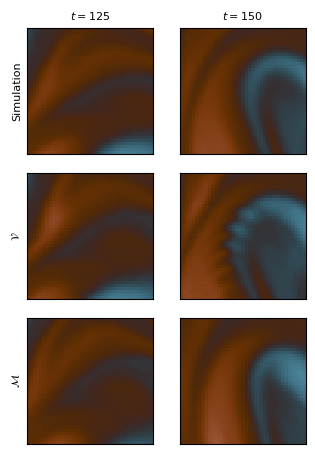}
    \end{subfigure}
    \caption{Rollout predictions. Left panel: snapshots of the ground-truth solution (vorticity, top row) against snapshots inferred by vanilla EDM diffusion model \Mvanilla\ (middle row) and by diffusion model on divergence-free manifold \Mmanifold\ (bottom row). Right panel: zoomed-in regions of vorticity.}
    \label{fig:curlRollout}
\end{figure}

\paragraph{Results} As shown in \fref{fig:curlRollout}, we observe the same rippling artifacts that are caused by the same methods (all but \Mmanifold). This behavior is expected and, to some extent, satisfactory, as the flow patterns is consistent across time and methods. The same qualitative and quantitative conclusions as in \sref{sec:ID_results} hold in this rollout setting. In particular, these ripples are likely caused by hard or previously unseen configurations. We deliberately choose to display such cases, noting that most samples do not exhibit these anomalies. We address this behavior in the next \sref{sec:rollout_long}, where a correction algorithm is introduced for even longer rollout, where \Mmanifold\ also suffers from the same fate for large enough time. Again, all methods (apart from this artifact) yield visually consistent samples, as shown in detail in the exhaustive study of \ref{APP:ROLL_results}.

We now turn to the full rollout regime results, where our primary interest lies in statistical properties rather than point-wise errors. Interestingly, the trend reverses (as expected from trends observed in \sref{sec:ID_results}): divergence-free methods enforced through the vanilla model \Mvanilla\ perform better (\emph{e.g.} \Mcorrector\ and \Mautoreg). This behavior is consistently observed in the empirical probability density functions shown in \fref{fig:pdfs} across multiple $\delta r$ scales (see \ref{APP:ROLL_results}). This result may be considered intuitive: as constraints are relaxed, relative to the unconstrained \Mvanilla\ method, the learned dynamics become statistically more accurate, since a model trained without constraints has more degrees of freedom. In contrast, the more stringent methods (built upon physical constraints) effectively operate at the interface between the training distribution and the out-of-distribution regime encountered toward the end of the rollout, where a softer enforcement of constraints is required to allow for greater flexibility. Overall, this represents a favorable trade-off. Notably, the method of predictor-corrector iterates with gradual Leray projections \Mcorrector\ achieves both perfect divergence-free behavior, the lowest statistical error (which we further examine in Table~\ref{tab:rollout}) and satisfying visual consistency (despite \Mmanifold\ being the only one free of any artifact).

Finally, we consider more point-wise error metrics. Although not our primary focus, these temporally averaged quantities remain highly informative. The results are summarized in Table~\ref{tab:rollout}. As previously noted, methods derived from \Mvanilla\ achieve excellent performance, exhibiting strong overall results with limited variance across approaches (without undermining the performance of \Mmanifold\ and \Mdenoiser). The $\text{r}A$ metric further supports the claim made in the previous paragraph: the best amplitude agreement is achieved by \Mvanilla, followed by \Mcorrector\ and \Mautoreg. This indicates that \Mvanilla\ better captures the local mean flow and that its predictions can be corrected at post-training using divergence-free methods. Nevertheless, a clear winner emerges: \Mcorrector\ consistently outperforms the vanilla baseline and all other approaches in terms of error, while simultaneously enforcing perfect divergence-free constraints. Conversely, \Mresidual\ performs poorly across all metrics, consistently under-performing both in in-distribution assessment and in the rollout evaluation.

\begin{table}[htbp] 
\centering
{\tiny
\begin{tabular}{lccccc} 
\toprule
Method & $\operatorname{MSE}$ & $\operatorname{MAE}$ & $e_2$ & $\varepsilon_\text{div}$ & $\text{r}A$ \\
\midrule
$\mathcal{V}$ & $ 1.41\cdot 10^{-4} \pm 1.59\cdot 10^{-4} $ & $ 7.02\cdot 10^{-3} \pm 4.69\cdot 10^{-3} $ & $ 9.09\cdot 10^{-2} \pm 6.30\cdot 10^{-2} $ & $ 8.07\cdot 10^{-3} \pm 2.52\cdot 10^{-4} $ & $ \mathbf{1.00\cdot 10^{0}} \pm 5.36\cdot 10^{-3} $ \\
\ref{eq:EDMdivfree} & $ 1.55\cdot 10^{-4} \pm 1.86\cdot 10^{-4} $ & $ 7.23\cdot 10^{-3} \pm 5.44\cdot 10^{-3} $ & $ 9.20\cdot 10^{-2} \pm 7.11\cdot 10^{-2} $ & $ \mathbf{0} \pm 0 $ & $ 9.70\cdot 10^{-1} \pm 2.01\cdot 10^{-2} $ \\
\ref{eq:denoiser-Leray} & $ 1.83\cdot 10^{-4} \pm 2.24\cdot 10^{-4} $ & $ 7.67\cdot 10^{-3} \pm 5.62\cdot 10^{-3} $ & $ 1.00\cdot 10^{-1} \pm 7.70\cdot 10^{-2} $ & $ \mathbf{0} \pm 0 $ & $ 9.73\cdot 10^{-1} \pm 1.51\cdot 10^{-2} $ \\
\ref{eq:PINN-loss} & $ 2.07\cdot 10^{-4} \pm 2.58\cdot 10^{-4} $ & $ 8.25\cdot 10^{-3} \pm 5.96\cdot 10^{-3} $ & $ 1.07\cdot 10^{-1} \pm 8.21\cdot 10^{-2} $ & $ 5.21\cdot 10^{-3} \pm 2.63\cdot 10^{-5} $ & $ 9.68\cdot 10^{-1} \pm 2.05\cdot 10^{-2} $ \\
\ref{eq:sample-correction} & $ 1.20\cdot 10^{-4} \pm 1.50\cdot 10^{-4} $ & $ 6.26\cdot 10^{-3} \pm 4.44\cdot 10^{-3} $ & $ 8.20\cdot 10^{-2} \pm 6.11\cdot 10^{-2} $ & $ 7.01\cdot 10^{-3} \pm 2.29\cdot 10^{-4} $ & $ 1.01\cdot 10^{0} \pm 6.07\cdot 10^{-3} $ \\
\ref{eq:PC-Leray} & $ \mathbf{1.15\cdot 10^{-4}} \pm 1.44\cdot 10^{-4} $ & $ \mathbf{6.18\cdot 10^{-3}} \pm 4.38\cdot 10^{-3} $ & $ \mathbf{8.04\cdot 10^{-2}} \pm 5.98\cdot 10^{-2} $ & $ \mathbf{0} \pm 0 $ & $ 1.01\cdot 10^{0} \pm 8.16\cdot 10^{-3} $ \\
\bottomrule
\end{tabular}
}
\caption{Rollout predictions averaged metrics.}
\label{tab:rollout}
\end{table}

\subsubsection{Long-horizon structural coherence} \label{sec:rollout_long}

Motivated by the observations done in \sref{sec:ID_results} and \sref{sec:rollout_results_short}, we address the small-scale rippling artifacts observed in the predicted velocity fields. These artifacts appear as spurious high-frequency oscillations on top of on otherwise accurate large-scale flow structures. They likely arise from local overfitting to high-frequency training components and imperfect score estimation at low noise levels. Such effects are amplified in spectral regions with low signal-to-noise ratio and accumulate during iterative rollouts, ultimately limiting prediction horizons. We therefore introduce a transport-based correction that suppresses these oscillations while preserving physically meaningful flow structures, enabling longer and more stable predictions (up to $j=250$ and possibly more). A detailed analysis of its impact on overall predictive performance is beyond the scope of this paper; instead, we focus on correcting the artifacts observed in \sref{sec:ID_results} and \sref{sec:rollout_results_short} for \Mmanifold, which has so far proven to be the most visually consistent method. The full ablation study, presented in \ref{APP:ROLL_results}, nevertheless shows that such artifacts do emerge for sufficiently long time horizons. For completeness, we provide the full construction and intuition in \ref{APP:HF_corrector}. It is controlled by two hyperparameters: $\alpha_\text{R}$, which sets how many times the correction is applied; and $\alpha_\text{T}$, which specifies how early in the backward diffusion process the correction begins (larger values apply it earlier, \emph{i.e.} more steps before the final timestep). Together, they control the strength and timing of the correction during the generative process.

\begin{figure}[h!]
    \centering
    \begin{subfigure}{0.8\linewidth}
        \centering
        \includegraphics[width=\linewidth]{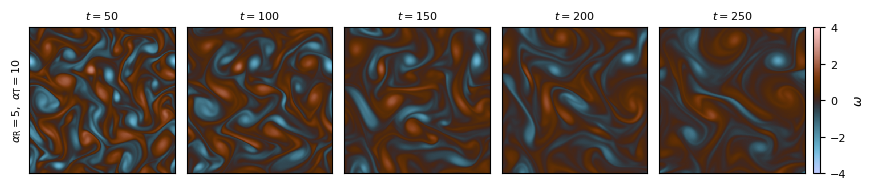}
    \end{subfigure}
    \hfill
    \begin{subfigure}{0.19\linewidth}
        \centering
        \includegraphics[width=\linewidth]{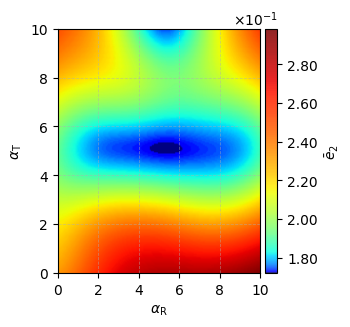}
        \vspace{-2em}
    \end{subfigure}
    \caption{Long rollout prediction. Left panel: snapshots of vorticity inferred by diffusion model on divergence-free manifold \Mmanifold\ using the transport-based corrector with hyperparameters $\alpha_\text{R}=5$ and $\alpha_\text{T}=10$. Right panel: $\alpha_\text{R}$ and $\alpha_\text{T}$ parameters and associated averaged $e_2$ metric.}
    \label{fig:long_roll}
\end{figure}

The hyperparameters $\alpha_\text{R}$ and $\alpha_\text{T}$ are optimized as follows. Due to the high computational cost of parameter optimization, we restrict the search to a coarse grid of $15$ configurations, as illustrated in \fref{fig:long_roll}, and report the result according to $e_2$. Although ground-truth trajectories are unavailable at such long time horizons, this strategy remains sufficient as clear trends emerge. Overall, the dominant factor governing performance is the step in the reverse ODE at which the correction is applied, controlled by $\alpha_\text{T}$. When $\alpha_\text{T}$ is set to an early or intermediate stage of the reverse process, the correction already induces most of the observable improvement, effectively correcting the post–turnover prediction. In this regime, increasing the number of correction iterations (high $\alpha_\text{R}$) yields diminishing returns, since the updates are driven by strong gradients and rapidly saturate.

Interestingly, an outlier configuration arises at larger $\alpha_\text{T}$, which yields the longest stable rollout (see \ref{APP:ROLL_results}) and is the one being shown in \fref{fig:long_roll}. This confirms that $\alpha_\text{R}$ is not a critical factor as long as a suitable (early or mid) $\alpha_\text{T}$ is chosen: the correction has already reoriented the reverse dynamics in a favorable direction. Indeed, most of the modeling precision comes from the late reverse dynamics (or density collapse) \cite{karras2022elucidating,biroli2024dynamical}. However, pushing the stability horizon further becomes more tricky. In this extreme regime, long rollouts remain achievable, but extending them requires a careful joint balance between an early $\alpha_\text{T}$ and the previously selected $\alpha_\text{R}$. Achieving long stability increasingly relies on both consistent internal dynamics and transport correction.

Finally, two additional regimes can be identified. When the overall correction strength is too weak (low $\alpha_\text{R}\times\alpha_\text{T}$), residual ripples persist in the generated trajectories and may even be amplified. Conversely, when the correction is excessively strong (high $\alpha_\text{R}\times\alpha_\text{T}$), artifacts re-emerge, as the transport term dominates and suppresses the intrinsic free dynamics of the statistical model (see again \ref{APP:ROLL_results}). At the end, we where able to push predictions up to $2.5*j_{\max}$ ($j=250$), with no apparent signs of instability.

\subsection{Out-of-distribution (OOD) tests} \label{sec:OOD_res}
In these last experiments, the proposed methods are tested on two OOD sets, using the same physical time horizon as the one used during the training phase. This approach enables a clean evaluation of the performances of those methods. Details on the choice of simulation parameters are left to the \ref{APP:SNSE_digest_PARAMS}.
\subsubsection{OOD generation} \label{sec:OOD_gen}
The two test cases considered to analyze the performances of generative methods for OOD configurations are the following.

\paragraph{Vortex superposition (VS)} The initial velocity field $\smash{\uv_\text{VS}}=(\uj,\vj)$ is imposed directly as pure sinusoids: 
\begin{equation*}
    \uv_\text{VS} = A_\text{VS}
    \begin{pmatrix}
    -\sin\left(2\frac{y}{L}\right) \\[4pt]
    \sin\left(3\frac{x}{L}\right)
    \end{pmatrix} \,.
\end{equation*}
This yields a solenoidal shear-type vortex field consisting of pure Fourier modes $(0,2)$ and $(3,0)$ for $\uj$ and $\vj$, respectively, without randomness but with perfect symmetry. It is commonly used as a simple analytical vortex initial condition for testing. This strong symmetry constraint ensures the model consistency with isotropic dynamics. OOD tests are constructed with arbitrary amplitudes $\smash{A_\text{VS}}$ that lie outside the range encountered during training. This deliberate amplitude mismatch makes prediction particularly challenging: even small autoregressive errors can amplify rapidly when the model is forced to operate in a dynamical regime it has never seen.

\paragraph{Perturbed Taylor–Green vortex (TGV)} This test case consists in a two-dimensional incompressible velocity field constructed by starting from a canonical Taylor–Green vortex and adding a small, smooth low-frequency perturbation:
\begin{equation*}
\uv_0 = 
    \begin{pmatrix}
    \sin(2k_0 x) \cos(2k_0 y) \\[4pt]
    -\cos(2k_0 x)\sin(2k_0 y)
    \end{pmatrix} \,, \quad k_0 = \frac{2 \pi}{L}\,.
\end{equation*}
Since the pure Taylor-Green vortex is a stationary solution of incompressible flows, we add a random, mostly high-frequency perturbation to the initial field:
\begin{equation*}
\uv_\text{TGV} = A_\text{TGV}\parenthesis{\uv_0 + \Leray \bs{\eta}} \,, \quad \bs{\eta} \sim \mathcal{N}(\bs{0},\Iv_\ndim) \,.
\end{equation*}
Here $\smash{A_\text{TGV}}>0$ is the amplitude of the vortex. This setting, apart from being standard in two-dimensional fluid tests, shows a clear regime transition and vortex-scale change, from low to large wavenumbers, creating localized coherent dynamics along the way. This well supports the OOD claim since it mixes a coherent TGV pattern with high-frequency noise, creating scale combinations the model was not trained on.

\subsubsection{OOD results}
\begin{figure}[h!]
    \centering
    \begin{subfigure}{0.45\textwidth}
        \centering
        \includegraphics[width=\textwidth]{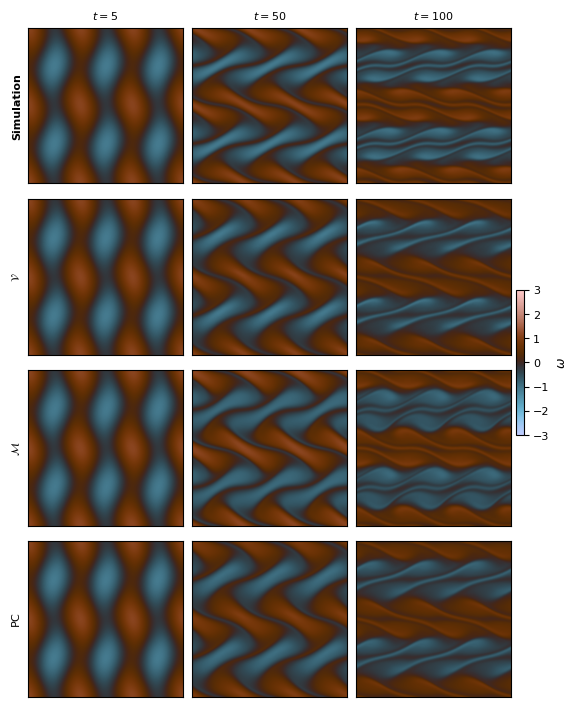}
    \end{subfigure}
    \begin{subfigure}{0.45\textwidth}
        \centering
        \includegraphics[width=\textwidth]{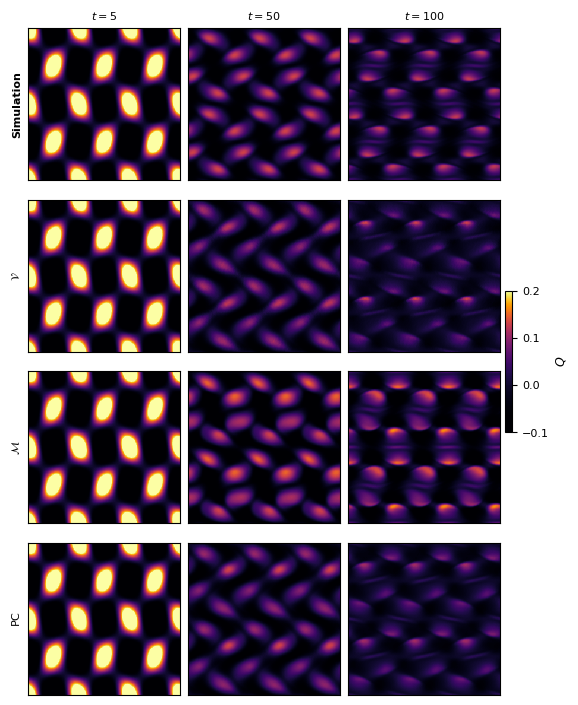}
    \end{subfigure}
    \hfill
    \caption{OOD tests on vortex superposition (VS): snapshots of the ground-truth solution (top row) against snapshots inferred by models \Mvanilla, \Mmanifold, and \Mcorrector. Left panel: vorticity field. Right panel: $Q$-criterion.}
    \label{fig:OOD_VS_curl}

    \centering
    \begin{subfigure}{0.45\textwidth}
        \centering
        \includegraphics[width=\textwidth]{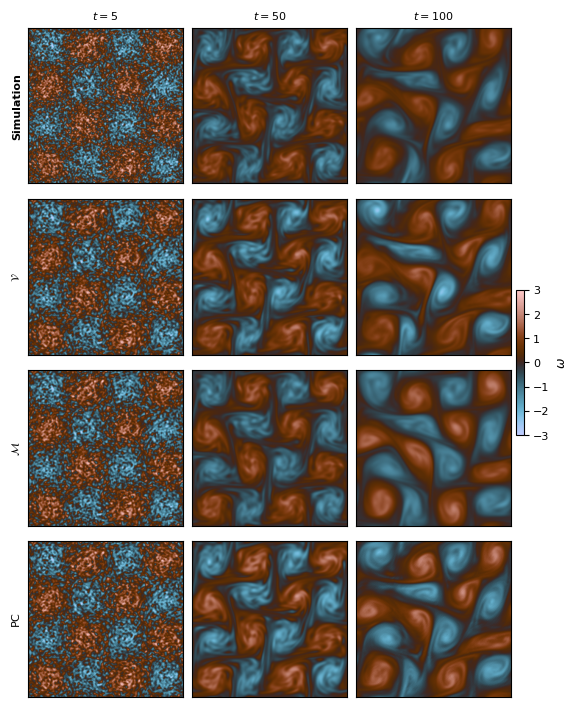}
    \end{subfigure}
    \begin{subfigure}{0.45\textwidth}
        \centering
        \includegraphics[width=\textwidth]{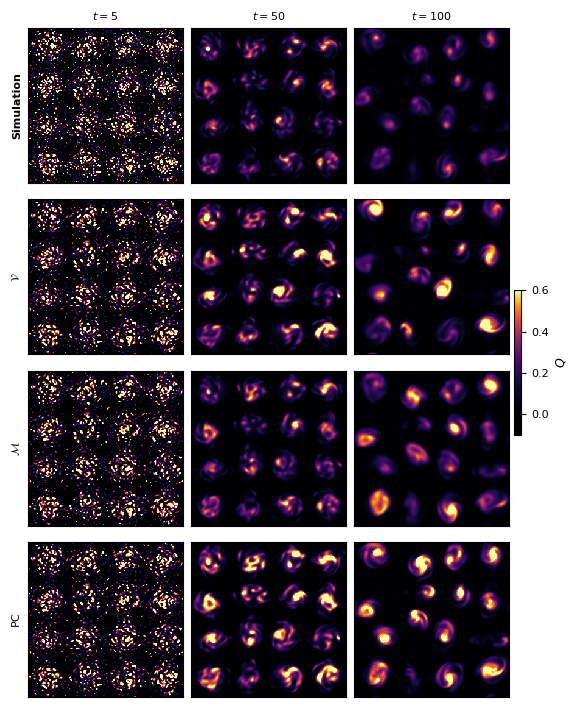}
    \end{subfigure}
    \hfill
    \caption{OOD tests on Taylor-Green vortex (TGV): snapshots of the ground-truth solution (top row) against snapshots inferred by models \Mvanilla, \Mmanifold, and \Mcorrector. Left panel: vorticity field. Right panel: $Q$-criterion.}
    \label{fig:OOD_TGV_curl}
\end{figure}

\paragraph{Metrics} While visual inspection of the vorticity field can provide a qualitative sense of the rotational structures present in the flow, OOD conditions may induce subtle artifacts or distortions. To obtain a more satisfying characterization, one typically turns to other visual vortex identification metrics, among which the $Q$-criterion \cite{DUB00}:
\begin{equation}\label{eq:critQ}
Q=\demi(\bft{\Omega}:\bft{\Omega}-\bft{D}:\bft{D})\,,
\end{equation}
where $\bft{D}=\nablav\otimes_s\uv=\smash{\demi(\nablav\otimes\uv+(\nablav\otimes\uv)^\itr)}$ is the symmetric velocity rate tensor, $\bft{\Omega}=\nablav\otimes_a\uv=\smash{\demi(\bnabla\otimes\uv-(\bnabla\otimes\uv)^\itr)}$ is the skew-symmetric rotation rate tensor, and $\bft{A}:\bft{B}=\smash{\sum_{i,j}A_{ij}B_{ij}}$. In two dimensions ($\ddim=2$) one has:
\begin{equation*}
\bft{D} = \displaystyle\demi
\begin{bmatrix}
2 \partial_x\uj & \partial_y\uj + \partial_x\vj \\
\partial_y\uj + \partial_x\vj & 2 \partial_y\vj
\end{bmatrix}
\,, \quad
\bft{\Omega} = \displaystyle\demi
\begin{bmatrix}
0 & -\vortj \\
\vortj & 0
\end{bmatrix}
\,.
\end{equation*}
The $Q$-criterion balances between the rotation rate and the strain rate and its positive iso-surfaces (or iso-values in two dimensions) display areas (or lines in two dimensions) where the strength of rotation overcomes the strain, highlighting the vortex envelopes. This criterion sharpens the depiction of vortex cores and enhances qualitative observations.

\begin{figure}[htbp] 
    \centering
    \begin{minipage}[t]{0.225\textwidth}
        \centering
        \includegraphics[width=\linewidth]{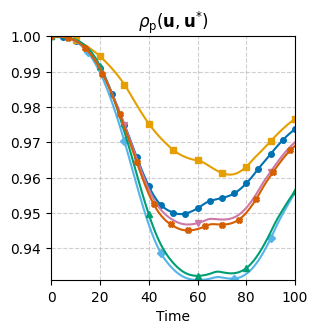}
        \vspace{-2em}
    \end{minipage}\hfill
    \begin{minipage}[t]{0.74\textwidth}
        \centering
        \vspace{-12em}
        \begin{tabular}{lccc}
\toprule
Method & $\operatorname{MSE}$ & $e_2$ & $\varepsilon_\text{div}$ \\ 
\midrule
$\mathcal{V}$ & $ 1.71\cdot 10^{-3} \pm 9.58\cdot 10^{-4} $ & $ 2.42\cdot 10^{-1} \pm 1.05\cdot 10^{-1} $ & $ 1.02\cdot 10^{-2} \pm 2.13\cdot 10^{-5} $ \\
$\mathcal{M}$ & $ \mathbf{1.44\cdot 10^{-3}} \pm 8.81\cdot 10^{-4} $ & $ \mathbf{2.20\cdot 10^{-1}} \pm 1.01\cdot 10^{-1} $ & $ \mathbf{0} \pm 0 $ \\
$\mathcal{D}$ & $ 2.74\cdot 10^{-3} \pm 1.64\cdot 10^{-3} $ & $ 3.04\cdot 10^{-1} \pm 1.40\cdot 10^{-1} $ & $ \mathbf{0} \pm 0 $ \\
$\mathcal{L}$ & $ 2.62\cdot 10^{-3} \pm 1.54\cdot 10^{-3} $ & $ 2.98\cdot 10^{-1} \pm 1.34\cdot 10^{-1} $ & $ 1.65\cdot 10^{-2} \pm 5.37\cdot 10^{-4} $ \\
$\text{SC}$ & $ 1.86\cdot 10^{-3} \pm 1.06\cdot 10^{-3} $ & $ 2.52\cdot 10^{-1} \pm 1.11\cdot 10^{-1} $ & $ 4.74\cdot 10^{-3} \pm 1.84\cdot 10^{-5} $ \\
$\text{PC}$ & $ 1.92\cdot 10^{-3} \pm 1.10\cdot 10^{-3} $ & $ 2.56\cdot 10^{-1} \pm 1.14\cdot 10^{-1} $ & $ \mathbf{0} \pm 0 $ \\
\bottomrule
\end{tabular}
        \vspace{1.225em}
    \end{minipage}
    \caption{OOD test on vortex superposition (VS). Left panel: Pearson correlation coefficient $\Pearson$ as a function of time. Right table: time-averaged mean-square error $\op{MSE}$, relative $\ell^2$ error $e_2$, and spectral divergence error $\varepsilon_\text{div}$. Legends:
    \Mvanilla~\textcolor[HTML]{0072B2}{$\bullet$},\, 
    \Mmanifold~\textcolor[HTML]{E69F00}{$\blacksquare$},\,
    \Mdenoiser~\textcolor[HTML]{56B4E9}{$\blacklozenge$},\,
    \Mresidual~\textcolor[HTML]{009E73}{$\blacktriangle$},\,
    \Mautoreg~\textcolor[HTML]{CC79A7}{$\blacktriangledown$},\,
    \Mcorrector~\textcolor[HTML]{F04000}{$\bft{\xv}$}.}\label{fig:VS_tab}

    \centering
    \begin{minipage}[t]{0.225\textwidth}
        \centering
        \includegraphics[width=\linewidth]{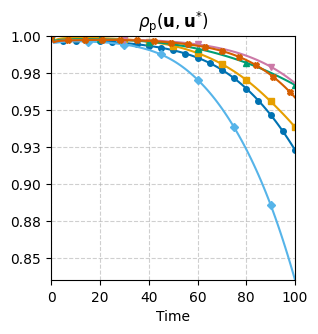}
        \vspace{-2em}
    \end{minipage}\hfill
    \begin{minipage}[t]{0.74\textwidth}
        \centering
        \vspace{-12em}
        \begin{tabular}{lccc}
\toprule
Method & $\operatorname{MSE}$ & $e_2$ & $\varepsilon_\text{div}$ \\
\midrule
$\mathcal{V}$ & $ 1.49\cdot 10^{-3} \pm 1.18\cdot 10^{-3} $ & $ 2.45\cdot 10^{-1} \pm 1.10\cdot 10^{-1} $ & $ 2.63\cdot 10^{-2} \pm 3.80\cdot 10^{-4} $ \\
$\mathcal{M}$ & $ 9.45\cdot 10^{-4} \pm 1.15\cdot 10^{-3} $ & $ 1.77\cdot 10^{-1} \pm 1.21\cdot 10^{-1} $ & $ \mathbf{0} \pm 0 $ \\
$\mathcal{D}$ & $ 1.56\cdot 10^{-3} \pm 1.73\cdot 10^{-3} $ & $ 2.33\cdot 10^{-1} \pm 1.48\cdot 10^{-1} $ & $ \mathbf{0} \pm 0 $ \\
$\mathcal{L}$ & $ \mathbf{4.53\cdot 10^{-4}} \pm 3.97\cdot 10^{-4} $ & $ \mathbf{1.34\cdot 10^{-1}} \pm 6.37\cdot 10^{-2} $ & $ 1.84\cdot 10^{-2} \pm 8.95\cdot 10^{-5} $ \\
$\text{SC}$ & $ 9.12\cdot 10^{-4} \pm 5.78\cdot 10^{-4} $ & $ 1.97\cdot 10^{-1} \pm 7.24\cdot 10^{-2} $ & $ 1.20\cdot 10^{-2} \pm 3.74\cdot 10^{-4} $ \\
$\text{PC}$ & $ 1.83\cdot 10^{-3} \pm 1.32\cdot 10^{-3} $ & $ 2.70\cdot 10^{-1} \pm 1.24\cdot 10^{-1} $ & $ \mathbf{0} \pm 0 $ \\
\bottomrule
\end{tabular}
        \vspace{1.225em}
    \end{minipage}
    \caption{OOD test on Taylor-Green vortex (TGV). Left panel: Pearson correlation coefficient $\Pearson$ as a function of time. Right table: time-averaged mean-square error $\op{MSE}$, relative $\ell^2$ error $e_2$, and spectral divergence error $\varepsilon_\text{div}$. Legends:
    \Mvanilla~\textcolor[HTML]{0072B2}{$\bullet$},\, 
    \Mmanifold~\textcolor[HTML]{E69F00}{$\blacksquare$},\,
    \Mdenoiser~\textcolor[HTML]{56B4E9}{$\blacklozenge$},\,
    \Mresidual~\textcolor[HTML]{009E73}{$\blacktriangle$},\,
    \Mautoreg~\textcolor[HTML]{CC79A7}{$\blacktriangledown$},\,
    \Mcorrector~\textcolor[HTML]{F04000}{$\bft{\xv}$}.}\label{fig:TGV_tab}
\end{figure}

\paragraph{Results} Only the most representative results are shown for compactness. Additional plots of the snapshots inferred with all methods of \tref{tab:methods} are displayed in \ref{APP:OOD} on \fref{fig:OOD_curl_ANNEX_VS} for the VS case, and on \fref{fig:OOD_curl_ANNEX_TGV} for the TGV case. At first for the VS case, the $Q$-criterion provides a clearer visual insight into the amplitude distribution of the predicted flow fields, as seen in \fref{fig:OOD_VS_curl}. Overall, our proposed diffusion model on divergence-free manifold \Mmanifold\ performs best, again. The vorticity field itself offers limited additional insight; however, it clearly demonstrates that symmetry is perfectly preserved, which is a very desirable geometry property, in addition to the divergence-free enforcement. Coherent structures are present across models, although they remain difficult to evaluate qualitatively. The Pearson correlation coefficient $\smash{\Pearson}$, displayed alongside temporal averages in \fref{fig:VS_tab}, breaks this apparent tie and identifies \Mmanifold\ as the best model, while also highlighting that other models such as \Mcorrector\ and \Mautoreg\ should not be overlooked. The atypical $\smash{\Pearson}$ trend, initially decreasing and then increasing, also highlights the peculiarity of this OOD setting, challenging our initial expectations. One might have expected mildly physically constrained methods, such as \Mcorrector, to dominate in this setting. Instead, the strict constrained model achieves the best performances, likely due to the complex symmetry patterns inherent to the VS configuration. In this setup, the vanilla EDM diffusion model \Mvanilla\ is only outperformed by \Mmanifold. 

This trend, however, does not hold for the TGV case, which instead supports the claim that softer divergence-free constraints are better suited for OOD generalization (also supported with the previous results), as they allow the model to satisfy physical constraints while retaining a sufficient degree of freedom. In this case, all models successfully capture the high to low wavenumber transition, as illustrated by the $Q$-criterion in \fref{fig:OOD_TGV_curl}. Nevertheless, a clear amplitude disparity across vortices is observed, which is likely responsible for most of the prediction error, rather than strict divergence-free conditioning alone. Consequently, \Mcorrector\ and \Mautoreg\ models are favored, as expected, and we finally observe a boost in accuracy in \Mresidual, as seen in \fref{fig:TGV_tab}. In this regime, \Mresidual\ achieves the best overall performance, although it is closely matched by \Mcorrector\ and \Mautoreg. Notably, \Mcorrector\ is the only method that perfectly enforces the divergence-free constraint, which is the primary objective. By contrast, the strict divergence-free models \Mmanifold\ and \Mdenoiser\ exhibit reduced variability, likely due to overfitting, an effect already discussed in \sref{sec:rollout_results}, but beyond the scope of the present OOD analysis.

\section{Conclusions and perspectives} \label{SEC:conclusion}

In this work, we studied how physical constraints should be embedded into generative diffusion models for incompressible flow prediction, using Kolmogorov turbulence as a controlled yet informative benchmark. Building upon the plain EDM diffusion framework, we systematically compared a spectrum of divergence-free enforcement strategies, ranging from strict manifold-based formulations to soft guidance and post hoc corrections, across in-distribution, rollout, and out-of-distribution regimes.

A first key finding is that enforcing incompressibility at the level of the generative manifold yields the most accurate and physically consistent reconstructions in-distribution. The proposed manifold-based diffusion model \Mmanifold\ consistently suppresses fine-scale numerical artifacts that arise in penalty-based or autoregressive formulations, achieving the lowest pointwise errors and the cleanest spectral behavior at short and moderate horizons. These results demonstrate that embedding physical constraints directly into the geometry of the generative process is more effective than enforcing them through auxiliary losses or penalties, particularly when local accuracy and structural fidelity are critical. At the same time, our experiments reveal a clear trade-off between strict and soft constraints as predictions are propagated in time or exposed to distributional and statistical metrics. In extended rollouts, the advantages of hard-constrained models gradually diminish as errors accumulate, whereas methods derived from the unconstrained vanilla EDM model, combined with divergence-free correction mechanisms, yield more accurate statistical behavior. In this context, predictor–corrector strategies such as \Mcorrector\ provide an interesting compromise, achieving perfect incompressibility while preserving amplitude statistics and visual coherence. Out-of-distribution evaluations further confirm that no single constraint strategy dominates across all regimes. For moderately shifted configurations with strong geometric or symmetry constraints, strict manifold-based enforcement remains competitive and can even be advantageous. In contrast, for more severe distribution shifts, softer constraints offer the flexibility required to adapt to unseen dynamics, while overly rigid formulations tend to overfit and lose variability. These observations highlight that the effectiveness of physical constraints comes in pair with the degree of distributional overset encountered during inference.

Taken together, our findings establish a design principle for diffusion-based flow modeling: strict constraints are more effective near the training distribution and at short horizons, where precision and smoothness are critical, while softer constraints, or unconstrained models augmented with principled corrections, become preferable for long-horizon prediction and robust out-of-distribution generalization. Rather than competing paradigms, these approaches are complementary and should be combined depending on the target regime. While the present study is restricted to periodic domains and single-field velocity prediction, the proposed framework is general and extensible. Extending it to non-periodic geometries, coupled multi-field prediction, and more complex physical settings represents a natural next step. More broadly, viewing diffusion models through the lens of probability flow ODEs reveals deep conceptual connections with fluid dynamics itself, offering a promising foundation for future diffusion-based approaches in computational fluid mechanics.

\clearpage
\appendix

\section{Numerical methodology} \label{APP:SNSE_digest}
\subsection{Galerkin-Fourier simulation of a periodic flow} \label{APP:SNSE_digest_FOURIER}

The simulation of periodic flows allow us to work with the Fourier exponentials basis $\stokesbasis_\ddim$ of \sref{sec:incompressible}, in pair with Fourier expansions of the operators and vector fields \cite{CAN88}. The projected Stokes operator $\stokes_\Ndim$ in \eref{eq:FourierNSE} being simply diagonal in $\stokesbasis_\ddim$, one may write:
\begin{equation*}
\viscosity\stokes_\Ndim \uv_\Ndim(t) = -\viscosity \sum_{\kv \in \kset_\Ndim} \norm{\kv}^2 \uv_\kv(t) \ef_\kv \,.
\end{equation*}
Besides, the finite-dimensional convective operator can be seen as a convolution within the Fourier domain with the Leray projection \eqref{eq:Leray-Fourier}:
\begin{equation*}
\convect_\Ndim(\uv_\Ndim)(t) = \sum_{\kv \in \kset_\Ndim}\leray_\kv \left(\sum_{\pv+\qv=\kv} 
(\ci\qv\cdot \uv_\pv(t))\uv_\qv(t) \right) \ef_\kv\,.
\end{equation*}
In this regard, we observe that such a convolution spans a set $\smash{\kset_{2\Ndim}}$, larger than the initial one $\smash{\kset_\Ndim}$ of cardinality $\smash{\#\kset_\Ndim} = \smash{\demi((2\Ndim+1)^\ddim-1)}$. Apart from setting $\Ndim$ below the Nyquist limit, \emph{e.g.} $\Ndim < \smash{\frac{L}{2\Delta\rj}}$ for $\Delta\rj$ being the spatial discretization step, we must also compute the convective term onto a zero-padded set to support the full-convolution without increasing the number of modes \cite[Chapter 3]{CAN88}. We define the zero-padded spectrum of the velocity field $\uv$ with Fourier coefficients $\smash{\{\tilde{\uv}_\kv\}_{\kv\in\kset_M}}$ with $\Ndim_\text{pad}=\Ndim+\demi\Ndim$ and $M=\left\lceil N_\text{pad}\right\rceil_\text{even}$ (rounded to the nearest even integer) by:
\begin{equation*}
\tilde{\uv}_\kv =
\begin{cases}
\uv_\kv & \text{if } \kv \in \kset_\Ndim\,, \\
\zerov & \text{if } \kv \in \kset_M \setminus \kset_\Ndim \,.
\end{cases}
\end{equation*}
Then the convective term is computed for the padded field $\tilde{\uv}_\Ndim(t)=\sum_{\kv\in\kset_M}\tilde{\uv}_\kv(t)\ef_\kv$ with formal discrete difference methods. The full time-marching scheme reads at time step $t_j=j\Delta t$, $j\in\Nset$:
\begin{equation}\label{eq:explicit_Euler}
\begin{cases}
\uv_\kv^{(j+1)} = \uv_\kv^{(j)} -
\left((\viscosity \norm{\kv}^2 + \friction_\kv)\uv^{(j)}_\kv +\scalar{\convect_\Ndim(\tilde{\uv}^{(j)}_\Ndim)}{\ef_\kv}\right) \Delta t\,,\\
\uv^{(0)}_\kv=\displaystyle\frac{\ci\kv^\perp}{\norm{\kv}^2}\FFT{\vortj}_F(\kv)\,,
\end{cases}
\end{equation}
\emph{i.e.} a simple (yet efficient) Euler scheme for $\smash{\uv_\kv^{(j)}}\simeq\smash{\uv_\kv(t_j)}$, provided a small timestep $0<\Delta t \ll \frac{\Delta\rj^2}{\viscosity}$ is considered. Here $\FFT{\vortj}_F(\kv)$ is the filtered initial vorticity field in Fourier space sampled according to \eref{eq:IC}. 
\subsection{Simulation and machine learning parameters} \label{APP:SNSE_digest_PARAMS}
In this appendix, we list the parameters used for the simulation and data-processing throughout the paper. The numerical simulation global parameters are gathered in \tref{tab:simu_param}. Parameters for the database generation are gathered in \tref{tab:IC_param}.

\begin{table}[h!] 
\centering 
    \begin{tabular}{llc}
    \hline
    \textbf{Symbol} & \textbf{Definition} & \textbf{Value} \\
    \hline
    $L$ & Domain side length & $2 \pi$ \\
    $\ndim$ & Spatial discretization & $512^2$ \\
    $\Ndim$ & Frequency cut-off ($\ell^\infty$ frequency bound) & 200 \\
    $\Delta t$ & Time discretization step & $1.0 \times 10^{-4}$ \\
    $\viscosity$ & Kinematic viscosity & $5.0 \times 10^{-4}$ \\
    $\kj_\text{d}$ & Drag force wavenumber (maximum range) & $2$ \\
    $\friction$ & Friction coefficient & $3.75 \times 10^{-4}$ \\
    \hline
    \end{tabular}
\caption{Numerical parameters of Fourier-Galerkin simulation scheme.}
\label{tab:simu_param}
\end{table}

\begin{table}[h!]
\centering
\begin{tabular}{lll}
\hline
\textbf{Symbol} & \textbf{Description} & \textbf{Value} \\
\hline
    $H \times W$ & Spatial resolution of velocity field (subsampled from simulation) & $256^2$ \\
    $\Ndim'$ & Frequency cut-off (subsampled from simulation) & $N/2$ \\
    $[k_\ikf, k_\nu]$ & Initial energy spectrum range & $[4,\,13]$ \\
    $\NIC$ & Number of random simulation seeds drawn for the training base & $10$ \\
    $\Nframes$ & Number of frames regularly sampled from a simulation (train set) & $100$ \\
    $\MIC$ & Number of random simulation seeds drawn for in-distribution and rollout test base & $3$ \\
    $\MS$ & Number of repeated diffusion model generation calls (different  noise seeds) for the same task & $5$ \\
    $A$ & Amplitude (in-distribution, rollout, long rollout) & $0.25$ \\
    $A_{\mathrm{VS}}$ & Vortex superposition amplitude & $0.45$ \\
    $A_{\mathrm{TGV}}$ & Perturbed Taylor-Green vortex amplitude & $0.30$ \\
\hline
\end{tabular}
\caption{Numerical parameters used for dataset and machine learning applications.}
\label{tab:IC_param}
\end{table}

\section{EDM framework: training objective and implementation details} \label{APP:EDM_loss}

Here we give further details on the implementation of the EDM pipeline in \sref{sec:EDM}. We basically follow the framework of \cite{karras2022elucidating}. The sampling distribution $\log\noise\sim \Normal(\sigma_\text{m},\Sigma^2)$ is considered in the EDM training loss \eqref{eq:EDM_loss} (focusing on intermediate noise levels with $\sigma_\text{m} = -1.2$ and $\Sigma = 1.2$). The weighting function is:
\begin{equation}\label{eq:weighting}
\lambda(\noise)=\frac{1}{\noise^2}+\frac{1}{\sigma_\idata^2}\,,
\end{equation}
for $\sigma_\idata^2$ being the variance of the data. The sampling procedure \eqref{eq:prediction} is implemented along with:
\begin{equation*}
\noise_i = \left( \noise_\imax^\frac{1}{\rho} + \frac{i}{\nstep-1} ( \noise_\imin^\frac{1}{\rho} - \noise_\imax^\frac{1}{\rho} ) \right)^\rho\,,
\end{equation*}
with $\smash{\noise_\nstep}=0$, fixing $\rho = 7$, $\smash{\sigma_\imin}=2\cdot10^{-3}$, and $\smash{\sigma_\imax}=80$ for normalized data. Here $\rho$ controls how much the steps near $\smash{\noise_\imin}=\smash{\noise_{\nstep-1}}$ (late backward iterates) are shortened compared to the longer ones near $\smash{\noise_\imax}=\smash{\noise_0}$ (early backward iterates).

With this setting, the number of generative discretization steps $\smash{\{\noise_i\}_{1\leq i\leq T}}$ is reduced to as few as $T=25$, which is about $40$ times fewer than some earlier implementations \cite{DDPM,Song2019,SMSDE}. To achieve accuracy and smoothness comparable to, or even surpassing, the classical approach, we introduce a small number of correction steps near the end of the backward iterates when the denoiser operates at the finest scales. Empirically, we found that the mean-shift corrector is by far the most efficient and robust among the classical options (\emph{e.g.} Langevin, annealed Langevin, \emph{etc}.). The following corrector is thus applied for a few iterations towards the end of the backward process:
\begin{equation}\label{eq:mean_shift_corrector}
\Ubwd_i \leftarrow \Ubwd_i + \eta\big(- \Ubwd_i + \denoiser_{\opt{\parav}}(\Ubwd_i;\noise_i)\big) \,,
\end{equation}
where $\eta=0.05\times\noise_i^2/2$, and it is repeated $\alpha_\text{R}$ times during the last $\alpha_\text{T}$ discrete noise scales $\noise_i$ of the backward ODE integration. The above corrector can be interpreted as a deterministic counterpart of the classical Langevin corrector \cite{hyvarinen2024noise}:
\begin{equation*}\label{eq:Langevin_corrector}
\Ubwd_i \leftarrow \Ubwd_i + \eta \nablav_\uv \log\pdf_{\dift_i}(\Ubwd_i) + \sqrt{2\eta}\,\lvec{Z} \,,\quad\lvec{Z}\sim\Normal(\zerov,\Iv)\,,
\end{equation*}
of which purpose is to nudge samples toward high-likelihood regions of the marginal distribution $\smash{\pdf_\dift(\uv)}$.


Alternatively the denoiser is parameterized by a preconditioned neural network $\bs{F}_\parav$ with a noise-dependent skip connection such that:
\begin{equation}\label{eq:skipped-denoiser}
\denoiser_\parav(\uv;\noise) = c_\text{skip}(\noise)\uv + c_\text{out}(\noise)\bs{F}_\parav(c_\text{in}(\noise)\uv;c_\text{noise}(\noise))\,.
\end{equation}
Skip connections modulate the denoiser output. Based on the value of the coefficients, either noise or input data can be predicted, as well as any noisy version in between. This choice helps to carefully modulate the amplitude of the noise to be removed at each diffusion time step. The training loss becomes:
\begin{equation*}
\Loss_\text{EDM}(\parav)=
\esp_{\uv_0\sim\measure_\idata}
\esp_{\noise\sim\pdf(\noise)}
\esp_{\zv\sim\Normal(\zerov,\Iv)}
\Big[
\lambda(\noise)c_\text{out}(\noise)^2\norm{\bs{F}_\parav\big(c_\text{in}(\noise)(\uv_0+\noise\zv);c_\text{noise}(\noise)\big)- \bs{F}_\text{target}(\uv_0,\zv;\noise)}^2
\Big]
\,
\end{equation*}
where the effective training target $\bs{F}_\text{target}$ is:
\begin{equation*}
\bs{F}_\text{target}(\uv_0,\zv;\noise) = \frac{1}{c_\text{out}(\noise)}\big(\uv_0-c_\text{skip}(\noise)(\uv_0+\noise\zv)\big)\,,
\end{equation*}
and the functions $c_\text{skip}$, $c_\text{out}$, $c_\text{in}$, and $c_\text{noise}$ are:
\begin{equation*}
c_\text{skip}(\noise) = \frac{\sigma_\text{data}^2}{\noise^2 + \sigma_\text{data}^2 }\,,
\quad c_\text{in}(\noise) = \frac{1}{\sqrt{\noise^2 + \sigma_\text{data}^2 }}\,, 
\quad c_\text{out}(\noise) = \frac{\noise\sigma_\text{data}}{\sqrt{\noise^2 + \sigma_\text{data}^2}}\,, 
\quad c_\text{noise}(\noise) = \frac{1}{4} \log\noise\,.
\end{equation*}
The weighting function $\lambda$ is the one given by \eref{eq:weighting}. These choices ensure that the training inputs of $\bs{F}_\parav$ have unit variance: $\Var[c_\text{in}(\noise)(\uv_0+\noise\zv)]=1$; the effective training target $\bs{F}_\text{target}$ has unit variance alike; $c_\text{skip}$ minimizes $c_\text{out}$, so that the errors of $\bs{F}_\parav$ are amplified as little as possible; and the weighting function $\lambda(\noise)c_\text{out}(\noise)^2=1$ for all noise levels $\noise$. In other words $\smash{c_\text{skip}}$ and $\smash{c_\text{noise}}$ weight the skip connection and noise levels, respectively, whereas $\smash{c_\text{in}}$ and $\smash{c_\text{out}}$ scale the input and output magnitudes. 

\section{Correctors for diffusion models}\label{APP:correctors}
\subsection{Mitigating stiff artifacts in long-horizon rollouts}\label{APP:HF_corrector}




As shown in \sref{sec:ID_results} and \sref{sec:rollout_results_short}, artifacts arise during backward integration and are amplified by autoregression. We therefore introduce a correction strategy based on predictor–corrector sampling, inspired by transport consistency and gradient filtering in numerical fluid modeling and signal processing. In score-based generative models, predictor–corrector schemes stabilize reverse-time sampling by enforcing consistency with backwards dynamics \cite{SMSDE}, a principle also supported by multi-step and transport-aware SDE solvers \cite{kloeden1977numerical}. Related ideas appear in semi-Lagrangian fluid schemes, where past state aligns updates with transport directions \cite{filbet2016high}. To suppress ripple artifacts, we draw inspiration from total variation denoising \cite{rudin1992nonlinear}, and gradient-domain filtering, which attenuate spurious high-frequency oscillations while preserving structure through time displacement \cite{brox2004high} (anisotropic diffusion/optical flow techniques). Together, these works motivate our use of past-state projection for transport alignment and gradient smoothing for artifact attenuation.

\paragraph{Observations and prior}

Empirically, we observe spatially localized artifacts (or patches) that appear visually \emph{opposed} to the flow, clashing with the expected continuous transport motion. These errors manifest as stripe-like oscillations, visually high-frequency compared to larger coherent structures, and tend to appear at unexpected times during autoregression, where they are further amplified once present. Although localized, they often span a non-negligible portion of the domain and thus cannot be regarded as simple transport errors, but rather as the result of imperfect evaluations. From a physical standpoint, transport equations generate spatially coherent dynamics, with no spurious localized events unless externally forced. In hyper-viscous and decaying regimes, the dynamics are dominated by large-scale structures (\emph{e.g.} the growth of vortices in \fref{fig:curlRollout}) and evolve smoothly through gradual increments; abrupt, localized stiff bursts violate this behavior. In the diffusion model setting, the learned score is inherently imperfect, and reverse-time integration and autoregressive sampling amplify errors over time, particularly at low noise levels where fine-scale details dominate. However, the combination of diffusion and autoregression provides access not only to past states but also to the underlying dynamics encoded by the score, enabling transport-consistent correction and selective suppression of dynamically inconsistent artifacts.

\paragraph{Key deductions and correction mechanism} 

If the model is (to a certain extent) accurate and reconstruction is objectively coherent, the instantaneous state increment:
$$
\lvec{\delta}^{(j)}_\noise
:= \big(\Ubwd_\noise \cond \opt{(\uv^{(j)})}\big) - \opt{(\uv^{(j-1)})}
$$
evaluated at low noise level $\noise$ is dominated by the transport direction, with a small residual error. In this regime, the dominant component is low-frequency, while residual noise from imperfect score estimation is typically high-frequency and amplified during reverse integration. Hence, projecting the score via:
$$
\bs{\Pi}_\delta^{(j)}:= \frac{\lvec{\delta}^{(j)}_\noise\otimes\lvec{\delta}^{(j)}_\noise}{\big\|\lvec{\delta}^{(j)}_\noise\big\|^2 + \varepsilon}\,,
$$
with some small $\varepsilon>0$, aligns the dynamics with the transport-consistent direction encoded by $\smash{\lvec{\delta}^{(j)}_\noise}$ and suppresses incoherent components. This projection is valid once the update becomes transport-dominated, \emph{i.e.}, at sufficiently low noise $\noise$ near the end of backward integration of \eref{eq:EDM-PF-ODE3-cond}. Furthermore, spurious high-frequency bursts along this direction are detected using the relative gradient growth rather than its absolute magnitude, since sharp structures may be physical. We thus define localized gradient norms:
$$
G^{(j)}_\noise = \norm{\bnabla\big(\Ubwd_\noise \cond \opt{(\uv^{(j)})}\big)   - \bnabla\opt{(\uv^{(j-1)})} }^2 \,, \quad 
G_\text{ref} = \norm{\bnabla\opt{(\uv^{(j-1)})}}^2 \,,
$$
where $\norm{\cdot}$ is taken pixel-wise, as well as a smoothed mask $\smash{ \bs{m} := \Gamma[\bft{1}(G^{(j)}_\noise > G_\text{ref})]}$, where $\bft{1}$ is a binary indicator and $\Gamma$ a smoothing kernel to avoid reintroducing sharp gradients \cite{brox2004high}. Then we introduce the second projector:
$$
\bs{\Pi}_G^{(j)} := \Iv - \gamma \bs{m} \,,
$$
which selectively attenuate (to a factor $0< 1-\gamma <1$, where $\gamma$ is set to $0.8$) spurious, non-transport high-frequency artifacts without harming coherent structures. Finally, as observed in \cite{karras2022elucidating,song2023consistency}, increasing task difficulty (such as forecasting a long rollout) amplifies the benefit of stochasticity during backward integration. Noise injection introduces controlled thermal energy that enables re-exploration, thereby improving diversity and robustness during this stiff regime. This stochasticity selectively restores effective degrees of freedom without reintroducing numerical instabilities. Overall, the correction step \eqref{eq:mean_shift_corrector} now reads (for the backward integration step $i$ on the physical frame $j$):
$$
\Ubwd_i \leftarrow \Ubwd_i + \eta\cdot \bs{\Pi}_G^{(j)} \circ \bs{\Pi}_\delta^{(j)} \circ \score_{\opt{\parav}}(\Ubwd_i ;\noise_i) + \sqrt{2\eta} \cdot \bs{\Pi}_\delta^{(j)} \lvec{Z} \,, \quad\lvec{Z}\sim\Normal(\zerov,\Iv)\,,
$$
and it is repeated, just like in \ref{APP:EDM_loss}, $\alpha_\text{R}$ times during the last $\alpha_\text{T}$ scales. 
Here $\eta \propto \noise_i^2/2$ \cite{hyvarinen2024noise}, the same parameter as the one used in the mean-shift corrector \eqref{eq:mean_shift_corrector}, and $\score_\parav(\uv ;\noise)=\smash{\frac{1}{\noise^2}(- \uv + \denoiser_\parav(\uv;\noise))}$.
In effect, the dynamics are constrained to a coherent, autoregressive evolution direction. Gradient-based masking selectively damps localized high-frequency tangent modes that persist after projection, while smoothing enforces a locality prior reflecting the localized nature of artifacts. Noise then recovers global flexibility without destabilizing the dynamics.

\subsection{Out-of-distribution energy regulation in autoregressive correction}\label{APP:Energy_corrector}
In order to prevent amplitude blow-up resulting from autoregressive error accumulation, a simple yet effective kinetic-energy corrector can be implemented. Apart from the governing PDE, no additional information is assumed when predicting the system behavior on OOD inputs. The corrector solely enforces consistency with the established physical laws, without introducing any external or non-physical priors. This design is implemented by first computing the energy dissipation rate, ignoring the bilinear convective operator $\Convect$ which is conservative and does not modify the total kinetic energy. One first has:
\begin{equation*}
\begin{split} 
\frac{\id\uv_\kv}{\id t} 
& = -\viscosity \norm{\kv}^{2} \uv_\kv(t) + \fv_\kv(t) + \text{(conservative convective terms)}\\
& = -(\viscosity \norm{\kv}^{2} + \friction_\kv)\uv_\kv(t) \\
& = -\id_\kv \uv_\kv(t)  \,,
\end{split}
\end{equation*}
where $\friction_\kv$ is the friction coefficient \eqref{eq:friction_k} in Fourier space. Then, take the modal energy $\smash{E_\kv} = \smash{\demi\norm{\uv_\kv}^2}$ and compute the instantaneous change in time:
\begin{equation*}
\begin{split}
\frac{\id E_\kv}{\id t} &= \uv_\kv \cdot \frac{\id\uv_\kv}{\id t}\\
&= -\id_\kv \norm{\uv_\kv}^2\\
&= -2 \id_\kv E_\kv \,,
\end{split}
\end{equation*}
and therefore, the target modal energy at the time step $j+1$ from the modal energy $\smash{E^j_\kv}=\smash{\opt{E}_\kv(j\Delta t)}$ at the time step $j$ is $\smash{E_\kv^\text{target}}=\smash{E_\kv^j\exp(-2 \id_\kv \Delta t)}$; here $\opt{E}_\kv=\demi\norm{\opt{\uv}_\kv}$ where $\opt{\uv}$ is again the predicted velocity field. Letting:
$$
\gamma_\kv=\sqrt{\frac{E_\kv^\text{target}}{E_\kv^{j+1}}}\,,
$$
the corrector step is define as follows (only targeting deviations of the energy increments with a $\pm 5\%$ rate, potentially keeping defect signature of each method):
\begin{equation}\label{eq:clip}
\opt{(\uv_\kv^{(j+1)})}\leftarrow\operatorname{clip}(\gamma_\kv, 0.95, 1.05)\times\opt{(\uv_\kv^{(j+1)})}\,.
\end{equation}
As a side note, the data being normalized prior to being input in the model \emph{i.e.}:
$$
\uv_\text{normalized} = \frac{\uv-\overline{\uv}}{\overline{\norm{\uv-\overline{\uv}}_2}}\,,
$$
where $\overline{\uv}$ stands for the empirical average of the data frames in $\smash{\Rset^\ndim}$, which is $\overline{\uv}\approx\zerov$ from isotropy and zero-mean field property, we apply the filtering \eqref{eq:clip} for frames $\uv$ in physical units. 

\section{Machine learning configuration and hyperparameters} \label{APP:ML_params}

Below is a list of the parameters used for the machine learning experiments, as well as the correctors. As a side note, whenever an optimization is required (training and sampling), the vanilla method \Mvanilla\ is always optimized on the first prediction frame. Once the vanilla method has reached its optimal parameters, all other methods retain these very parameters. This approach ensures that all comparisons are conducted fairly by prioritizing the baseline model and allows discrepancies relative to the vanilla method to be clearly distinguished.

\paragraph{Machine learning, training phase} \leavevmode
\begin{table}[htbp]
\centering
\begin{tabular}{lll}
\hline
\textbf{Variable name (numerical)} & \textbf{Description} & \textbf{Value} \\
\hline
$\texttt{epochs}$ & Number of training epochs & $750$ \\
$\texttt{lr}$ & Initial learning rate & $3.0 \times 10^{-4}$ \\
$\texttt{batch\_size\_train}$ & Training batch size & $12$ \\
$\texttt{batch\_size\_valid}$ & Validation batch size & $5$ \\
$\texttt{stop}$ & Early stopping patience (iterations) & $120$ \\
$\texttt{patience}$ & Learning rate scheduler patience & $25$ \\
$\texttt{factor}$ & Learning rate decay factor & $0.75$ \\
$\texttt{precision}$ & Numerical precision (bits) & $32$ \\
$\texttt{clipVal}$ & Gradient clipping threshold & $1.0$ \\
\hline
\end{tabular}
\caption{Training hyperparameters and optimization settings used for experiments.}
\end{table}
While training can be performed on a typical workstation, equipped with a multi-core CPU, 16–32 GB of system RAM, and a single consumer-grade GPU, the training phase for the multiple models is conducted on a high-performance computing node. This node is equipped with 96 GB of system RAM and two NVIDIA HGX A100 GPUs, each providing 80 GB of HBM2e VRAM.

\paragraph{Machine learning, sampling phase} \leavevmode
\begin{table}[htbp]
\centering
\begin{tabular}{lcc}
\hline
\textbf{Experiment section} & $\alpha_\text{R}$ (repetitions) & $\alpha_\text{T}$ (tail) \\
\hline
In-distribution (\sref{sec:ID_results}) & $2$ & $5$ \\
Short-horizon rollout (\sref{sec:rollout_results_short}) & $2$ & $5$ \\
Long-horizon rollout (\sref{sec:rollout_long}) & $5$ & $10$ \\
Out-of-distribution (\sref{sec:OOD_res}) & $4$ & $8$ \\
\hline
\end{tabular}
\caption{Corrector parameters specific to evaluation regimes.}
\end{table}
The sampling phase is executed on a standalone computer, featuring 32GB of system RAM, an AMD Ryzen 7 3700X processor, and an ASUS NVIDIA GeForce RTX 3090 GPU (24GB of GDDR6X VRAM). The sampling process can also be performed on a more modest workstation, provided that a GPU and at least 16 GB of system RAM are available.

\section{Model design and architecture} \label{APP:model_design}

In this appendix, we describe the main architectural choices of the neural network which has been implemented, emphasizing its differences from the standard U-Net \cite{UNet,UNet3D} and its use in an EDM-style framework.

\subsection{Conditional generation}

The autoregressive setup enables the generation of a complete, coherent flow time series by requiring each predicted step $j$ to be conditioned on the previous step $j-1$. This is achieved through the use of a concatenated input:
\begin{equation*}
\xv_\text{in} = \op{Concat}\big[\Ufwd_\dift\cond\Ufwd_0=\uv^{(j)}, \uv^{(j-1)}\big] \in \Rset^{B\times 2C\times H \times W} \,,
\end{equation*}
where $B$ is the batch size, $C$ is the channel size (or state dimension, $C=2$ in our case since the fluid medium is the torus $\torus^2$), $H$ and $W$ stand for the height and width of the discretized field, respectively (with $2\times H \times W=\ndim$ since $\smash{\Ufwd_0} \in \smash{\Rset^\ndim}$).

\subsection{EDM noise embedding and low-$\noise$ refinement}

The noise level $\noise$ being key to the sample's inner diffusion step and gradient strength, we embed it using log-Fourier features:
\begin{equation*}
\begin{aligned}
    & \bft{e}_t  = \op{MLP}
    \Big(
    \big[ \log\noise,
    (\log\noise)^2,
    \sin(\omega_k \log\noise),
    \cos(\omega_k \log\noise)
    \big]_{k=1,\dots,K}
    \Big)\,, \\
    & \omega_k = \exp{\parenthesis{\op{linspace}(\log 0.1, \log 1000, K})} \,,
\end{aligned}
\end{equation*}
where $\op{MLP}$ stands for Multi-Layer Perceptron, which provides stable conditioning across the full noise range. To improve predictions in the low-noise limit, we also apply a gated refinement on the earliest high-resolution skip:
\begin{equation*}
    \yv_{\text{ref}}=\yv_{\text{out}} + g(\bft{e}_t) \cdot \op{CU}(\xv_{\text{skip}}, \bft{e}_t)\,,
\end{equation*}
where $\smash{\yv_\text{out}}$ is the output of the entire U-Net before refinement; $\smash{\xv_\text{skip}}$ is the tensor taken before downsampling, at the earliest stage of processing (U-net's main down/up-sampling pipeline); $g$ is a learned scalar gate controlling refinement strength; and $\op{CU}$ is a small conditional U-Net block. This approach introduces a $\noise$-dependent correction mostly absent from standard U-Net architectures.

\subsection{Condition embedding and periodic boundaries}

The noise embedding $\bft{e}_t$ (conditioning vector) obtained from $\noise$ gets input in each residual block (U-Net's up and down encoder/decoder pipeline), which in turn, are modulated via:
\begin{equation*}
\begin{aligned}
& \op{Block}(\xv \lvert \bft{e}_t) = \xv + \op{Rem} \left(
\op{DSConv}\left(
\op{SiLU}\left(
\op{FiLM}(\xv,\bft{e}_t)
\right)\right), \bft{e}_t \right) \,, \\
& \op{FiLM}(\xv,\bft{e}_t) = \op{GN}(\xv) \cdot (1+\gamma(\bft{e}_t)) + \beta(\bft{e}_t) \,,
\end{aligned}
\end{equation*}
where $\op{Rem}$ encapsulate basic linear operations (convolutions and activations), $\op{DSConv}$ is a Depthwise Separable Convolution, $\op{SiLU}$ is a Sigmoid Linear Unit, $\op{GN}$ is a Group Normalization (proven efficient for diffusion) and $\gamma$, $\beta$ are the scale and shift coefficients from the $\op{FiLM}$ (Feature-wise Linear Modulation \cite{FiLM17}) module learned from $\bft{e}_t$. Classical U-Net blocks use plain $\op{Conv}$–$\op{ReLU}$–$\op{Conv}$, while EDM blocks use $\op{FiLM}$-modulated $\op{GroupNorm}$-$\op{SiLU}$-$\op{DSConv}$. Furthermore, the data being periodic at its boundaries, all $3{\times}3$ depthwise (\emph{i.e.}, each channel) convolutions are built with periodic spatial padding:
\begin{equation*}
\op{DSConv}(\xv)_{b,c,i,j}:=\yv_{b,c}(i,j) = \sum_{u=-1}^{1}  \sum_{v=-1}^{1} w_{c,u,v} \cdot \xv_{b,c}((i+u)\op{mod}[H], (j+v)\op{mod}[W]) \,,
\end{equation*}
followed by a $1{\times}1$ pointwise projection. 
This ensures wrap-around continuity and avoids boundary artifacts typical of zero padding.

\section{Divergence-free diffusion process} \label{APP:div free manifold}

In this section, we provide a simple illustration of how classical diffusion processes can be adapted to the divergence-free manifold, thereby enabling the derivation of the EDM setup of \sref{sec:EDMincompressible}. To this end, we start from the standard infinite-dimensional Ornstein–Uhlenbeck (OU) process and modify it to fit the incompressible flow setting of \sref{sec:incompressible}. This adaptation naturally leads to the formulation of a diffusion framework constrained to divergence-free fields, from which the EDM construction follows.

Let $\Uilbert$ be a separable Hilbert space with scalar product $\scalar{\cdot}{\cdot}$; for instance, 
$\Uilbert=L^2(\medium,\Rset^d)$ the set of square-integrable $\Rset^d$-valued vector fields on the torus $\medium=\torus^d$, which is precisely the functional setting considered in the framework of \sref{sec:Fourier_incompressible}.
Let $\mathcal{G}:\text{Dom}(\mathcal{G}) \subset\Uilbert \to \Uilbert$ be a linear operator, we assume that $\mathcal{G}$ generates a strongly continuous semigroup $\{\iexp^{t\mathcal{G}}\}_{t \geq 0}$ on $\Uilbert$. Also let $\mathcal{Q}\in\mathcal{L}(\Uilbert)$ be a bounded, symmetric, positive non negative operator that is trace-class on $\Uilbert$, then $\mathcal{S}:=\smash{\sqrt{\mathcal{Q}}}$ is a Hilbert-Schmidt operator on $\Uilbert$, so it is an integral operator. Let $(\smash{\Wfwd_t}, t\geq 0)$ denote a cylindrical Wiener process on $\Uilbert$ defined in some filtered probability space $(\Omega,\filtration,\{\filtration_t\}_{t\geq 0},\PP)$. The infinite-dimensional Ornstein-Uhlenbeck (OU) process $(\smash{\Ufwd_t},t\geq 0)$ with values in $\Uilbert$ is then defined by the SDE:
\begin{equation}
\begin{cases}
\id\Ufwd_t = \mathcal{G} \Ufwd_t\id t + \sqrt{\mathcal{Q}}\id\Wfwd_t\,,\\
\Ufwd_0\in\Uilbert \,,
\end{cases}
\end{equation}
which admits a mild solution \cite{da2014stochastic,prevot2007concise}:
\begin{equation}\label{eq:OUinU}
\Ufwd_t = \iexp^{t \mathcal{G}} \Ufwd_0 + \int_0^t \iexp^{(t-s)\mathcal{G}}\sqrt{\mathcal{Q}}\id\Wfwd_s \,.
\end{equation}
It is a Gaussian process in $\Uilbert$ conditionally on $\Ufwd_0$, with conditional mean and variance:
\begin{equation}
\esp[\smash{\Ufwd_t\cond\Ufwd_0}] = \iexp^{t \mathcal{G}} \Ufwd_0\,,
\quad 
\Var[\smash{\Ufwd_t\cond\Ufwd_0}] = \int_0^t\iexp^{(t-s)\mathcal{G}} \, \mathcal{Q} \, \iexp^{(t-s)\mathcal{G}^{*}} \id s \,,
\end{equation}
respectively. Now introduce the closed subspace $\Hdiv$ of $\Uilbert$ of divergence-free fields defined by:
\begin{equation*}
\Hdiv:=\set{\uv \in \Uilbert;\, \nablav\cdot\uv =0} \subset\Uilbert\,,
\end{equation*}
and denote by $\Leray:\Uilbert\to\Hdiv$ the Leray projection operator of \eref{eq:Leray}, which is the orthogonal projection in $\Uilbert$ onto divergence-free fields. 
It satisfies $\Leray^2=\Leray$ as a projector and is bounded, with range space $\text{Ran}(\Leray)=\Hdiv$. For $\uv \in \text{Dom}(\mathcal{G})\cap\Hdiv$ we then define the projected drift operator $\Drift:\text{Dom}(\Drift)\subset\Hdiv\to\Hdiv$ such that:
\begin{equation*}
\Drift\uv := \Leray(\mathcal{G}\uv)\,,
\end{equation*}
and the projected covariance operator $\Diff:\Uilbert\to\Hdiv$ such that:
\begin{equation*}
\sqrt{\Diff}\uv := \Leray(\sqrt{\mathcal{Q}}\uv)\,. 
\end{equation*}
The OU process constrained to $\Hdiv$ thus reads:
\begin{equation}\label{eq:OUdivfree}
\begin{cases}
\id\Ufwd_t = \Drift\Ufwd_t\id t + \sqrt{\Diff}\id\Wfwd_t\,, \\
\Ufwd_0\in\Hdiv\,.
\end{cases}
\end{equation}
It follows that $\smash{\Ufwd_t\cond\Ufwd_0}$ is Gaussian with the same semigroup transition operator as in \eref{eq:OUinU} but restricted to $\Hdiv$, namely $\smash{\{\iexp^{t\Drift}\}_{t\geq 0}}$, and:
\begin{equation}\label{eq:OUinH}
\Ufwd_t = \iexp^{t \Drift} \Ufwd_0 + \int_0^t \iexp^{(t-s)\Drift}\sqrt{\Diff}\id\Wfwd_s \,.
\end{equation}
If $\smash{\Ufwd_0}\in\Hdiv$, then $\smash{\Ufwd_t}\in\Hdiv$ for all $t \geq 0$ almost surely, since both the drift and covariance operators $\Drift$ and $\smash{\Diff}$, respectively, remain tangent to $\Hdiv$.

Now let $\smash{\{\ev_i\}_{i \in \Nset}}$ be an orthonormal basis of $\Hdiv$, and introduce the Galerkin projector $\projector_\Ndim:\Hdiv\to\Hdiv_\Ndim$ onto the finite dimensional subset $\smash{\Hdiv_\Ndim}=\smash{\text{span}\{\ev_{i_1},\dots\ev_{i_\Ndim}\}}$ of $\Hdiv$ defined by:
\begin{equation*}
\projector_\Ndim := \sum_{n=1}^\Ndim \ev_{i_n} \otimes \ev_{i_n}\,,\quad (i_1,\dots i_\Ndim)\in\Nset^\Ndim\,.
\end{equation*}
The $N$-th order Galerkin approximation of the divergence-free OU process is then given by:
\begin{equation}
\begin{cases}
\id\Uv_t^\Ndim = \bft{\drift}_\Ndim\Uv_t^\Ndim\id t + \displaystyle\sqrt{\bft{\diff}_\Ndim}\id\Wfwd_t\,, \\
\Uv_0^\Ndim \in\Hdiv_\Ndim\,,
\end{cases}
\end{equation}
where $\smash{\Uv_t^\Ndim}:=\smash{\projector_\Ndim\Ufwd_t}$, and the drift operator $\bft{\drift}_\Ndim:\Hdiv_\Ndim\to\Hdiv_\Ndim$ and covariance operator $\bft{\diff}_\Ndim:\Uilbert\to\Hdiv_\Ndim$ are defined by:
\begin{equation*}
\bft{\drift}_\Ndim = \projector_\Ndim\Drift\projector_\Ndim^*\,, \quad \sqrt{\bft{\diff}_\Ndim} = \projector_\Ndim\sqrt{\Diff}\,. 
\end{equation*}
The $\Rset^\Ndim$-valued stochastic process $\smash{(\Uv_t^\Ndim,t\geq 0)}$ is again an OU process on $\smash{\Hdiv_\Ndim}$, which is conditionally Gaussian with explicit conditional mean and covariance. 
Galerkin projections have been used to prove the existence of (weak) martingale solutions of SDEs of the form of \eref{eq:OUdivfree} with continuous non-linear drift and additive or multiplicative noise; see \emph{e.g.} \cite{FLA08} or \cite[Chapter 8]{da2014stochastic}. Strong solutions may be constructed alike, which may be approximated by Galerkin expansions in the mean-square sense; see \emph{e.g.} \cite{BRE00}. In summary, the Leray projection ensures that both the drift and diffusion operators remain divergence-free, and that the Galerkin truncation provides a consistent finite-dimensional approximation of the divergence-free OU process. This construction provides with a rigorous basis for the formulation of (latent) diffusion models constrained to incompressible fields.



\section{Additional results: in-distribution assessment}\label{APP:IN_results}
\subsection{Additional metrics} \label{APP:ID_metrics}

Following the notations introduced in \sref{sec:ID_results}, we further consider additional standard error metrics to complement the analysis. In particular, we employ the mean-squared error ($\smash{\op{MSE}}$) and the mean absolute error ($\smash{\op{MAE}}$). For completeness, we also report write their relative variants ($\smash{\op{rMSE}}$ and $\smash{\op{rMAE}}$), which normalize the error with respect to the magnitude of the true values, as well as their relative-root counterparts ($\smash{\op{rRMSE}}$ and $\smash{\op{rRMAE}}$), providing scale-adjusted measures that facilitate comparisons across models and datasets. These metrics are defined in terms of the $\ell^1$ semi-norm $\abs{\cdot}_1$ and the $\ell^2$ norm $\norm{\cdot}_2$ of vectors of $\smash{\Rset^\ndim}$ by:
\begin{equation*}
\begin{aligned}
& \op{MSE}  = \frac{1}{\ndim}\norm{\uv - \opt{\uv}}_2^2 \,,
&&\op{rMSE}  = \frac{1}{\ndim}\frac{\norm{\uv - \opt{\uv}}_2^2}{\norm{\uv}_2^2} \,,
&&&\op{rRMSE} = \sqrt{\op{rMSE}} \,, \\
& \op{MAE} = \frac{1}{\ndim}\abs{\uv - \opt{\uv}}_1 \,, 
&&\op{rMAE}  = \frac{1}{\ndim}\frac{\abs{\uv - \opt{\uv}}_1}{\abs{\uv}_1} \,,
&&&\op{rRMAE} = \sqrt{\op{rMAE}}\,.
\end{aligned}
\end{equation*}
The $\smash{\op{MSE}}$ measures the average squared deviation between predictions and true values, penalizing large errors more strongly. In contrast, the $\smash{\op{MAE}}$ computes the average absolute deviation and deals with all errors uniformly, making it less sensitive to outliers. In addition to classical pointwise metrics such as the MSE, we report the relative amplitude $\text{r}A$, which quantifies the agreement between predicted and reference fields in terms of their global $\ell^2$ norm:
$$
\text{r}A = \frac{\norm{\opt{\uv}}_2}{\norm{\uv}_2 + \varepsilon} \,,
$$
where values closer to $1$ indicate better agreement between the predicted and reference amplitudes.

\subsection{Additional plots} \label{APP:ID_results}

\begin{figure}[h!]
    \centering
    \begin{subfigure}{0.24\textwidth}
        \centering
        \includegraphics[width=\linewidth]{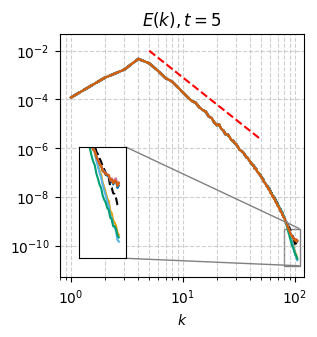}
    \end{subfigure}
    \hfill
    \begin{subfigure}{0.24\textwidth}
        \centering
        \includegraphics[width=\linewidth]{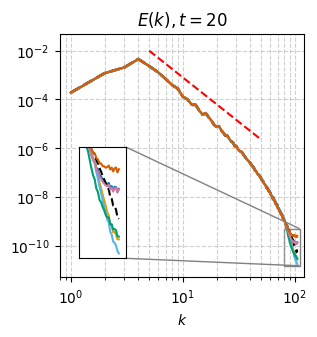}
    \end{subfigure}
    \hfill
    \begin{subfigure}{0.24\textwidth}
        \centering
        \includegraphics[width=\linewidth]{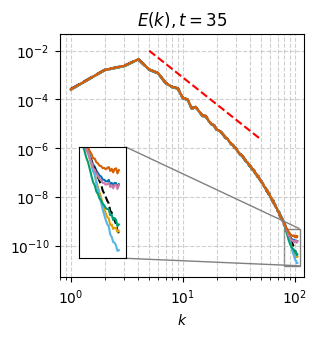}
    \end{subfigure}
    \hfill
    \begin{subfigure}{0.24\textwidth}
        \centering
        \includegraphics[width=\linewidth]{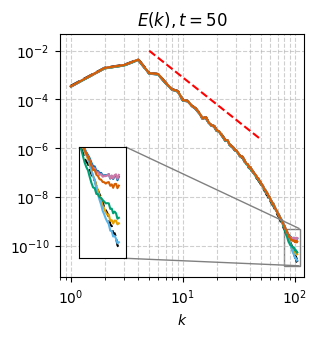}
    \end{subfigure}

    \caption{In-distribution assessment: TKE spectra $\kj\mapsto\TKES(\kj)$ at multiple time instances. Legends:
    \textcolor[HTML]{0072B2}{\rule[1.5pt]{0.3cm}{1pt}} \Mvanilla, 
    \textcolor[HTML]{E69F00}{\rule[1.5pt]{0.3cm}{1pt}} \Mmanifold,
    \textcolor[HTML]{56B4E9}{\rule[1.5pt]{0.3cm}{1pt}} \Mdenoiser,
    \textcolor[HTML]{009E73}{\rule[1.5pt]{0.3cm}{1pt}} \Mresidual,
    \textcolor[HTML]{CC79A7}{\rule[1.5pt]{0.3cm}{1pt}} \Mautoreg,
    \textcolor[HTML]{F04000}{\rule[1.5pt]{0.3cm}{1pt}} \Mcorrector,
    \blackdash\ ground truth spectrum, \reddash\ $\kj^{-3-\delta}$ slope.}
    \label{fig:four_images}
\end{figure}

\begin{figure}[h!]
    \centering
    \includegraphics[width=1.\linewidth]{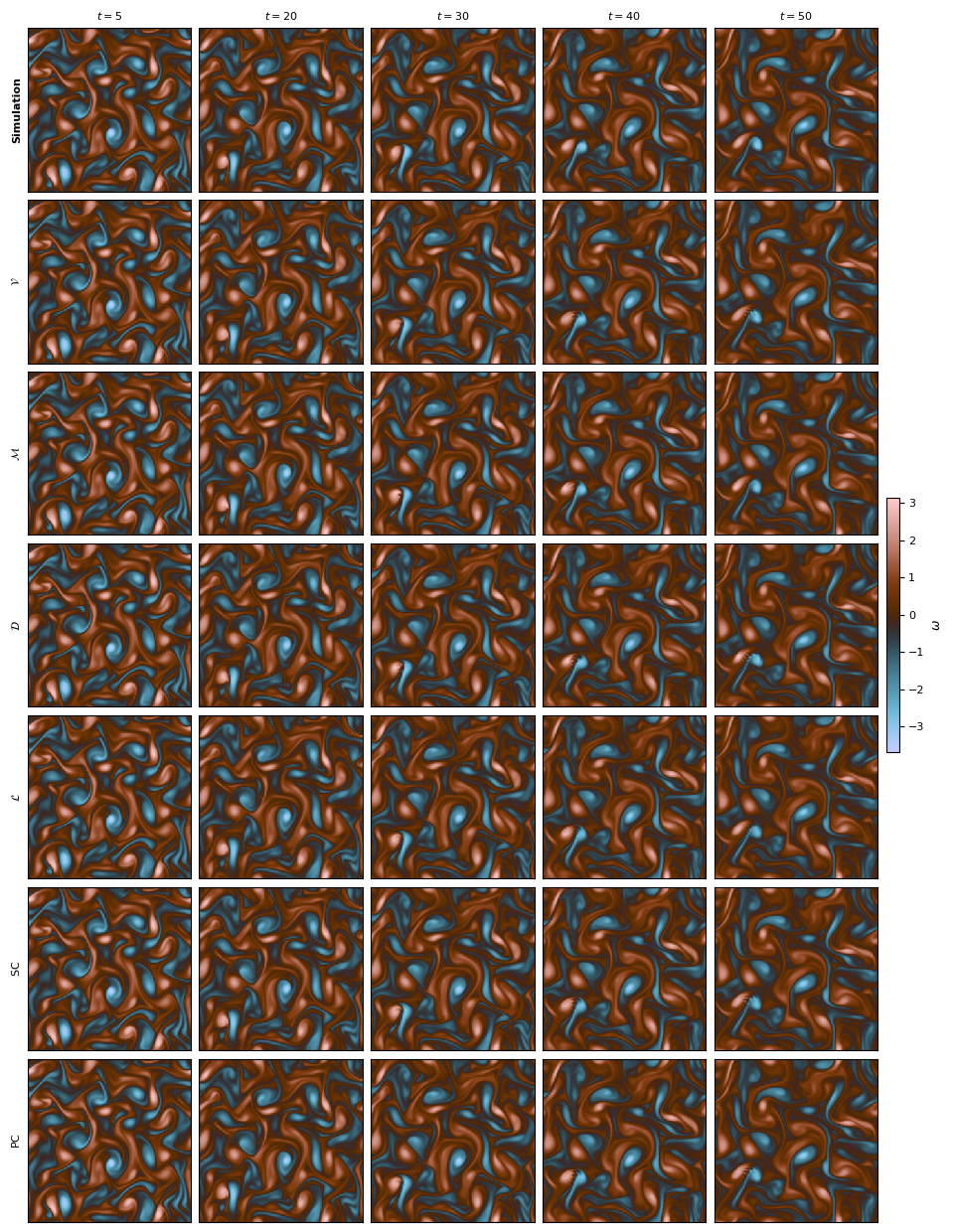}
    \vspace{-2em}
    \caption{In-distribution assessment: snapshots of the ground-truth solution (vorticity, top row) against snapshots inferred by all models \Mvanilla, \Mmanifold, \Mdenoiser, \Mresidual, \Mautoreg, and \Mcorrector.}
    \label{fig:ID_curl_ANNEX}
\end{figure}

In this section, we add a variety of additional plots for the in-distribution results, see \fref{fig:four_images}. We also add the full predictions, see \fref{fig:ID_curl_ANNEX}.

\section{Additional results: rollout predictions} \label{APP:ROLL_results}

\begin{figure}[h!]
    \centering
    \includegraphics[width=\linewidth]{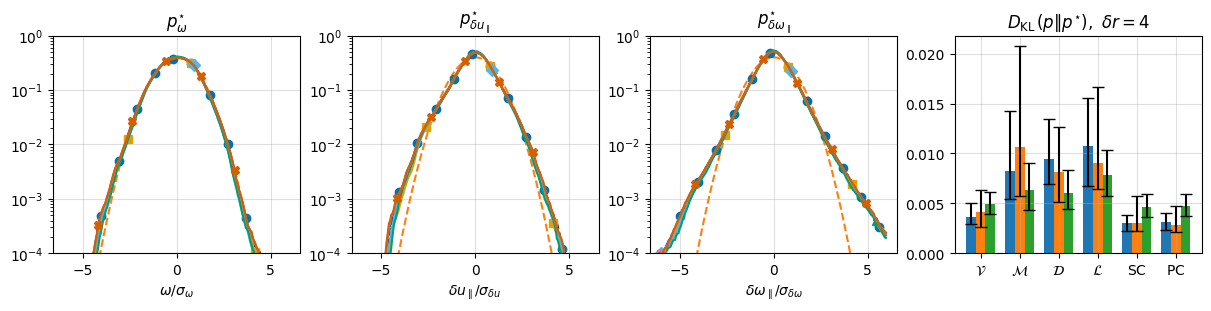}

    \includegraphics[width=\linewidth]{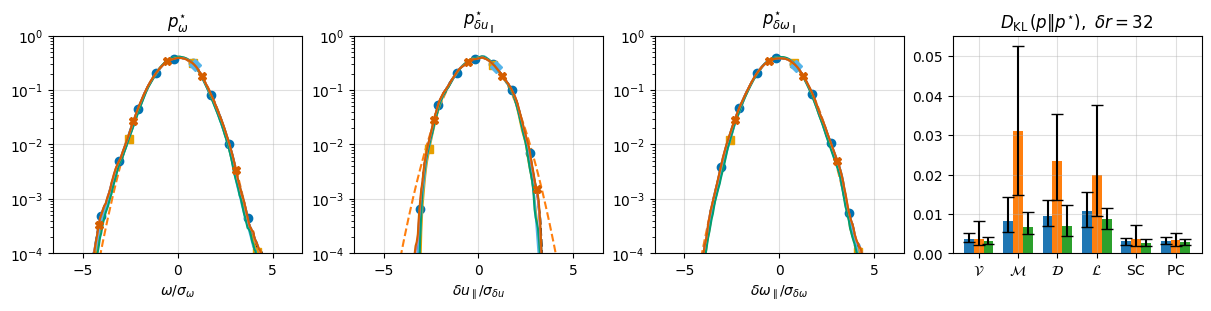}

    \includegraphics[width=\linewidth]{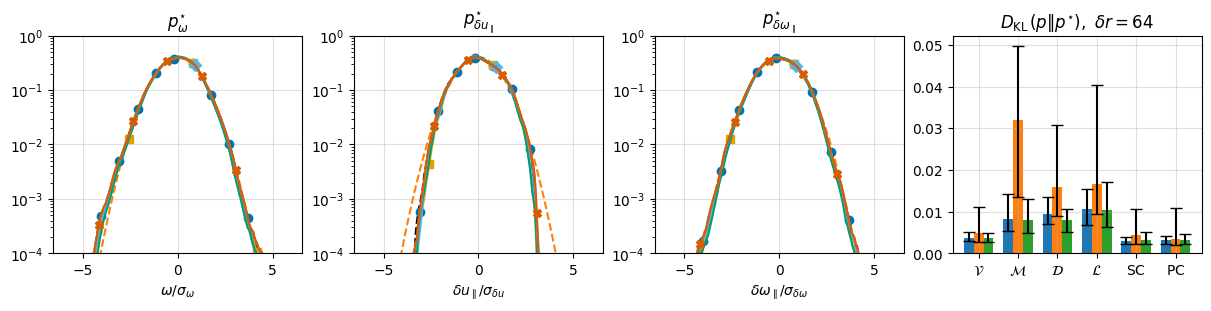}

    \vspace{-.5em}
    \caption{Rollout predictions: empirical densities for $\delta\rj=4$ (top row), $\delta\rj=32$ (middle row), and $\delta\rj=64$ (bottom row). First panel from left to right: empirical density $\opt{\pdf}_\vortj(z;\delta\rj)$. Second panel: empirical density $\opt{\pdf}_{\delta\uj_\parallel}(z;\delta\rj)$. Third panel: empirical density $\opt{\pdf}_{\delta\vortj_\parallel}(z;\delta\rj)$. Fourth panel: Kullback-Leibler divergences. Density plots legends:
    \blackdash\ ground truth, \orangedash\ $\Normal(0,1)$,\,
    \Mvanilla\ \textcolor[HTML]{0072B2}{$\bullet$},\, 
    \Mmanifold\ \textcolor[HTML]{E69F00}{$\blacksquare$},\,
    \Mdenoiser\ \textcolor[HTML]{56B4E9}{$\blacklozenge$},\,
    \Mresidual\ \textcolor[HTML]{009E73}{$\blacktriangle$},\,
    \Mautoreg\ \textcolor[HTML]{CC79A7}{$\blacktriangledown$},\,
    \Mcorrector\ \textcolor[HTML]{F04000}{$\bft{\xv}$}.
    Box plot legends: 
    $D_{\op{KL}}(\pdf_\vortj\lVert\opt{\pdf}_\vortj;\delta\rj)$ \textcolor[HTML]{0072B2}{$\blacksquare$},\,
    $D_{\op{KL}}(\pdf_{\delta\uj_\parallel} \lVert \opt{\pdf}_{\delta\uj_\parallel};\delta\rj)$ \textcolor[HTML]{E69F00}{$\blacksquare$},\,
    $D_{\op{KL}}(\pdf_{\delta\vortj_\parallel} \lVert \opt{\pdf}_{\delta\vortj_\parallel};\delta\rj)$ \textcolor[HTML]{009E73}{$\blacksquare$}. Whiskers indicate the interquartile range.}
    \label{fig:ID_spectras_ANNEX}
\end{figure}

\begin{figure}[h!]
    \centering
    \includegraphics[width=1.\linewidth]{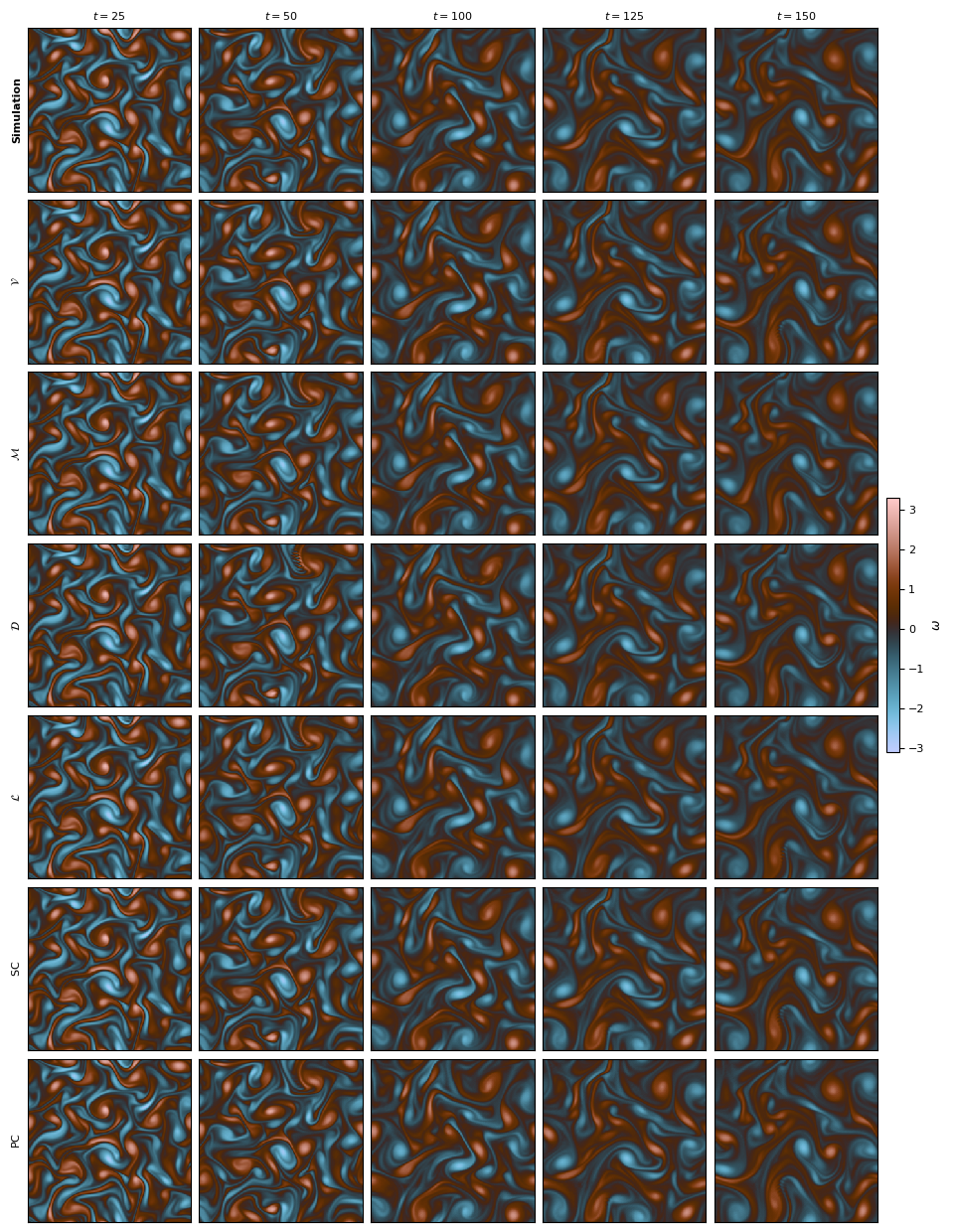}
    \vspace{-2em}
    \caption{Rollout predictions: snapshots of the ground-truth solution (vorticity, top row) against snapshots inferred by all models \Mvanilla, \Mmanifold, \Mdenoiser, \Mresidual, \Mautoreg, and \Mcorrector.}
    \label{fig:ROLL_curl_ANNEX}
\end{figure}

\begin{figure}[h!]
    \centering
    \includegraphics[width=1.\linewidth]{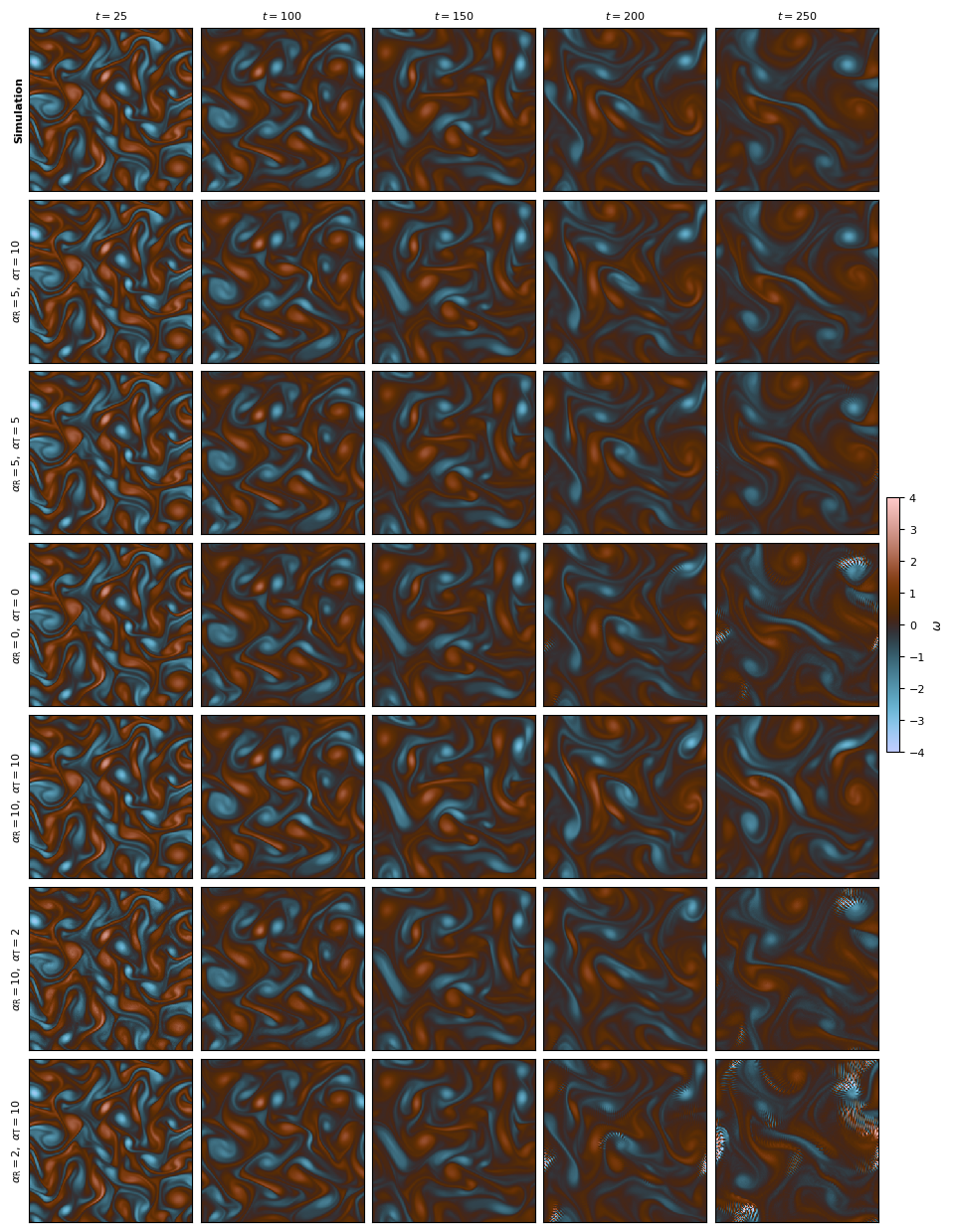}
    \vspace{-2em}
    \caption{Long rollout predictions: snapshots of the ground-truth solution (vorticity, top row) against snapshots inferred by diffusion model on divergence-free manifold \Mmanifold\ using the transport-based corrector with different hyperparameters $\alpha_\text{R}$ and $\alpha_\text{T}$. Small-scale rippling artifacts are observable at long horizons for $\alpha_\text{R}=\alpha_\text{T}=0$ (forth row), $\alpha_\text{R}=10$ and $\alpha_\text{T}=2$ (sixth row), and $\alpha_\text{R}=2$ and $\alpha_\text{T}=10$ (seventh row).}
    \label{fig:ROLL_curl_LONG_ANNEX}
\end{figure}

In this section, we add a variety of plots for the rollout results, see \fref{fig:ID_spectras_ANNEX} and \fref{fig:ROLL_curl_ANNEX}. We also provide the long rollout test curls (vorticities); see \fref{fig:ROLL_curl_LONG_ANNEX}.

\section{Additional results: OOD tests}\label{APP:OOD}

\begin{figure}[h!]
    \centering
    \includegraphics[width=1.\linewidth]{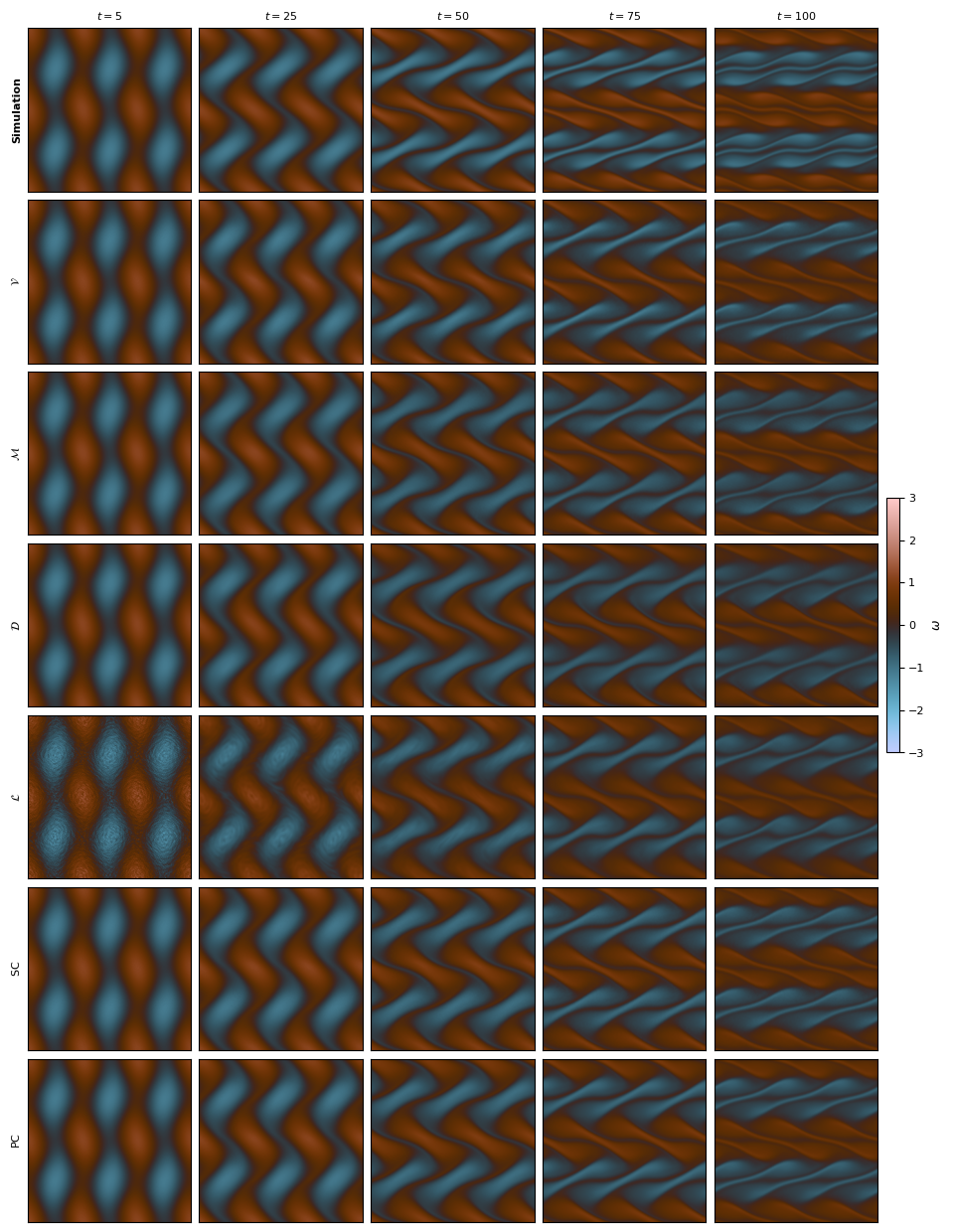}
    \vspace{-2em}
    \caption{OOD tests, VS case: snapshots of the ground-truth solution (vorticity, top row) against snapshots inferred by all models \Mvanilla, \Mmanifold, \Mdenoiser, \Mresidual, \Mautoreg, and \Mcorrector.}
    \label{fig:OOD_curl_ANNEX_VS}
\end{figure}

\begin{figure}[h!]
    \centering
    \includegraphics[width=1.\linewidth]{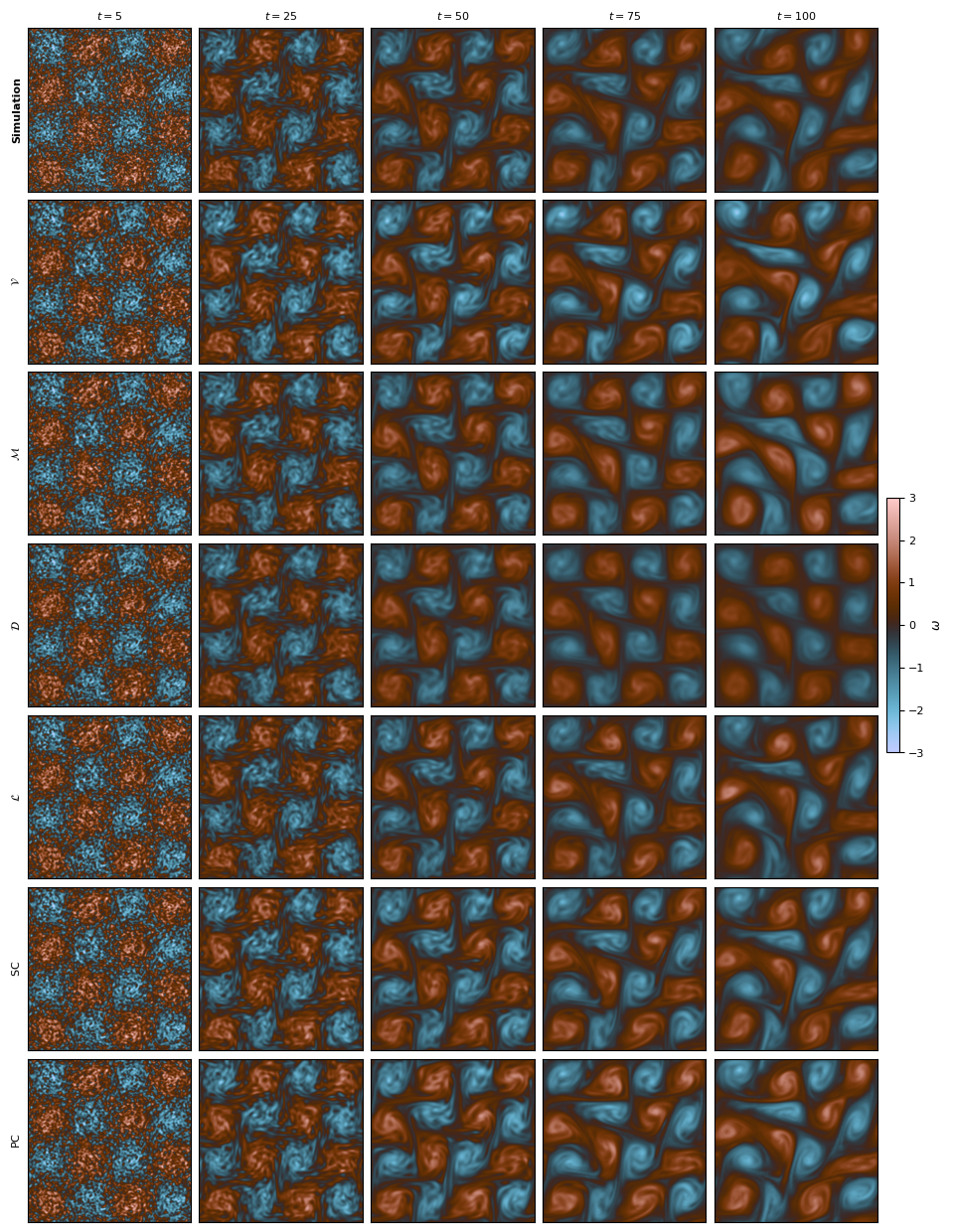}
    \vspace{-2em}
    \caption{OOD tests, TGV case: snapshots of the ground-truth solution (vorticity, top row) against snapshots inferred by all models \Mvanilla, \Mmanifold, \Mdenoiser, \Mresidual, \Mautoreg, and \Mcorrector.}
    \label{fig:OOD_curl_ANNEX_TGV}
\end{figure}

In this section, we add a variety of plots for OOD results, see \fref{fig:OOD_curl_ANNEX_VS} for the VS case and \fref{fig:OOD_curl_ANNEX_TGV} for the TGV case. Interestingly, in this setting, we can observe a small instability at the beginning of the flow prediction for \Mresidual, as seen in \fref{fig:OOD_curl_ANNEX_VS}, which is probably due to a lack of model accuracy (and thus, residual noise).

\clearpage  






\bibliographystyle{unsrtnat} 
\bibliography{external/references}


\end{document}